\documentclass[twocolumn, resetfootnote]{aastex7}

\newcommand{\msun}{M$_{\sun}$}
\newcommand{\rsun}{R$_{\sun}$}
\newcommand{\rstar}{R$_{\star}$}

\newcommand{\mstar}{M$_{\star}$}
\newcommand{\lsun}{L$_{\sun}$}
\newcommand{\lstar}{L$_{\star}$}
\newcommand{\lacc}{L$_{\rm acc}$}

\newcommand{\msunyr}{\msun\,yr$^{-1}$}

\newcommand{\teff}{$T_{\rm eff}$}
\newcommand{\tshock}{$T_{\rm shock}$}
\newcommand{\tmax}{$T_{\rm max}$}
\newcommand{\escm}{{\rm erg/s/cm}$^2$}
\newcommand{\Lhalpha}{L$_{H\alpha}$}
\newcommand{\halpha}{H$\alpha$}

\newcommand{\brgamma}{Br$\gamma$}
\newcommand{\mdot}{$\dot{M}$}
\newcommand{\ri}{$R_{\rm i}$}
\newcommand{\rw}{$W_{\rm r}$}

\def\curf{{\cal F}}
\newcommand{\vff}{$v_{\rm ff}$}
\newcommand{\Fmean}{$\curf_{\rm mean}$}

\newcommand{\vsini}{$v\sin i$}
\newcommand{\hst}{HST} 

\newcommand{\rco}{R$_{\rm co}$}

\newcommand{\omegas}{$\omega_s$}

\newcommand{\prot}{P$_{\rm rot}$}
\newcommand{\av}{$A_V$}
\newcommand{\civ}{\ion{C}{4}}
\newcommand{\siiv}{\ion{Si}{4}}
\newcommand{\heii}{\ion{He}{2}}

\newcommand{\degrees}{$^\circ$}
\newcommand{\chisq}{$\chi^2$}

\newcommand\be {\begin{equation}}
\newcommand\en{\end{equation}}
\def\micron{$\mu$m}

\def\fp1{$f_{p1}$}

\usepackage{tikz}
\usepackage[percent]{overpic}
\usepackage{subcaption}

\usepackage[utf8]{inputenc}
\usepackage[T1]{fontenc}
\usepackage{lmodern}  
\usepackage{graphicx}
\usepackage{tabularx}
\usepackage{amsmath,amssymb}
\usepackage{breqn}
\usepackage{booktabs}
\usepackage{lipsum}
\usepackage{amsmath}
\usepackage{float}
\usepackage{comment}

\usepackage[acronym,nosuper]{glossaries}
\newacronym{odysseus}{ODYSSEUS}{Outflows and Disks around Young Stars: Synergies for the Exploration of ULLYSES Spectra}
\newacronym{tts}{TTS}{T Tauri Star}
\newacronym{ullyses}{ULLYSES}{Hubble UV Legacy Library of Young Stars as Essential Standards}

\usepackage{colortbl}

\shorttitle{ODYSSEUS Magnetospheric Accretion Survey}
\shortauthors{Pittman et al.}

\begin{document}

\title{The ODYSSEUS Survey. Characterizing magnetospheric geometries and hotspot structures in T Tauri stars}

\correspondingauthor{Caeley V. Pittman}
\email{cpittman@bu.edu}

\author[0000-0001-9301-6252]{Caeley V. Pittman}\altaffiliation{NSF Graduate Research Fellow}
\affiliation{Department of Astronomy, Boston University, 725 Commonwealth Avenue, Boston, MA 02215, USA}
\affiliation{Institute for Astrophysical Research, Boston University, 725 Commonwealth Avenue, Boston, MA 02215, USA}
\email{cpittman@bu.edu}

\author[0000-0001-9227-5949]{Catherine C. Espaillat}
\affiliation{Department of Astronomy, Boston University, 725 Commonwealth Avenue, Boston, MA 02215, USA}
\affiliation{Institute for Astrophysical Research, Boston University, 725 Commonwealth Avenue, Boston, MA 02215, USA}
\email{cce@bu.edu}

\author[0000-0003-1639-510X]{Connor E. Robinson}
\affiliation{Division of Physics and Astronomy, Alfred University, 1 Saxon Drive, Alfred, NY 14802, USA}
\email{robinsonc@alfred.edu}

\author[0000-0003-4507-1710]{Thanawuth Thanathibodee}
\affiliation{Department of Physics, Faculty of Science, Chulalongkorn University, 254 Phayathai Road, Pathumwan, Bangkok 10330, Thailand}
\email{Thanawuth.T@chula.ac.th}

\author[0009-0003-8647-672X]{Sophia Lopez}
\affiliation{Department of Astronomy, Boston University, 725 Commonwealth Avenue, Boston, MA 02215, USA}
\affiliation{Institute for Astrophysical Research, Boston University, 725 Commonwealth Avenue, Boston, MA 02215, USA}
\email{lopezsophia172@yahoo.com}

\author[0000-0002-3950-5386]{Nuria Calvet}
\affiliation{Department of Astronomy, University of Michigan, 311 West Hall, 1085 S. University Avenue, Ann Arbor, MI 48109, USA}
\email{ncalvet@umich.edu}

\author[0000-0003-3616-6822]{Zhaohuan Zhu}
\affil{Department of Physics and Astronomy, University of Nevada, Las Vegas, 4505 S. Maryland Pkwy, Las Vegas, NV 89154, USA} 
\affil{Nevada Center for Astrophysics, University of Nevada, 4505 S. Maryland Pkwy., Las Vegas, NV 89154-4002, USA} 
\email{zhaohuan.zhu@unlv.edu}

\author[0000-0001-7796-1756]{Frederick M. Walter}
\affiliation{Department of Physics and Astronomy, Stony Brook University, Stony Brook NY 11794-3800, USA}
\email{frederick.walter@stonybrook.edu}

\author[0000-0002-6808-4066]{John Wendeborn}
\affiliation{Department of Astronomy, Boston University, 725 Commonwealth Avenue, Boston, MA 02215, USA}
\affiliation{Institute for Astrophysical Research, Boston University, 725 Commonwealth Avenue, Boston, MA 02215, USA}
\email{jwendeborn@gmail.com}

\author[0000-0003-3562-262X]{Carlo F. Manara}
\affiliation{European Southern Observatory, Karl-Schwarzschild-Strasse 2, 85748 Garching, Germany}
\email{cmanara@eso.org}


\author[0000-0002-3913-3746]{Justyn Campbell-White}
\affiliation{European Southern Observatory, Karl-Schwarzschild-Strasse 2, 85748 Garching, Germany}
\email{Justyn.Campbell-White@eso.org}

\author[0000-0001-8194-4238]{Rik Claes}
\affiliation{European Southern Observatory, Karl-Schwarzschild-Strasse 2, 85748 Garching, Germany}
\email{rikclaes@live.be}

\author[0000-0001-8060-1321]{Min Fang}
\affiliation{Purple Mountain Observatory, Chinese Academy of Sciences, 10 Yuanhua Road, Nanjing 210023, People’s Republic of China}
\email{mfang.cn@gmail.com}

\author[0000-0002-0474-0896]{Antonio Frasca}
\affiliation{INAF -- Osservatorio Astrofisico di Catania, via S. Sofia, 78, 95123 Catania, Italy}
\email{antonio.frasca@inaf.it}

\author[0000-0002-1970-7001]{Jorge F. Gameiro}
\affiliation{Instituto de Astrof\'{\i}sica e Ci\^encias do Espa\c co, Universidade do Porto, CAUP, Rua das Estrelas, P-4150-762 Porto,
Portugal}
\affiliation{Departamento de F\'{\i}sica e Astronomia, Faculdade de Ci\^encias, Universidade do Porto, Rua do Campo Alegre 687,
P-4169-007 Porto, Portugal}
\email{jgameiro@astro.up.pt}

\author[0000-0002-8364-7795]{Manuele Gangi}
\affiliation{ASI, Italian Space Agency, Via del Politecnico snc, 00133, Rome, Italy}
\email{gangiorama@gmail.com}

\author[0000-0001-9797-5661]{Jesus Hern{\'a}ndez}
\affil{Instituto de Astronom\'{i}a, Universidad Aut\'{o}noma de M\'{e}xico Ensenada, B.C, M\'{e}xico}
\email{hernandj@astro.unam.mx}

\author[0000-0001-7157-6275]{\'Agnes K\'osp\'al}
\affiliation{Konkoly Observatory, HUN-REN Research Centre for Astronomy and Earth Sciences, MTA Centre of Excellence, Konkoly-Thege Mikl\'os \'ut 15-17, 1121 Budapest, Hungary}
\affiliation{Institute of Physics and Astronomy, ELTE E\"otv\"os Lor\'and University, P\'azm\'any P\'eter s\'et\'any 1/A, 1117 Budapest, Hungary}
\affiliation{Max Planck Institute for Astronomy, K\"onigstuhl 17, 69117 Heidelberg, Germany}
\email{kospal.agnes@csfk.org}

\author[0000-0001-8284-4343]{Karina Mauc\'o}
\affiliation{European Southern Observatory, Karl-Schwarzschild-Strasse 2, 85748 Garching, Germany}
\email{Karina.MaucoCoronado@eso.org}


\author[0000-0002-5943-1222]{James Muzerolle}
\affil{Space Telescope Science Institute, 3700 San Martin Drive, Baltimore, MD 21218, USA}
\email{muzerol@stsci.edu}

\author[0000-0001-5018-3560]{Michał Siwak}
\affiliation{Mt. Suhora Astronomical Observatory, University of the National Education Commission, ul. Podchor\k{a}\.zych 2, 30-084 Krak{\'o}w, Poland}
\affiliation{Konkoly Observatory, HUN-REN Research Centre for Astronomy and Earth Sciences, MTA Centre of Excellence, Konkoly-Thege Mikl\'os \'ut 15-17, 1121 Budapest, Hungary}
\email{michal.siwak@gmail.com}


\author[0000-0002-9470-2358]{Łukasz Tychoniec}
\affiliation{Leiden Observatory, Leiden University, PO Box 9513, 2300 RA Leiden, The Netherlands}
\email{lukasz.tychoniec@gmail.com}

\author[0000-0002-4115-0318]{Laura Venuti}
\affiliation{SETI Institute, 339 Bernardo Ave, Suite 200, Mountain View, CA 94043, USA}
\email{lvenuti@seti.org}

\submitjournal{ApJ}
\received{March 25, 2025}
\revised{June 9, 2025}
\accepted{June 29, 2025}

\begin{abstract}
Magnetospheric accretion is a key process that shapes the inner disks of T Tauri stars, controlling mass and angular momentum evolution.
It produces strong ultraviolet and optical emission that irradiates the planet-forming environment. 
In this work, we characterize the magnetospheric geometries, accretion rates, extinction properties, and hotspot structures of 67 T Tauri stars in the largest and most consistent study of ultraviolet and optical accretion signatures to date.
To do so, we apply an accretion flow model to velocity-resolved \halpha\ profiles for T Tauri stars from the \hst\ ULLYSES program with consistently-derived stellar parameters. We find typical magnetospheric truncation radii to be almost half of the usually-assumed value of 5 stellar radii.
We then model the same stars' \hst/STIS spectra with an accretion shock model, finding a diverse range of hotspot structures.
Phase-folding multi-epoch shock models reveals rotational modulation of observed hotspot energy flux densities, indicative of hotspots that persist for at least 3 stellar rotation periods.
For the first time, we perform a large-scale, self-consistent comparison of accretion rates measured using accretion flow and shock models, finding them to be consistent within $\sim$0.16~dex for contemporaneous observations.
Finally, we find that up to 50\% of the total accretion luminosity is at short wavelengths accessible only from space, highlighting the crucial role of ultraviolet spectra in constraining accretion spectral energy distributions, hotspot structure, and extinction. 

\end{abstract}


\section{Introduction} \label{sec:intro}

Young, low mass ($\rm M_\star<2$~\msun) star systems (T Tauri stars; TTS) are an ideal laboratory for studying the origins of planetary systems in our galaxy. They are more common, long-lived, and Sun-like than their high mass counterparts, making them particularly useful for inferring the history of our own solar system. The T Tauri phase is the earliest point at which the central star is optically visible due to the dispersal of the protostellar envelope, revealing the early radiative and magnetic structure of the star \citep{ALS1987}.

Characterizing stellar properties is important for understanding exoplanet formation and evolution. Short-period exoplanets have been detected in abundance around Sun-like stars at orbital radii peaking around 0.1~au \citep[so-called ``Kepler planets'';][]{Mulders2018}.
This is just outside the region in which the strong magnetic field of the central TTS \citep[on the order of a kG; e.g.,][]{JohnsKrull2007,Donati2010} interacts with the disk through magnetospheric accretion, typically spanning from the stellar surface to $\sim0.05$~au \citep[e.g.,][]{bouvier20,gravity2023CITau} and even out to 0.1~au \citep[][]{gravity2023survey}.
Thus, the structure of the accretion system may influence the final exoplanetary system by controlling mass and angular momentum evolution in the inner disk.

Two defining features of actively-accreting TTS (classical TTS; CTTS) are broad hydrogen recombination line emission and ultraviolet--optical continuum emission in excess of the intrinsic stellar emission. Both features are well described by the magnetospheric accretion paradigm \citep[][]{koenigl91,shu94,hartmann94,hartmann16}. In this process, the stellar magnetic field truncates the hydrogen-dominated inner gas disk and causes the  material to flow along the field lines and crash onto the stellar surface at near-freefall velocity.
As the material flows, it emits atomic hydrogen emission lines with broad high-velocity components, which have been confirmed interferometrically to originate from the magnetospheric accretion region
\citep[e.g.,][]{gravity20TWHya, bouvier20}.
Using these emission lines, accretion flow models \citep[][]{hartmann94,muzerolle98b,muzerolle01} can estimate the geometry and mass flux of the accretion flow. 

Once the accreting material reaches the photosphere, it forms a strong shock that emits line and continuum emission at far-ultraviolet (FUV) through optical and near-infrared wavelengths \citep{Gullbring1998,cg98,Lamzin1998}. 
Because CTTS have effective temperatures below $\sim$5000~K, the higher-energy accretion shock emission dominates the ultraviolet spectral energy distribution for moderate to strong accretors \citep[][]{cg98,ingleby13}. To characterize the full accretion shock spectrum, it is therefore necessary to use space telescopes such as the Hubble Space Telescope (HST).

Studying the inner regions of CTTS disks provides vital information on the evolution of mass and angular momentum in these systems. As disk material accretes inwards through the disk and onto the star, it loses angular momentum through turbulent stress from disk instabilities or magnetic stress at the disk surface from the magnetocentrifugal winds \citep{BaiStone2013,BaiStone2017}. In the central region where the stellar magnetic field truncates the inner gas disk, the disk is sufficiently ionized for the magnetorotational instability to operate \citep{DeschTurner2015} and magnetically launch disk winds and jets \citep{Lovelace1995,Romanova2009,Zhu2025}.

Observing the inner regions of CTTS is very challenging in all but the nearest and brightest systems. Infrared interferometry of \brgamma\ has provided spatially-resolved observations of the magnetospheric region of a number of TTS \citep[e.g.,][]{gravity20TWHya,gravity2023CITau,gravity2024SCrAN}, but it is distance-limited and requires advanced instrumentation. Conversely, velocity-resolved optical and infrared spectra of hydrogen recombination lines can be obtained with small telescopes for more distant and faint targets. Modeling lines such as \halpha\ with accretion models is a powerful tool for measuring magnetospheric properties of a much larger sample of CTTS.

This paper will present results from modeling 1) the accretion flow from the disk to the star
and 2) the accretion shock on the stellar surface for a sample of 67 CTTS from the \hst\ ULLYSES DDT program. Performing the accretion flow and accretion shock analyses in concert is valuable for building a complete picture of the TTS systems. 
The flow model constrains the region over which the stellar magnetic field has a strong effect on the inner disk. Applying the model to continuum-normalized \halpha\ profiles permits the detection of very low accretion rates \citep[e.g.,][]{thanathibodee23}, and it is effectively extinction-independent because the differential effect of extinction is negligible across the small wavelength range \citep{cardelli89,whittet04}. The shock model then measures the surface coverage and spectral energy distribution of the hotspot at the base of the accretion flow, constraining the higher-energy radiation field that irradiates the disk.

In Section~\ref{sec:SampleObs}, we describe the ULLYSES sample, the contributions from the ODYSSEUS and PENELLOPE collaborations, and the data sources and reduction. In Section~\ref{sec:models}, we present the accretion flow and shock models. In Section~\ref{sec:results}, we report the results from applying the models to the ULLYSES sample. In Section~\ref{sec:discussion}, we discuss the importance of the ultraviolet for accretion studies. Finally, in Section~\ref{sec:summary}, we summarize our results.

 \section{Sample \& Observations} \label{sec:SampleObs}

The Hubble UV Legacy Library of Young Stars as Essential Standards Director's Discretionary Time program \citep[ULLYSES;][]{ullyses20,RomanDuval2025} produced the most comprehensive UV observations of TTS to date. It obtained single observations of 58 new ``survey'' TTS between 2020-2022 using a uniform observing strategy from the ultraviolet (UV) through near-infrared (NIR), as well as $\sim$24 epochs of observation each of four ``monitoring targets'' in the UV through optical.
ULLYSES also consistently re-reduced all archival data for 78 TTS and intermediate-mass young stellar objects (YSOs) to produce the full ULLYSES sample of 140 YSOs,\footnote{\url{https://ullyses.stsci.edu/ullyses-targets-ttauri.html}} which includes both accretors and non-accretors. It enabled a global coordinated effort to obtain contemporaneous observations of each system across a broad wavelength range (see Section~\ref{subsec:odysseuspenellope} below). This ideal dataset enabled the present work, in which we utilize 2000-10000~\AA\ \hst/STIS G230L-G430L-G750L spectra for the accretion shock modeling and medium- to high-resolution ground-based \halpha\ spectra for the accretion flow modeling (discussed further in Sections~\ref{subsec:odysseuspenellope} and \ref{subsec:data}).

The final sample presented in this work consists of 67 accreting TTS from Taurus, Chamaeleon, Lupus, Orion, and Corona Australis from ULLYSES DR7. The data can be found in MAST at DOI \dataset[10.17909/t9-jzeh-xy14]{http://dx.doi.org/10.17909/t9-jzeh-xy14} \citep[][]{ULLYSESdoi}.
Fifty of these are new ULLYSES observations, and seventeen are archival re-reduced data.
These CTTS were chosen because they have stellar parameter measurements from PENELLOPE \citep[][2025 in preparation]{manara13methods,manara14,manara16,manara17acc,manara21}, sufficient data quality and wavelength coverage, and broad \halpha\ profiles and ultraviolet continuum excesses indicative of accretion. 

Table~\ref{tab:params} gives the sample and their stellar parameters used in our work. The median host region ages range from 1-2~Myr in Taurus to 8-14~Myr in $\eta$~Cha \citep{Michel2021}. Stellar masses range from 0.05~\msun\ in the M6-type brown dwarf RECX~16 to 2.37~\msun\ in the G9-type Sz~19. In order to ensure detection of UV accretion emission, the ULLYSES Working Group recommended a sample with low extinction and accretion rates (\mdot), with $\rm \log_{10}(\dot{M}/M_\odot/yr)>-9.7$.\footnote{\url{https://www.stsci.edu/files/live/sites/www/files/home/stsci-research/research-topics-and-programs/ullyses/_documents/HSTUV-report-ULLYSES.pdf}} As we will discuss in Section~\ref{subsec:MdotvMstar}, this introduces a slight bias towards higher accretors in the sample.

Twelve of the archival CTTS lack NIR HST spectra, which help ensure that the shock model fit to the continuum is compatible with the expected photospheric emission. For these, we check whether the HST continua match their existing X-Shooter spectra in the region of overlap between 3100--5700~\AA. Six of the CTTS, indicated in Table~\ref{tab:params}, have consistent HST and X-Shooter continua, and for these we stitch the X-Shooter spectrum to the HST spectrum beyond 5700~\AA\ to provide stronger constraints for the model fitting.
 
\subsection{ODYSSEUS and PENELLOPE}  \label{subsec:odysseuspenellope}
The Outflows and Disks around Young Stars: Synergies for the Exploration of ULLYSES Spectra (ODYSSEUS)\footnote{\url{https://sites.bu.edu/odysseus/}} \citep{espaillat22} and PENELLOPE\footnote{\url{https://sites.google.com/view/cfmanara/penellope}} \citep{manara21} collaborations provide the ancillary data and consistent stellar parameter derivation required to maximize the scientific value of ULLYSES. ODYSSEUS acquired multi-epoch 4000--9000~\AA\ spectra for a subset of ULLYSES targets using Chiron on CTIO/SMARTS \citep{Tokovinin2013} on the nights surrounding the \hst\ observations (taken in fiber mode with $4\times4$ on-chip binning, resulting in $R\sim27,800$). The ESO PENELLOPE Large Programme at the Very Large Telescope (VLT; Program ID 106.20Z8) obtained 3000--24000~\AA\ medium-resolution spectra from X-Shooter for each TTS, as well as multiple epochs of high-resolution ($R>70,000$) optical spectra using UVES and ESPRESSO taken on the three nights surrounding the \hst\ observations. 
The median time difference between ground-based observations and \hst\ observations is six days for the entire ULLYSES sample, which includes archival observations. This decreases to two days when considering only the new ULLYSES observations. We discuss these time differences in more detail in Section~\ref{subsec:MdotflowMdotshock}.

The X-Shooter spectra were used to derive stellar masses (\mstar), luminosities (\lstar), and spectral types (SpT) for the entire sample in a consistent manner, as described in \cite{manara21}. Stellar effective temperatures (\teff) come from the SpT-\teff\ relation for 5--30~Myr stars in \cite{pm13}, which is consistent within 3\% of the \cite{hh14} relation. This SpT-\teff\ relation extends to SpT M5; for our CTTS with spectral types M5.5--M6, we perform an extrapolation using the \cite{pm13} relation for main sequence M-dwarfs as a scaling guide. The UVES and ESPRESSO spectra provide \halpha\ observations and allow us to measure the veiling ($r_{\lambda_0}$) of photospheric absorption lines, indicative of the amount of continuum excess due to accretion. In cases when UVES and ESPRESSO observations are unavailable, we use X-Shooter to measure veiling instead. Distances ($d$) are calculated as the inverse parallax of Gaia Data Release 3 \citep{GaiaDR3} for objects with reliable astrometric solutions (where RUWE$<$1.4 and/or the distance is consistent with that of the host stellar association). 
The fractional parallax uncertainties are below 5\% in 94\% of this sample, and in all cases they are below the 20\% limit recommended by \cite{BailerJones2015}. For those without reliable solutions, the mean distance to the association is used assuming an uncertainty of 10\%.

\subsection{\halpha\ observations and data reduction}  \label{subsec:data}
\halpha\ profiles come from four sources: the PENELLOPE Large Programme, the VLT archive (Program IDs 090.C-0050(A), 093.C-0658(A), 094.C-0805(A), 108.22M8.001), Chiron (from ODYSSEUS\footnote{\url{https://www.astro.sunysb.edu/fwalter/SMARTS/Odysseus/odysseus.html}} and Proposal ID 782), and LCOGT/NRES. Most data were contemporaneous to the \hst\ observations, but in an attempt to capture the ``typical'' accretion structures of the sample, we include data from longer baselines when available. Most TTS have four observations: one from X-Shooter and three from UVES or ESPRESSO.
These spectra have resolutions ranging from 16~km/s in X-Shooter's optical arm to 2~km/s in ESPRESSO. Since the freefall velocities of the flows are around 300~km/s, these resolutions are sufficient for resolving the \halpha\ profiles.
 
Custom-reduced PENELLOPE spectra are publicly available on Zenodo \citep{zenodo1,zenodo2,zenodo3,zenodo6,zenodo5,zenodo4,zenodo7,zenodo9,zenodo8},\footnote{\url{https://zenodo.org/communities/odysseus/} PENELLOPE data are flux-calibrated and corrected for telluric absorption as described in \cite{manara21}. In some cases, multiple exposures are combined to increase signal-to-noise.} the PENELLOPE data collection on the ESO archive,\footnote{\url{https://doi.eso.org/10.18727/archive/88}} and in the ULLYSES HLSP collection \citep{ULLYSESdoi}.\footnote{\url{https://mast.stsci.edu/search/ui/\#/ullyses}}
Reduced 1D spectra are available for all archival VLT observations on the ESO archive.
The Chiron data were reduced using a custom IDL pipeline,\footnote{\url{http://www.astro.sunysb.edu/fwalter/SMARTS/CHIRON/ch_reduce.pdf}} and the NRES data were reduced according to \cite{Espaillat2024}. Absolute flux calibration of the spectra is not necessary because the accretion flow models are fit to continuum-normalized profiles. We perform radial velocity and photospheric absorption correction for all spectra using a PHOENIX model spectrum \citep{husser13} with solar metallicity and the same \teff\ and surface gravity of the CTTS. After convolving the model spectrum to the instrumental resolution, we use a cross-correlation function in a region with strong photospheric absorption lines (6420--6520~\AA\ for most targets) to determine the required wavelength shift. Then, we perform photospheric subtraction using the model photosphere broadened by a hand-tuned \vsini\ using the \texttt{rotBroadInt} routine of \cite{rotBroadInt}.

\subsection{Multiplicity} \label{subsec:multiplicity}
A number of CTTS in our sample are in multiple systems. The binaries CVSO~109~AB (0\farcs64/270~au separation) and CVSO~165~AB (0\farcs30/94~au separation) are resolved in their \hst/STIS observations, but not in \hst/COS or ground-based spectroscopy \citep{proffitt21}.
CVSO~109~B shows no evidence of accretion or ejection activity, and its \halpha\ equivalent width is fainter than $-$2~\AA\ \citep{espaillat22}, so we model the \halpha\ profile of CVSO~109~AB
as a single star with the stellar parameters derived in \cite{espaillat22}. While CVSO~165~B is 1.9 magnitudes fainter than CVSO~165~A in V-band, the resolved STIS spectra show that its \halpha\ flux is equal to that of CVSO~165~A. Using the unresolved ESPRESSO spectrum of CVSO~165~AB taken 6.5~hours after the STIS spectrum, we confirm that CVSO~165~B's \halpha\ flux is half of the total unresolved flux in that observation as well. 
We estimate CVSO~165~B's contribution as its observed continuum plus a Gaussian with a standard deviation of 150~km/s and an integrated flux equal to that observed for CVSO~165~B in the low-resolution STIS spectrum.
We subtract this contribution from the unresolved ESPRESSO spectrum to approximate CVSO~165~A's profile, then model it using the parameters of CVSO~165~A from \cite{pittman22}.

CVSO~104 is a double-lined spectroscopic binary with an orbital period $\sim5$~days, 
a mass ratio of 0.92, spectral types of M0 and M2,
and a flux ratio of 1.7 \citep{Frasca2021}. 
It is unresolved in all observations, so we cannot disentangle the two contributions to \halpha. For this reason, we do not include CVSO~104 in the flow modeling.
For the accretion shock modeling, we shift the unresolved HST spectrum down according to the flux ratio, then make an approximation by treating it as a single object with the stellar parameters of the primary component.
CS~Cha~A is an SB1-type spectroscopic binary with a separation of $31.6\pm1.3$~mas and a flux ratio of 0.31 in $R$-band \citep{Ginski2024}, and it has a companion (CS~Cha~B) at 1\farcs3 that is 10 magnitudes fainter in the optical \citep{Haffert2020}. These are unresolved in all observations, but CS~Cha~Aa should dominate the continuum and line emission and is hereafter referred to as CS~Cha.
Sz~19~A, hereafter referred to as Sz~19, is the primary component of a quadruple system, and its emission dominates at all wavelengths \citep{Schmidt2013,Juhasz2022}. We therefore model the HST and \halpha\ spectra using the stellar parameters of the primary component.
DG~Tau~A, DK~Tau~A, UX~Tau~A, and HD~104237~E are resolved in all observations.

\section{Models} \label{sec:models}

We constrain the magnetospheric accretion properties of the ULLYSES sample using an accretion flow model fit to \halpha\ profiles (Section~\ref{subsec:flowmodel}) and an accretion shock model fit to UV--NIR \hst/STIS spectra (Section~\ref{subsec:shockmodel}). Section~\ref{subsec:procedure} describes the modeling procedure. In short, we employ an iterative process consisting of an initial run of the shock model (\emph{Step 1}), then a run of the flow model (\emph{Step 2}), followed by two additional rounds of fitting with the shock model (\emph{Steps 3} and \emph{4}). Finally, we confirm that the flow model results remain valid when using the final results from the shock model (\emph{Step 5}). This procedure ensures consistency between the flow and shock models and adjusts for the extinction measurements in the shock model fit, as described below.

\subsection{Magnetospheric accretion flow model} \label{subsec:flowmodel}
We model \halpha\ emission profiles of the ULLYSES sample using a magnetospheric accretion flow model from \cite{hartmann94} and \cite{muzerolle98b,muzerolle01}, which provides insight into the geometry and mass flux of the accreting systems. The model assumes a dipolar magnetic field aligned with the stellar rotation axis and an accretion flow that lands on the stellar surface in an axisymmetric ring with temperature \tshock.
The magnetic field truncates the disk at radius \ri, and the truncation region extends radially outward over a width \rw\ at the disk midplane.
The flow has an accretion rate \mdot\ and a maximum temperature \tmax. The final \halpha\ line profile is determined for a given viewing inclination $i$ using the ray-by-ray method as described in \cite{muzerolle01}.

\startlongtable
\begin{deluxetable*}{>{\raggedright\arraybackslash}p{2.1cm}p{1.55cm}lllp{1.6cm}p{2.4cm}p{2.4cm}ll}
\tablewidth{\textwidth}
\tabletypesize{\small}
\tablecaption{Stellar parameters of sample \label{tab:params}}
\tablehead{
\colhead{Object} & \colhead{HST/STIS obs} & \colhead{SpT} & \colhead{\mstar} & \colhead{\teff} & \colhead{$d_{\rm Gaia}$} & \colhead{L$_{\star,i}\rightarrow$L$_{\star,f}$} &\colhead{R$_{\star,i}\rightarrow$R$_{\star,f}$}  
& \colhead{$r_{\lambda_0}$} & \colhead{$\lambda_{0}$} 
\\
\colhead{} & \colhead{(UT)} & \colhead{} & \colhead{(\msun)} & \colhead{(K)} & \colhead{(pc)} & \colhead{(\lsun)} & \colhead{(\rsun)} 
& \colhead{} & \colhead{($\mu m$)} 
}
\startdata
AA Tau & 2011-01-07 T03:16:58\tablenotemark{\textdagger\textdaggerdbl} & K7 & 0.80 & 3970 & $134.7\pm1.6$ & $0.93\rightarrow0.47^{+0.23}_{-0.24}$ & $2.04\rightarrow1.45^{+0.36}_{-0.37}$ & $1.0^{1.2}_{0.0}$ & 0.5500 \\
DE Tau & 2010-08-20 T17:40:07\tablenotemark{\textdagger\textdaggerdbl} & M2 & 0.35 & 3490 & $128.0\pm0.4$ & $0.65\rightarrow1.00^{+1.03}_{-0.98}$ & $2.21\rightarrow2.74^{+1.40}_{-1.34}$ & $0.64^{2.22}_{0.0}$ & 0.5380 \\
DG Tau A\tablenotemark{a} & 2001-02-20 T10:05:31\tablenotemark{\textdagger} & K5 & 1.40 & 4140 & $125.3\pm1.9$ & $1.24\rightarrow0.76^{+0.46}_{-0.41}$ & $2.17\rightarrow1.70^{+0.51}_{-0.46}$ & $2.33^{4.02}_{0.64}$ & 0.4410 \\
DK Tau A\tablenotemark{a} & 2010-02-04 T22:51:38\tablenotemark{\textdagger} & K7.5 & 0.63 & 3870 & $132.0\pm0.9$ & $0.74\rightarrow0.27^{+0.04}_{-0.04}$ & $1.91\rightarrow1.16^{+0.09}_{-0.08}$ & $0.82^{0.92}_{0.73}$ & 0.4410 \\
DM Tau & Multiple\tablenotemark{\textdagger}\tablenotemark{*} & M2 & 0.38 & 3490 & $144.0\pm0.5$ & $0.25\rightarrow0.46^{+0.03}_{-0.03}$ & $1.37\rightarrow1.85^{+0.06}_{-0.06}$ & $1.41^{1.46}_{1.36}$ & 0.5500 \\
DN Tau & 2011-09-10 T22:09:44\tablenotemark{\textdagger} & M0 & 0.58 & 3770 & $128.6\pm0.4$ & $0.66\rightarrow0.44^{+0.05}_{-0.05}$ & $1.90\rightarrow1.56^{+0.09}_{-0.09}$ & $0.12^{0.23}_{0.05}$ & 0.4410 \\
UX Tau A\tablenotemark{a} & 2011-11-10 T13:16:06\tablenotemark{\textdagger} & G8 & 1.50 & 5520 & $144.5\pm1.5$ & $2.60\rightarrow3.48^{+4.03}_{-3.96}$ & $1.80\rightarrow2.04^{+1.18}_{-1.16}$ & $0.25^{1.67}_{0.25}$ & 0.4410 \\
\hline \multicolumn{10}{c}{\textbf{Cha I}} \\
2MASS  J11432669-7804454 & 2021-06-12 T17:29:49 & M5.5 & 0.14 & 2830 & $190\pm19$ & $0.09\rightarrow0.11^{+0.08}_{-0.08}$ & $1.25\rightarrow1.42^{+0.52}_{-0.51}$ & $2.03^{2.5}_{0.0}$ & 0.5505 \\
CHX18N & 2021-04-29 T23:13:45 & K6 & 0.73 & 4020 & $191.6\pm0.4$ & $1.23\rightarrow0.83^{+0.07}_{-0.06}$ & $2.29\rightarrow1.88^{+0.08}_{-0.07}$ & $0.3^{0.35}_{0.26}$ & 0.4410 \\
CS Cha & 2015-04-23 T03:34:10\tablenotemark{\textdagger} & K2 & 1.22 & 4760 & $190\pm19$ & $2.01\rightarrow2.08^{+0.73}_{-0.71}$ & $2.08\rightarrow2.12^{+0.37}_{-0.36}$ & $0.0^{0.26}_{0.0}$ & 0.6067 \\
CV Cha & 2011-04-13 T21:41:17\tablenotemark{\textdagger\textdaggerdbl} & K0.5 & 1.40 & 4975 & $191.8\pm0.5$ & $3.43\rightarrow4.20^{+0.49}_{-0.46}$ & $2.49\rightarrow2.76^{+0.16}_{-0.15}$ & $0.68^{0.81}_{0.55}$ & 0.4410 \\
Hn 5 & 2021-05-31 T07:31:45 & M5 & 0.15 & 2880 & $194.7\pm0.9$ & $0.20\rightarrow0.12^{+0.03}_{-0.03}$ & $1.80\rightarrow1.40^{+0.17}_{-0.17}$ & $0.37^{0.67}_{0.07}$ & 0.7128 \\
IN Cha & 2021-06-05 T01:43:33 & M5 & 0.15 & 2880 & $192.6\pm0.8$ & $0.18\rightarrow0.20^{+0.02}_{-0.02}$ & $1.70\rightarrow1.80^{+0.08}_{-0.08}$ & $0.42^{0.5}_{0.35}$ & 0.4410 \\
SY Cha & 2022-03-23 T21:32:22 & K7 & 0.70 & 3970 & $180.7\pm0.4$ & $0.60\rightarrow0.34^{+0.31}_{-0.30}$ & $1.64\rightarrow1.23^{+0.56}_{-0.55}$ & $1.25^{3.2}_{0.0}$ & 0.5630 \\
SZ Cha\tablenotemark{a} & 2015-03-15 T04:36:17\tablenotemark{\textdagger} & K2 & 1.22 & 4760 & $190.2\pm0.9$ & $1.17\rightarrow0.84^{+0.25}_{-0.25}$ & $1.59\rightarrow1.35^{+0.20}_{-0.20}$ & $0.30^{1.21}_{0.28}$ & 0.4430 \\
Sz 10 & 2021-05-01 T14:51:47 & M5 & 0.17 & 2880 & $184.2\pm1.3$ & $0.20\rightarrow0.15^{+0.10}_{-0.10}$ & $1.80\rightarrow1.53^{+0.53}_{-0.52}$ & $0.8^{1.98}_{0.0}$ & 0.6067 \\
Sz 19 & 2022-03-12 T04:46:50 & G9 & 2.37 & 5120 & $189.0\pm0.5$ & $7.04\rightarrow7.63^{+0.65}_{-0.65}$ & $3.37\rightarrow3.51^{+0.15}_{-0.15}$ & $0.72^{0.82}_{0.61}$ & 0.4410 \\
Sz 45 & Multiple\tablenotemark{\textdagger}\tablenotemark{*} & M0 & 0.60 & 3770 & $188.9\pm0.5$ & $0.48\rightarrow0.26^{+0.09}_{-0.09}$ & $1.62\rightarrow1.19^{+0.20}_{-0.20}$ & $0.27^{1.05}_{0.0}$ & 0.5500 \\
VZ Cha & 2022-05-05 T03:06:05 & K7.5 & 0.64 & 3870 & $191.1\pm0.6$ & $0.63\rightarrow0.22^{+0.03}_{-0.03}$ & $1.77\rightarrow1.04^{+0.07}_{-0.07}$ & $4.93^{5.61}_{4.24}$ & 0.5235 \\
WZ Cha & 2022-05-04 T22:34:16 & M3 & 0.29 & 3360 & $193.2\pm0.6$ & $0.21\rightarrow0.12^{+0.07}_{-0.06}$ & $1.35\rightarrow1.03^{+0.28}_{-0.27}$ & $1.33^{2.48}_{0.18}$ & 0.6067 \\
XX Cha & 2021-05-29 T01:34:33 & M3 & 0.29 & 3360 & $192.1\pm0.8$ & $0.31\rightarrow0.25^{+0.16}_{-0.15}$ & $1.64\rightarrow1.47^{+0.47}_{-0.45}$ & $0.28^{1.03}_{0.0}$ & 0.6474 \\
\hline \multicolumn{10}{c}{\textbf{Lupus}} \\
MY Lup & 2017-09-08 T05:42:44\tablenotemark{\textdagger\textdaggerdbl} & K0 & 1.06 & 5030 & $157.2\pm0.9$ & $0.85\rightarrow0.96^{+0.09}_{-0.09}$ & $1.21\rightarrow1.29^{+0.06}_{-0.06}$ & $0.18^{0.21}_{0.16}$ & 0.4410 \\
RX J1556.1-3655 & 2022-06-17 T02:31:50 & M1 & 0.50 & 3630 & $158.0\pm0.6$ & $0.26\rightarrow0.30^{+0.44}_{-0.43}$ & $1.29\rightarrow1.38^{+1.03}_{-1.00}$ & $1.0^{3.89}_{0.57}$ & 0.5380 \\
RY Lup & 2016-03-16 T06:54:34\tablenotemark{\textdagger} & K2 & 1.40 & 4760 & $158\pm16$ & $1.60\rightarrow0.37^{+0.09}_{-0.09}$ & $1.86\rightarrow0.89^{+0.11}_{-0.11}$ & $0.43^{0.5}_{0.36}$ & 0.4410 \\
SSTc2d J160000.6-422158 & 2021-07-23 T05:10:25 & M4.5 & 0.15 & 3020 & $159.4\pm0.8$ & $0.08\rightarrow0.09^{+0.02}_{-0.02}$ & $1.03\rightarrow1.07^{+0.10}_{-0.10}$ & $0.06^{0.3}_{0.0}$ & 0.5235 \\
SSTc2d J160830.7-382827 & 2017-07-29 T15:38:43\tablenotemark{\textdagger} & K2 & 1.38 & 4760 & $153.4\pm0.7$ & $1.40\rightarrow0.89^{+0.46}_{-0.45}$ & $1.74\rightarrow1.39^{+0.36}_{-0.35}$ & $0.0^{0.49}_{0.0}$ & 0.5630 \\
SSTc2d J161243.8-381503 & 2022-04-28 T07:04:09 & M0.5 & 0.57 & 3700 & $159.9\pm0.5$ & $0.31\rightarrow0.27^{+0.03}_{-0.03}$ & $1.36\rightarrow1.26^{+0.06}_{-0.06}$ & $0.04^{0.1}_{0.0}$ & 0.6067 \\
SSTc2d J161344.1-373646 & 2022-05-03 T06:14:32 & M4.5 & 0.13 & 3020 & $158.6\pm1.3$ & $0.03\rightarrow0.03^{+0.02}_{-0.02}$ & $0.63\rightarrow0.62^{+0.20}_{-0.20}$ & $0.41^{1.37}_{0.0}$ & 0.6474 \\
Sz 66 & 2021-08-17 T16:55:17 & M4.5 & 0.19 & 3020 & $155.9\pm0.7$ & $0.26\rightarrow0.26^{+0.15}_{-0.18}$ & $1.86\rightarrow1.87^{+0.54}_{-0.65}$ & $0.24^{1.24}_{0.0}$ & 0.5380 \\
Sz 68 & 2017-07-26 T14:28:13\tablenotemark{\textdagger\textdaggerdbl} & K2 & 1.83 & 4760 & $152.7\pm4.3$ & $5.36\rightarrow4.50^{+0.49}_{-0.46}$ & $3.40\rightarrow3.12^{+0.17}_{-0.16}$ & $0.37^{0.46}_{0.29}$ & 0.4410 \\
Sz 69 & 2021-07-23 T21:00:40 & M4.5 & 0.15 & 3020 & $152.6\pm1.6$ & $0.12\rightarrow0.04^{+0.01}_{-0.01}$ & $1.27\rightarrow0.70^{+0.14}_{-0.14}$ & $0.75^{1.39}_{0.11}$ & 0.7128 \\
Sz 71 & 2021-05-05 T12:32:46 & M2 & 0.37 & 3490 & $155.2\pm0.4$ & $0.40\rightarrow0.26^{+0.03}_{-0.03}$ & $1.73\rightarrow1.39^{+0.08}_{-0.08}$ & $0.44^{0.55}_{0.33}$ & 0.4410 \\
Sz 72 & 2021-05-04 T12:43:09 & M2 & 0.37 & 3490 & $156.7\pm0.5$ & $0.29\rightarrow0.19^{+0.18}_{-0.18}$ & $1.47\rightarrow1.20^{+0.57}_{-0.56}$ & $1.05^{2.93}_{0.0}$ & 0.6474 \\
Sz 75 & 2021-05-03 T14:28:50 & K6 & 0.82 & 4020 & $154.1\pm0.7$ & $1.34\rightarrow0.77^{+0.52}_{-0.52}$ & $2.39\rightarrow1.81^{+0.61}_{-0.61}$ & $0.52^{1.52}_{0.0}$ & 0.6474 \\
Sz 77 & 2021-05-07 T18:33:26 & K7 & 0.75 & 3970 & $155.3\pm0.4$ & $0.54\rightarrow0.53^{+0.04}_{-0.04}$ & $1.55\rightarrow1.54^{+0.06}_{-0.06}$ & $0.3^{0.35}_{0.26}$ & 0.4410 \\
Sz 82 & 2022-06-16 T02:42:09 & K7 & 0.68 & 3970 & $155.8\pm0.5$ & $0.73\rightarrow0.40^{+0.22}_{-0.21}$ & $1.81\rightarrow1.33^{+0.37}_{-0.36}$ & $0.67^{1.56}_{0.0}$ & 0.6067 \\
Sz 84 & 2022-05-11 T19:05:05 & M5 & 0.16 & 2880 & $155.6\pm1.1$ & $0.15\rightarrow0.15^{+0.06}_{-0.06}$ & $1.56\rightarrow1.54^{+0.31}_{-0.31}$ & $0.2^{0.7}_{0.0}$ & 0.5630 \\
Sz 97 & 2022-05-12 T18:53:23 & M4 & 0.19 & 3160 & $157.3\pm0.6$ & $0.14\rightarrow0.11^{+0.01}_{-0.02}$ & $1.25\rightarrow1.11^{+0.06}_{-0.09}$ & $0.59^{0.69}_{0.48}$ & 0.4410 \\
Sz 98 & 2022-05-04 T02:54:06 & K7.5 & 0.61 & 3870 & $156.3\pm0.6$ & $0.76\rightarrow0.31^{+0.03}_{-0.03}$ & $1.94\rightarrow1.23^{+0.06}_{-0.06}$ & $0.56^{0.65}_{0.47}$ & 0.4410 \\
Sz 99 & 2022-06-03 T06:46:08 & M4.5 & 0.13 & 3020 & $158.3\pm1.1$ & $0.03\rightarrow0.04^{+0.03}_{-0.03}$ & $0.63\rightarrow0.77^{+0.28}_{-0.27}$ & $0.37^{1.34}_{0.0}$ & 0.6474 \\
Sz 100 & 2022-07-27 T04:36:53 & M5.5 & 0.13 & 2830 & $158\pm16$ & $0.10\rightarrow0.10^{+0.08}_{-0.08}$ & $1.32\rightarrow1.37^{+0.52}_{-0.51}$ & $0.69^{1.87}_{0.0}$ & 0.6067 \\
Sz 103 & 2022-04-30 T05:08:01 & M4.5 & 0.15 & 3020 & $157.2\pm1.0$ & $0.09\rightarrow0.12^{+0.10}_{-0.09}$ & $1.10\rightarrow1.27^{+0.51}_{-0.49}$ & $0.92^{2.37}_{0.0}$ & 0.5630 \\
Sz 104 & 2022-06-19 T05:21:58 & M5 & 0.12 & 2880 & $159.8\pm1.1$ & $0.09\rightarrow0.08^{+0.07}_{-0.07}$ & $1.21\rightarrow1.17^{+0.46}_{-0.45}$ & $0.62^{1.84}_{0.0}$ & 0.6361 \\
Sz 110 & 2022-05-23 T15:15:14 & M3 & 0.29 & 3360 & $157.5\pm0.6$ & $0.18\rightarrow0.12^{+0.06}_{-0.06}$ & $1.25\rightarrow1.04^{+0.27}_{-0.25}$ & $0.52^{1.21}_{0.0}$ & 0.6067 \\
Sz 111 & 2021-08-14 T15:55:36 & M0.5 & 0.58 & 3700 & $158.4\pm0.5$ & $0.21\rightarrow0.16^{+0.02}_{-0.01}$ & $1.12\rightarrow0.98^{+0.05}_{-0.04}$ & $0.28^{0.32}_{0.25}$ & 0.4410 \\
Sz 117 & 2022-05-30 T07:33:32 & M3 & 0.29 & 3360 & $156.9\pm0.5$ & $0.19\rightarrow0.14^{+0.02}_{-0.02}$ & $1.29\rightarrow1.12^{+0.06}_{-0.06}$ & $0.6^{0.7}_{0.51}$ & 0.4410 \\
Sz 129 & 2022-05-01 T03:23:24 & K7 & 0.72 & 3970 & $160.1\pm0.4$ & $0.50\rightarrow0.21^{+0.05}_{-0.05}$ & $1.49\rightarrow0.96^{+0.11}_{-0.11}$ & $1.29^{1.77}_{0.81}$ & 0.5235 \\
Sz 130 & 2021-08-25 T04:25:21 & M3.5 & 0.24 & 3260 & $159.2\pm0.5$ & $0.20\rightarrow0.17^{+0.03}_{-0.03}$ & $1.40\rightarrow1.31^{+0.10}_{-0.10}$ & $0.23^{0.37}_{0.08}$ & 0.4410 \\
\hline \multicolumn{10}{c}{\textbf{$\sigma$ Ori}} \\
TX Ori & 2020-12-01 T01:42:28 & K5 & 1.09 & 4140 & $385\pm39$ & $3.61\rightarrow3.22^{+1.09}_{-0.81}$ & $3.69\rightarrow3.49^{+0.59}_{-0.44}$ & $1.05^{1.17}_{0.93}$ & 0.4410 \\
V505 Ori & 2020-12-01 T03:17:47 & K7 & 0.81 & 3970 & $399.0\pm4.0$ & $0.24\rightarrow0.56^{+0.08}_{-0.09}$ & $1.04\rightarrow1.58^{+0.12}_{-0.13}$ & $1.35^{1.66}_{1.03}$ & 0.5380 \\
V510 Ori & 2021-02-12 T13:09:47 & K7 & 0.76 & 3970 & $396.6\pm4.2$ & $0.17\rightarrow0.43^{+0.55}_{-0.57}$ & $0.87\rightarrow1.38^{+0.89}_{-0.92}$ & $1.18^{3.87}_{0.0}$ & 0.6223 \\
\hline \multicolumn{10}{c}{\textbf{Orion OB1b}} \\
CVSO 58 & 2020-12-01 T23:56:20 & K7 & 0.81 & 3970 & $354.2\pm2.9$ & $0.32\rightarrow0.66^{+0.12}_{-0.12}$ & $1.20\rightarrow1.72^{+0.15}_{-0.16}$ & $0.89^{0.89}_{0.54}$ & 0.5505 \\
CVSO 90 & 2020-12-16 T15:02:44 & M0.5 & 0.62 & 3700 & $343.6\pm3.9$ & $0.13\rightarrow0.25^{+0.20}_{-0.27}$ & $0.88\rightarrow1.21^{+0.48}_{-0.65}$ & $0.84^{2.22}_{0.0}$ & 0.7128 \\
CVSO 104\tablenotemark{b} & 2020-11-27 T05:32:50 & M0 & 0.57 & 3770 & $366.4\pm4.0$ & $0.21\rightarrow0.40^{+0.05}_{-0.05}$ & $1.07\rightarrow1.48^{+0.10}_{-0.09}$ & $0.2^{0.3}_{0.1}$ & 0.6474 \\
CVSO 107 & 2020-12-05 T01:01:03 & M0.5 & 0.53 & 3700 & $335.1\pm2.5$ & $0.32\rightarrow0.24^{+0.03}_{-0.02}$ & $1.38\rightarrow1.20^{+0.07}_{-0.06}$ & $1.89^{2.15}_{1.64}$ & 0.4410 \\
CVSO 109 A\tablenotemark{c} & 2020-11-28 T05:22:53 & M0 & 0.50 & 3768 & $400\pm40$ & $0.59\rightarrow1.20^{+0.54}_{-0.53}$ & $1.80\rightarrow2.57^{+0.58}_{-0.57}$ & $1.59^{2.57}_{0.61}$ & 0.5235 \\
CVSO 146 & 2020-12-09 T20:59:12 & K6 & 0.86 & 4020 & $336.7\pm1.7$ & $0.80\rightarrow0.49^{+0.08}_{-0.07}$ & $1.84\rightarrow1.45^{+0.11}_{-0.10}$ & $0.66^{0.79}_{0.52}$ & 0.4410 \\
CVSO 165 A\tablenotemark{c} & 2020-12-14 T20:09:05 & K5.5 & 0.84 & 4221 & $400\pm40$ & $0.90\rightarrow0.82^{+0.28}_{-0.27}$ & $1.77\rightarrow1.69^{+0.29}_{-0.28}$ & $0.3^{0.51}_{0.0}$ & 0.4410 \\
CVSO 165 B\tablenotemark{d} & 2020-12-14 T20:09:05 & M1 & 0.58 & 3829 & $400\pm40$ & $0.47\rightarrow0.36^{+0.13}_{-0.13}$ & $1.56\rightarrow1.36^{+0.25}_{-0.25}$ & $0.21^{0.51}_{0.0}$ & 0.4410 \\
CVSO 176 & 2020-11-29 T21:06:36 & M3.5 & 0.25 & 3260 & $306.8\pm3.0$ & $0.34\rightarrow0.40^{+0.05}_{-0.05}$ & $1.83\rightarrow1.98^{+0.13}_{-0.12}$ & $0.77^{0.91}_{0.63}$ & 0.4410 \\
\hline \multicolumn{10}{c}{\textbf{CrA}} \\
RX J1842.9-3532 & 2022-06-23 T20:50:36 & K1 & 1.04 & 4920 & $151.0\pm0.4$ & $0.59\rightarrow0.40^{+0.14}_{-0.14}$ & $1.06\rightarrow0.87^{+0.15}_{-0.15}$ & $0.41^{0.87}_{0.0}$ & 0.4410 \\
RX J1852.3-3700 & 2022-07-02 T19:08:42 & K6 & 0.82 & 4020 & $147.1\pm0.5$ & $0.51\rightarrow0.44^{+0.03}_{-0.03}$ & $1.47\rightarrow1.37^{+0.04}_{-0.04}$ & $0.18^{0.22}_{0.14}$ & 0.4410 \\
\hline \multicolumn{10}{c}{\textbf{$\epsilon$ Cha}} \\
HD~104237 E\tablenotemark{a} & 2022-03-22 T01:07:59 & K5.5 & 0.86 & 4080 & $100.2\pm0.3$ & $0.48\rightarrow0.18^{+0.03}_{-0.03}$ & $1.39\rightarrow0.86^{+0.08}_{-0.08}$ & $0.27^{0.49}_{0.05}$ & 0.4410 \\
\hline \multicolumn{10}{c}{\textbf{$\eta$ Cha}} \\
RECX 9 & 2022-01-26 T02:57:58 & M5 & 0.12 & 2880 & $99\pm10$ & $0.10\rightarrow0.10^{+0.04}_{-0.04}$ & $1.27\rightarrow1.27^{+0.28}_{-0.28}$ & $0.26^{0.84}_{0.0}$ & 0.5380 \\
RECX 11 & 2009-12-12 T06:20:48\tablenotemark{\textdagger\textdaggerdbl} & K6 & 0.78 & 4020 & $98.8\pm0.1$ & $0.67\rightarrow0.49^{+0.12}_{-0.11}$ & $1.69\rightarrow1.44^{+0.17}_{-0.16}$ & $0.17^{0.43}_{0.0}$ & 0.5380 \\
RECX 15 & 2010-02-05 T21:58:46\tablenotemark{\textdagger} & M4 & 0.19 & 3160 & $99\pm10$ & $0.09\rightarrow0.07^{+0.05}_{-0.05}$ & $1.00\rightarrow0.89^{+0.33}_{-0.33}$ & $0.74^{1.92}_{0.0}$ & 0.5700 \\
RECX 16 & 2021-06-11 T02:01:40 & M6 & 0.05 & 2770 & $99.6\pm0.4$ & $0.01\rightarrow0.01^{+0.00}_{-0.00}$ & $0.49\rightarrow0.49^{+0.08}_{-0.07}$ & $0.00^{0.33}_{0.00}$ & 0.7128 \\
\bottomrule
 \enddata
\tablenotetext{}{Stellar parameters for the sample. Objects are sorted by region in order of region median age. SpT, \mstar, and L$_{\star,i}$ come from PENELLOPE \citet[2025 in preparation]{manara13methods,manara14,manara16,manara17acc,manara21} except for AA~Tau \citep{ingleby13}, RX J1556.1-3655 \citep{alcala17} and UX~Tau~A \citep{espaillat10}. \teff\ comes from the SpT-\teff\ conversion for 5-30~Myr stars in \cite{pm13}. The stellar radius column shows both the initial stellar radius R$_{\star,i}$ (calculated using the stellar luminosity L$_{\star,i}$ from PENELLOPE) and the final stellar radius R$_{\star,f}$ (from the recalculated L$_{\star,f}$ using the extinction derived from our shock model fitting).
 The $r_{\lambda_0}$ column gives the veiling at $\lambda_{0}$ measured in this work. Note that the $r_{\lambda_0}$ uncertainties are given as upper and lower limits, whereas the previous columns are upper and lower uncertainties.
\textit{a}: Binary resolved in both \hst\ and \halpha\ spectra, \textit{b}: binary not resolved in \hst\ or \halpha\ but modeled with the shock model using the stellar parameters of the primary component, \textit{c}: binary resolved in only \hst\ but included in the flow modeling using the stellar parameters of the primary component, \textit{d}: binary resolved in only \hst\ and not included in the flow modeling.\\
$^{*}$DM~Tau and Sz~45 were observed by \cite{re19} over multple dates: DM~Tau on 2011-09-08T03:12:33, 2011-09-15T22:08:48, and 2012-01-04T12:16:22; and Sz~45 on 2016-05-14T21:04:48, 2016-05-17T03:03:27, 2016-05-18T18:43:45, 2016-05-20T23:04:08, 2016-07-06T01:59:30, and 2021-05-25T02:05:48.
\textit{\textdagger}: Archival observation re-reduced by the ULLYSES team. \textit{\textdaggerdbl}: X-Shooter spectrum stitched to HST spectrum beyond 5700~\AA.
 }
\end{deluxetable*}

\subsection{Magnetospheric accretion shock model} \label{subsec:shockmodel}

\cite{pittman22} performed accretion shock modeling of the first 9 CTTS observed through ULLYSES.
They used the \cite{cg98} accretion shock model (updated according to \citealt{re19}) to calculate spectra produced by three accretion ``columns'' (with energy flux densities of $\curf_{\rm low,med,high}=10^{10}$, $10^{11}$, $10^{12}$~\escm) flowing in from the magnetospheric truncation radius (\ri, assumed to be 5~\rstar) at freefall velocity \vff. Each column has an associated filling factor $f_{\rm low,med,high}$.
These discrete accretion columns approximate the observed density gradient in accretion hotspots \citep[e.g.,][]{espaillat21,Singh2024}. Additionally, multi-column models can explain the discrepancy between the low accretion rates derived from X-rays (which require low densities) and
the higher accretion rates found in UV-NIR studies \citep[][]{Orlando2010,Schneider2018}.

The model uses Cloudy to perform radiative transfer in the pre- and post-shock regions (see \citealt{re19} for details), and the total
emission from each column is a sum of the output fluxes from the pre-shock, post-shock, and heated photosphere.
In this work, we use the accretion flow model results to select \ri\ and $\curf_{\rm low,med,high}$ for each target individually, rather than assuming the same values for all targets. This is important given the large range in \mstar, \mdot, and stellar age in the ULLYSES sample. Because $\curf\propto v_{\rm ff}^3$, tailoring the $\curf$ values to each system enables a more natural comparison of hotspot surface coverage between different stars, as a star with lower \vff\ will have a lower typical $\curf$ than a star with higher \vff.

The \hst/STIS data are prepared by subtracting a non-accreting T Tauri star (a weak-lined TTS; WTTS) spectrum of the closest-available spectral type, scaled to the CTTS according to the contemporaneously-measured veiling: $F_{{\rm WTTS},\lambda_0}=F_{{\rm CTTS},\lambda_0}/(1+r_{\lambda_0})$. We measure veiling in regions with strong photospheric absorption lines by minimizing the difference between the CTTS and a HARPS template of the same SpT, adjusted by varying veiling and \vsini. The HARPS templates and veiling measurement procedures are the same as in \citet[see their Table C.1]{manara21}.
This produces a spectrum of the accretion excess above the stellar photosphere.

Following \cite{pittman22}, the shock model spectra are fit to the data by varying the surface coverage of each column ($f_{\rm low,med,high}$) and the $V$-band extinction (\av) using a Markov-Chain Monte Carlo (MCMC) procedure. \cite{pittman22} found the best fits for the Orion OB1b targets when using the Taurus-specific extinction law from \cite{whittet04}, which has a shallower 2200~\AA\ bump that avoided over-correcting for polycyclic aromatic hydrocarbon absorption. Here, we also use the \cite{whittet04} law for targets in Taurus and Orion, and we use the \cite{cardelli89} general ISM law for the remaining targets. \av\ is restricted between 0--2~mag for all stars, as the analysis by \citet[2025 in preparation]{manara21} finds a maximum extinction of 1.5~mag for this sample. This aligns with the ULLYSES Working Group's aim to select minimally-extincted sources for the sample. The total filling factor $f_{\rm tot}$ is restricted between 0--40\%, as higher surface coverage is rarely found by modeling or observations \citep[see, for example, Figures 9 and 10 in][]{venuti15}.

\subsection{Modeling procedure} \label{subsec:procedure}
\begin{deluxetable*}{lcc}
\tablecaption{Iterative modeling procedure\label{tab:procedure}}
\tablehead{
\colhead{Step} & \colhead{Inputs} & \colhead{Outputs}
}
\startdata
1. Shock model & \mstar, R$_{\star,i}$, $r_{\lambda_0}$, $R_{\rm i}=3.5$~R$_{\star,i}$, $d_{\rm Gaia}$, & \mdot$_{\rm shock}$, \tshock \\
 & $\curf_{\rm low,med,high}=10^{10},10^{11},10^{12}$~\escm & \\ \hline
2. Flow model & \mstar, R$_{\star,i}$, \teff, \tshock, \mdot$_{\rm shock}$ range & \ri, \rw, \mdot$_{\rm flow}$, \tmax, $i$, $A$, $\sigma$, $\curf_{\rm flow}$ \\ \hline
3. Shock model & \mstar, R$_{\star,i}$, $r_{\lambda_0}$, \ri, $d_{\rm Gaia}$ & R$_{\star,f}$ \\
 & $\curf_{\rm low,med,high}=0.1\curf_{\rm flow},\curf_{\rm flow},10\curf_{\rm flow}$ & \\ \hline
4. Shock model & \mstar, R$_{\star,f}$, $r_{\lambda_0}$, \ri, $d_{\rm Gaia}$ & \mdot$_{\rm shock}$, \tshock, \av, $f_{\rm low,med,high}$ \\
 & $\curf_{\rm low,med,high}=0.1\curf_{\rm flow},\curf_{\rm flow},10\curf_{\rm flow}$ & \\ \hline
5. Flow model & \mstar, R$_{\star,f}$, \teff, \tshock, & Updated  \mdot$_{\rm flow}$, \\
  &  Weighted means of \ri, \rw, \tmax, & Visual confirmation that using R$_{\star,f}$ still \\
    & \mdot$_{\rm flow}\times$(R$_{\star,f}$/R$_{\star,i}$), $i$, $A$, and $\sigma$ & consistent with results that used R$_{\star,i}$ \\
\enddata
 \tablenotetext{}{
Summary of the iterative modeling procedure and associated inputs and relevant outputs. Stellar parameter notation is the same as in Table~\ref{tab:params}.
}
\end{deluxetable*}
Here we describe each step of the iterative modeling procedure, which is summarized in Table~\ref{tab:procedure}.

\paragraph{Step 1. Shock model}
The modeling procedure begins with an initial iteration of the multicolumn accretion shock model. 
Stellar parameter inputs are mass (\mstar), radius (R$_{\star,i}$), effective temperature (\teff), and distance ($d_{\rm Gaia}$).
For this first run, we assume $\curf_{\rm low,med,high}=10^{10},10^{11},10^{12}$~\escm\ and \ri=3.5~R$_{\star,i}$ \citep[which we found to be more typical than 5~\rstar\ in an exploratory analysis of stellar magnetic field strengths and accretion rates in the literature, e.g.,][]{JohnsKrull2007}.
We then fit the models to the HST/STIS continua following the MCMC procedure of \cite{pittman22}. From this initial iteration, we derive \mdot$_{\rm shock}$ and the average temperature of the accretion hotspot, \tshock, for each CTTS. The average shock temperature is calculated by weighting the approximate blackbody temperature of each column's accretion shock spectrum by its best-fit filling factor: $T_{\rm shock} = [(f_{\rm low}T_{\rm low}^4 + f_{\rm med}T_{\rm med}^4 + f_{\rm high}T_{\rm high}^4)/f_{\rm total}]^{1/4}$.

\paragraph{Step 2. Flow model}
Next, we calculate the flow models using the following parameters as inputs: \mstar, R$_{\star,i}$, \teff, and \tshock.
Initial grids of $\sim$150,000 flow models are calculated for each CTTS. \mdot$_{\rm flow}$ spans a factor of 2 of \mdot$_{\rm shock}$ from \emph{Step~1}. The factor of 2 was chosen to sample the expected variability on timescales of a few days while limiting the computational cost. \tmax\ is inversely correlated with \mdot\ as shown in Figure~16 of \cite{muzerolle01}, so \tmax\ ranges over lower values for higher accretors, and higher values for lower accretors. In all cases, \tmax\ is chosen in steps of 350~K. For targets with existing disk inclination estimates, $i$ ranges from $\pm30$\degrees\ of the estimate in steps of 6\degrees. Otherwise, inclination ranges from 10--80\degrees\ in steps of 7\degrees. Finally, \ri\ ranges from 2--8~R$_{\star,i}$ in steps of 0.4~R$_{\star,i}$, and \rw\ ranges from 0.2--2~R$_{\star,i}$ in steps of 0.2~R$_{\star,i}$. 
While pairs of \ri\ and \rw\ could be estimated from the measured filling factor in \emph{Step~1}, we do not do this because the hotspot will rotate in and out of view with the stellar rotation, causing relatively rapid variability of the surface coverage observed from the continuum excess (see Section~\ref{subsec:hotspot_variability} for further discussion).

These grids are fit in velocity space to the wavelength-corrected and photosphere-subtracted \halpha\ profiles (produced as described in Section~\ref{subsec:data}). For most of the sample, the broad magnetospheric emission is much stronger than the narrow chromospheric emission, so the profiles are fit using the flow model alone. In some cases of lower accretors, however, the chromospheric component is not negligible. For those, we follow \cite{thanathibodee23} to include a Gaussian component in the fit, with a fixed center at 0~km/s and amplitude $A$ and width $\sigma$ as free parameters (up to $\sigma=35$~km/s). 

The regions in velocity space considered in the fitting are hand-selected for each observation. In general, when there are no blueshifted absorption features indicative of winds, the full region between $\pm500$~km/s is included in the fit. When such absorption is present, it is masked out because our model does not include winds.
Profiles that appear saturated at the line center are fit only at their higher-velocity broad wings.\footnote{This affects only two targets in the sample: DE~Tau and Sz~72. We see no evidence of non-linearity in the other CTTS observations.}

We use the \chisq\ statistic to measure the goodness-of-fit of each model for every individual epoch of observation. We choose the 100 best-fitting models for each observation, then ensure that none of the parameter distributions for those best fits are at the edge of the model grid range. If any parameter's distribution is at the edge, we expand the model grid and redo the fitting. When needed, we allow \ri\ to go down to 1.4~\rstar, expand the \rw\ range to 0.05--3~\rstar, and allow $i$ to go below 10\degrees\ or above 80\degrees. Finally, we calculate the median and standard deviation of each parameter to determine the results for each observation, weighting each model by exp(-\chisq/2). While the top 100 models are a small portion of the entire 150,000+ size grids, we find that increasing this number does not change the results because the exp(-\chisq/2) weights quickly reduce the contribution from worse-fitting models. 

The accretion shock models in \emph{Steps 3} and \emph{4} require two inputs from the accretion flow model: \ri\ and the energy flux density of the flow at the stellar surface, $\curf_{\rm flow}=\rho v_{\rm ff}^3/2$, where $\rho$ is the flow density and $v_{\rm ff}$ is the freefall velocity. This can be found from Equation 9 of \cite{hartmann94}, simplifying such that the angle at which the flow reaches the star is the average of the angles given by the innermost and outermost magnetic field lines dictating the flow of the material. To obtain \ri\ and $\curf_{\rm flow}$ for each system, we determine a ``typical'' accretion structure by taking the mean of the parameters measured for individual epochs of observation, weighted by the standard deviations of the parameters (effectively weighting the results from each epoch by the precision of its fit results). We then calculate a final accretion flow model profile using the weighted mean values for \mdot, \ri, \rw, \tmax, and $i$. We compare this final model profile to the observed \halpha\ profiles to confirm that they are consistent. In most cases, the weighted mean model effectively captures the typical observed profile. In the cases that it does not, we instead choose the results from the individual epoch that appears most typical given the available observations.\footnote{This is the case for DM~Tau, RXJ 1556.1-3655, Sz~69, Sz~77, Sz~98, and XX~Cha.}

\paragraph{Step 3. Shock model}
Next, we recalculate the accretion shock models using \ri\ from the flow model and choosing $\curf_{\rm low}=0.1\curf_{\rm flow}$, $\curf_{\rm med}=\curf_{\rm flow}$, and $\curf_{\rm high}=10\curf_{\rm flow}$. 
For particularly low $\curf_{\rm flow}$ ($\lesssim 10^{9}$~\escm), the lower-$\curf$ models are not successfully calculated. This is true for 11 out of the total 67 CTTS, and in those cases the exact values of $\curf_{\rm low,med,high}$ are typically increased by $\sim$2 orders of magnitude to produce successful model calculations.
We then fit the recalculated shock model spectra to the HST/STIS data using an MCMC procedure with 10000 steps, 200 walkers, and a burn-in of 1000. We visually inspect the walker chains to ensure convergence. 
Changing the extinction correction in our sample from that used by \citet[2025 in preparation]{manara21} when deriving $R_{\rm \star,i}$ changes the inferred stellar radii. Therefore, after correcting the CTTS spectra using the best-fit \av\ from the MCMC fit, we derive the final stellar radii $R_{\rm \star,f}$ based on the scaled flux of the WTTS template photosphere.

\paragraph{Step 4. Shock model}
Here we perform a final iteration of the shock modeling that recalculates the shock models using the stellar radius measured in \emph{Step 3}, $R_{\rm \star,f}$. This step provides the final values reported for the accretion shock model. 

\paragraph{Step 5. Flow model}
Finally, we re-calculate the weighted-mean accretion flow models using $R_{\rm \star,f}$ and the \mdot$_{\rm flow}$ adjusted by a factor of $R_{\rm \star,f}/R_{\rm \star,i}$ and confirm that the flow models still fit the \halpha\ profiles. These updated weighted-mean models are typically of equal or better fit quality than those calculated with $R_{\rm \star,i}$. This step provides the final values reported for the accretion flow model.

\section{Results} \label{sec:results}
Here we present the results from our accretion flow modeling (Section~\ref{subsec:flowresults}), accretion shock modeling (Section~\ref{subsec:shockresults}), and empirical correlation measurements (Section~\ref{subsec:correlations}).

\subsection{Accretion flow model} \label{subsec:flowresults}
\begin{figure}
    \begin{tikzpicture}
        \node at (0,0) {\includegraphics[width=0.49\linewidth]{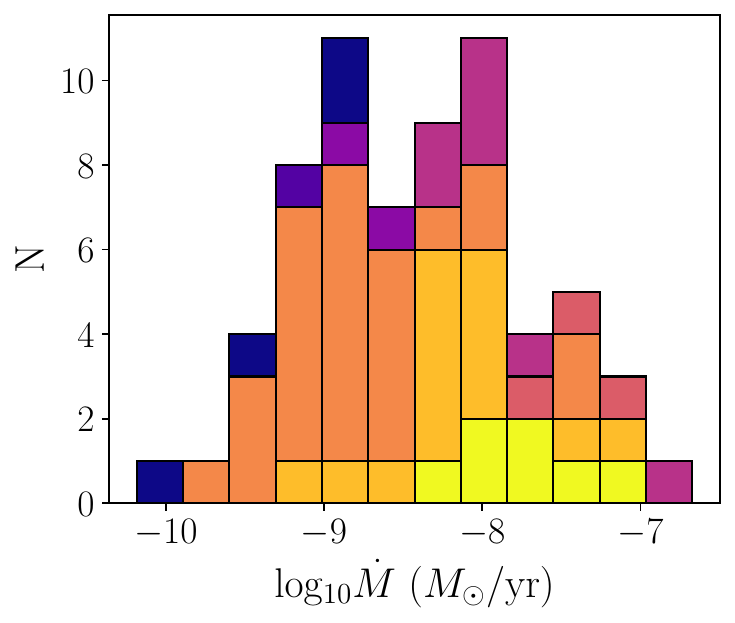}};
        \node[anchor=north east, font=\Large] at (2,1.7) {\textbf{a}};
        \node at (4.2,0) {\includegraphics[width=0.49\linewidth]{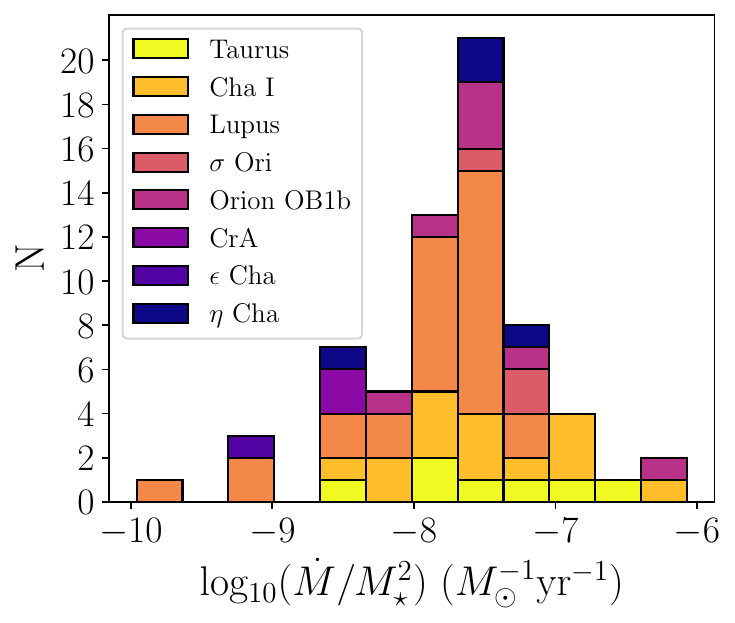}};
        \node[anchor=north east, font=\Large] at (6.2, 1.7) {\textbf{b}};
    \end{tikzpicture}
    \begin{tikzpicture}
        \node at (0,0) {\includegraphics[width=0.49\linewidth]{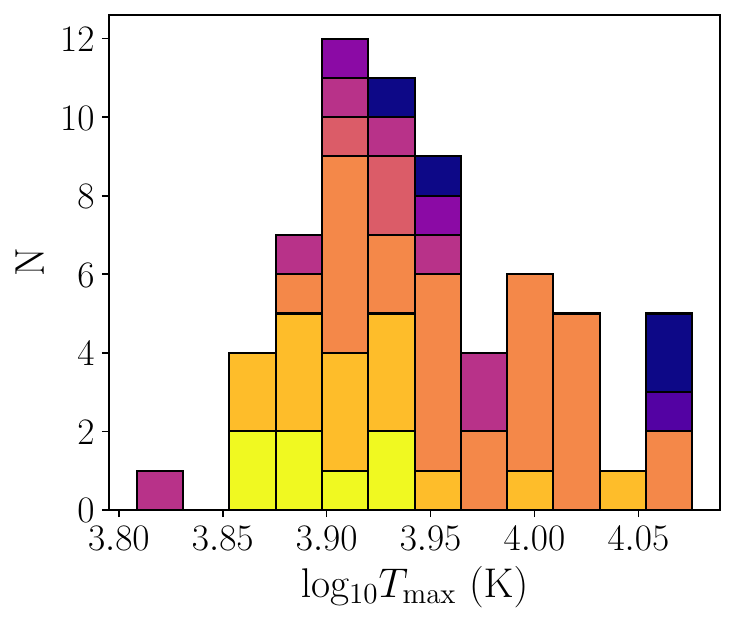}};
        \node[anchor=north east, font=\Large] at (2,1.7) {\textbf{c}};
        \node at (4.2,0) {\includegraphics[width=0.49\linewidth]{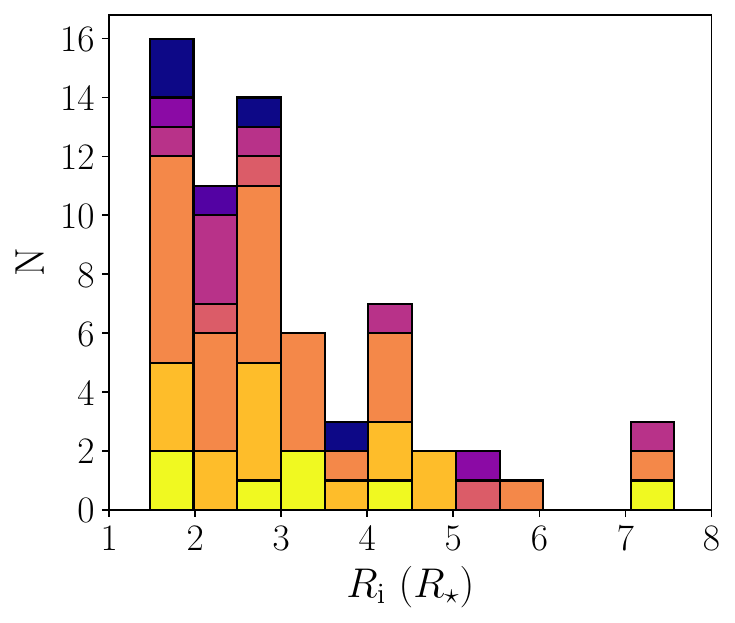}};
        \node[anchor=north east, font=\Large] at (6.2, 1.7) {\textbf{d}};
    \end{tikzpicture}
    \begin{tikzpicture}
        \node at (0,0) {\includegraphics[width=0.49\linewidth]{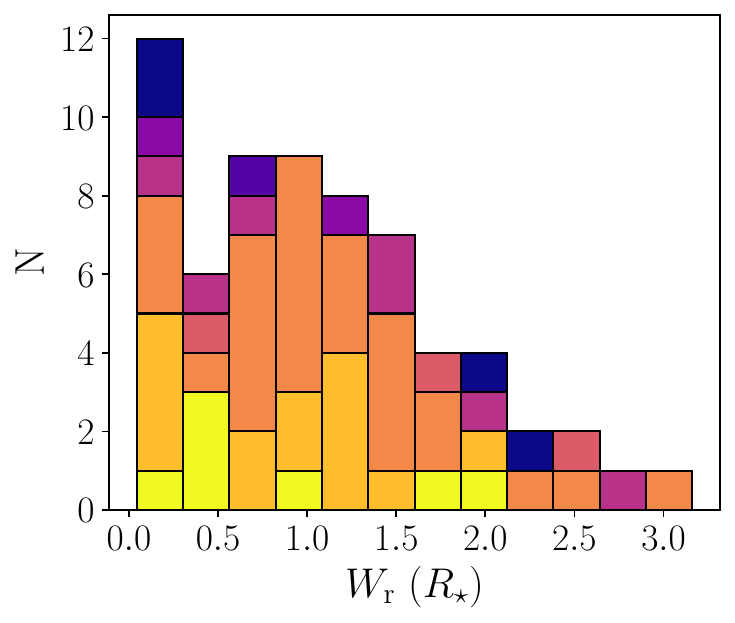}};
        \node[anchor=north east, font=\Large] at (2,1.7) {\textbf{e}};
        \node at (4.2,0) {\includegraphics[width=0.49\linewidth]{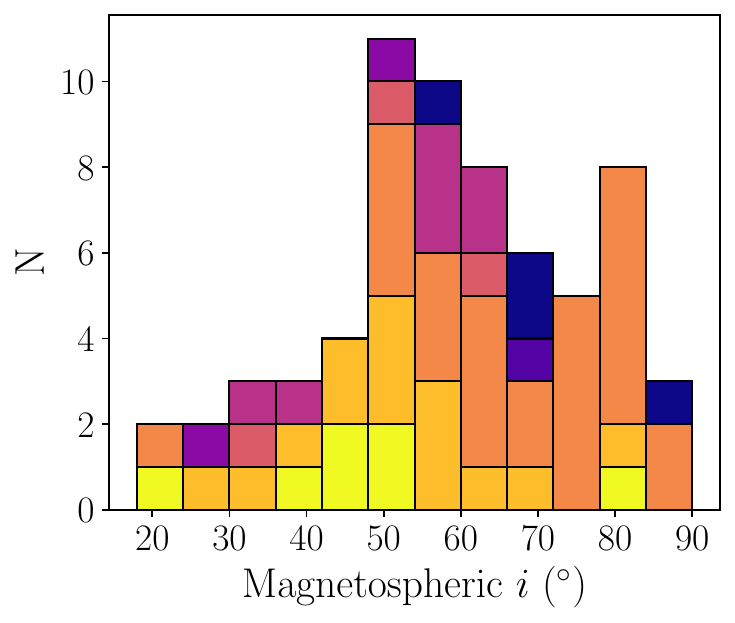}};
        \node[anchor=north east, font=\Large] at (6.2, 1.7) {\textbf{f}};
    \end{tikzpicture}
    \caption{Histograms of accretion flow model and stellar rotation results colored by region in order of approximate region age from youngest (yellow) to oldest (blue). Panels show a) accretion rate, b) accretion rate normalized by stellar mass according to $\dot{M}\propto M_\star^2$, c) maximum flow temperature, d) inner magnetospheric truncation radius, e) width of truncation region, and f) magnetospheric inclination. 
    }
    \label{fig:flow_hists}
\end{figure}
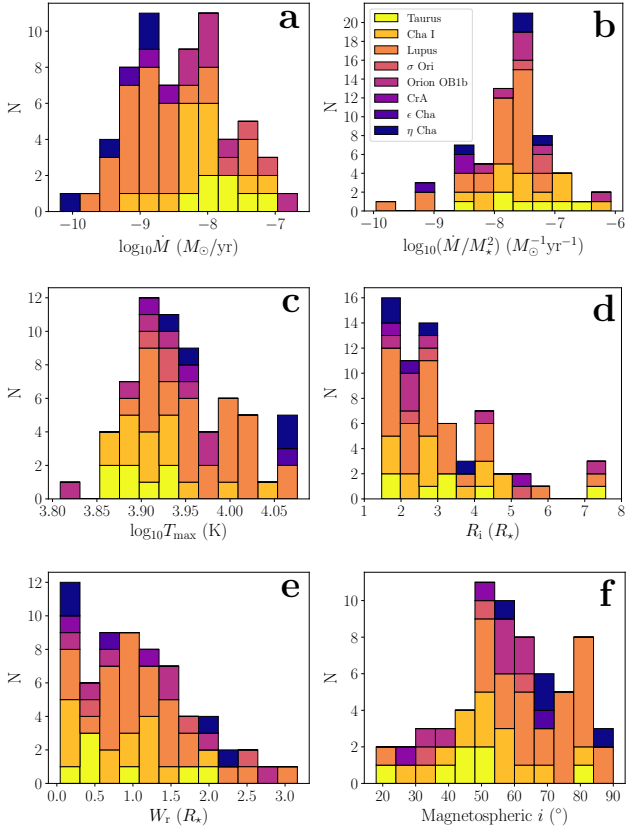

Table~\ref{tab:flowresults} in Appendix~\ref{Appsec:app_flowresults} provides the accretion flow model measurements of the inner magnetospheric truncation radius \ri, the radial width \rw\ between the inner and outer boundaries of the flow at the disk midplane, the accretion rate \mdot, the maximum flow temperature \tmax, the magnetospheric inclination $i$, and, for low accretors, the chromospheric Gaussian amplitude $A$ and width $\sigma$. 
It gives the weighted-mean results from combining the individual epochs as described in Section~\ref{subsec:flowmodel}, and the online machine-readable version contains results for every individual epoch of observation of each object. Figure~\ref{fig:flowfits} in Appendix~\ref{Appsec:app_flowresults} shows the \halpha\ profiles and weighted-mean model fits, and the online figure set contains the fits to individual epochs.
Figure~\ref{fig:flow_hists} shows the distributions of the weighted-mean model parameters for the sample.

Upon first glance at the distribution of accretion rates across different regions (Figure~\ref{fig:flow_hists}a), it might appear that the younger regions (Taurus and Chamaeleon I) have higher accretion rates than the older regions (Corona Australis, $\epsilon$ Cha, and $\eta$ Cha).
Additionally, the distribution appears slightly bimodal.
However, these are primarily stellar mass effects that disappear when the accretion rates are normalized by the approximate relationship $\dot{M}\propto M_\star^2$ \citep[][and references therein]{hartmann16,manara23PPVII}. As shown in Figure~\ref{fig:flow_hists}b, the $\log_{10}(\dot{M}/M_\star^2)$ distribution is unimodal and shows no clear delineation between older and younger regions.
The ULLYSES survey's intentional selection of targets with clear accretion signatures introduced a bias against low accretors, which are generally expected to be in older star forming regions. This bias, combined with the limited number of TTS in each region, might mask any \mdot-age correlations found by larger surveys \citep[e.g.,][]{SiciliaAguilar2010,Antoniucci2011,hartmann16,Betti2023}.
The maximum temperature of the accretion flow (Figure~\ref{fig:flow_hists}c) is inversely correlated with the accretion rate \citep{muzerolle01}, so its hint of age dependence is likely also a result of the stellar mass distribution.

Most previous magnetospheric accretion modeling has assumed that the magnetosphere truncates the inner disk at $\rm R_{\rm i}=5$~\rstar\ \citep[e.g.,][]{Gullbring1998,cg98,hh08,manara13,alcala17,re19,pittman22}. In contrast, we find a median of 2.8~\rstar\ and a mean of 3.0~\rstar, indicating a strong preference for magnetospheres smaller than 5~\rstar\ (Figure~\ref{fig:flow_hists}d). This agrees with the $R_{\rm i}<5$~\rstar\ values found for the ULLYSES monitoring targets by \cite{wendeborn24c}.

For a fixed accretion luminosity in an individual CTTS, $\dot{M}\propto(1-R_\star/R_{\rm i})^{-1}$. Thus, decreasing \ri\ from 5~\rstar\ to 2.8~\rstar\ would increase the accretion rate by a factor of 1.25. This 0.1~dex increase is relatively small compared to the 0.35~dex uncertainty associated with typical \mdot\ calculations \citep[e.g.,][]{alcala14,alcala17,manara23PPVII}. Therefore, existing \mdot\ measurements made assuming $R_{\rm i}=5$~\rstar\ should not be significantly affected. However, the accretion stability regimes predicted from 3D magneto-hydrodynamic (MHD) models are strongly dependent on \ri\ \citep{blinova16}. We will discuss this in detail in C. Pittman et al. (2025b, in preparation).
We note that while \cite{thanathibodee23} found a global inverse correlation between the measured \ri\ and \mdot\ for a sample of M-type low accretors (see their Figure~5), we do not find any global correlation in our sample of higher accretors. This may reflect differences in magnetic field strengths and configurations at different evolutionary stages.

The width of the magnetospheric flow region has a median and mean of 1.0~\rstar, and the distribution peaks at smaller values in its allowed range (Figure~\ref{fig:flow_hists}e).
The effect of \rw\ being small is to make the emitting region small. Supporting this, we find a moderate correlation between \rw\ from the flow model and the total filling factor from the shock model (Pearson correlation coefficient $r$ of 0.35). Similarly, \rw\ is moderately anticorrelated ($r$ of $-$0.36) with \tshock. Perhaps the flow model's assumption of axisymmetry causes it to choose small \rw\ as an approximation of a wedge-shaped flow that covers a smaller range of $\phi$, especially for systems with high accretion shock temperatures.

The median magnetospheric inclination for the entire sample is 58\degrees\ (Figure~\ref{fig:flow_hists}f), remarkably consistent with the expectation of median 60\degrees\ ($\cos i=0.5$) for random orientations. 
Individual regions have too few targets to sample the full range in inclinations, but the results suggest a potential tendency towards narrower ranges of inclinations within a given region. This, as well as comparisons to disk inclinations, will be discussed in detail in C. Pittman et al. (2025c, in preparation).
Overall, the standard deviations of the best-fit flow model parameters are low. For individual observations, the median standard deviations (in dex) of the top 100 model fits are: 0.04 (\ri), 0.12 (\rw), 0.16 (\mdot), 0.02 (\tmax), and 0.03 ($i$). Most of these uncertainties are comparable to the model grid spacing, indicating that the grid resolution drives most of the dispersion. However, the \mdot\ standard deviation is 0.10~dex larger than its grid spacing, suggesting a larger contribution from fit uncertainties. Combining the multi-epoch observations using a weighted mean halves the standard deviations, resulting in: 0.02 (\ri), 0.06 (\rw), 0.07 (\mdot), 0.01 (\tmax), and 0.01 ($i$).

\subsection{Accretion shock model} \label{subsec:shockresults}

\begin{figure*}
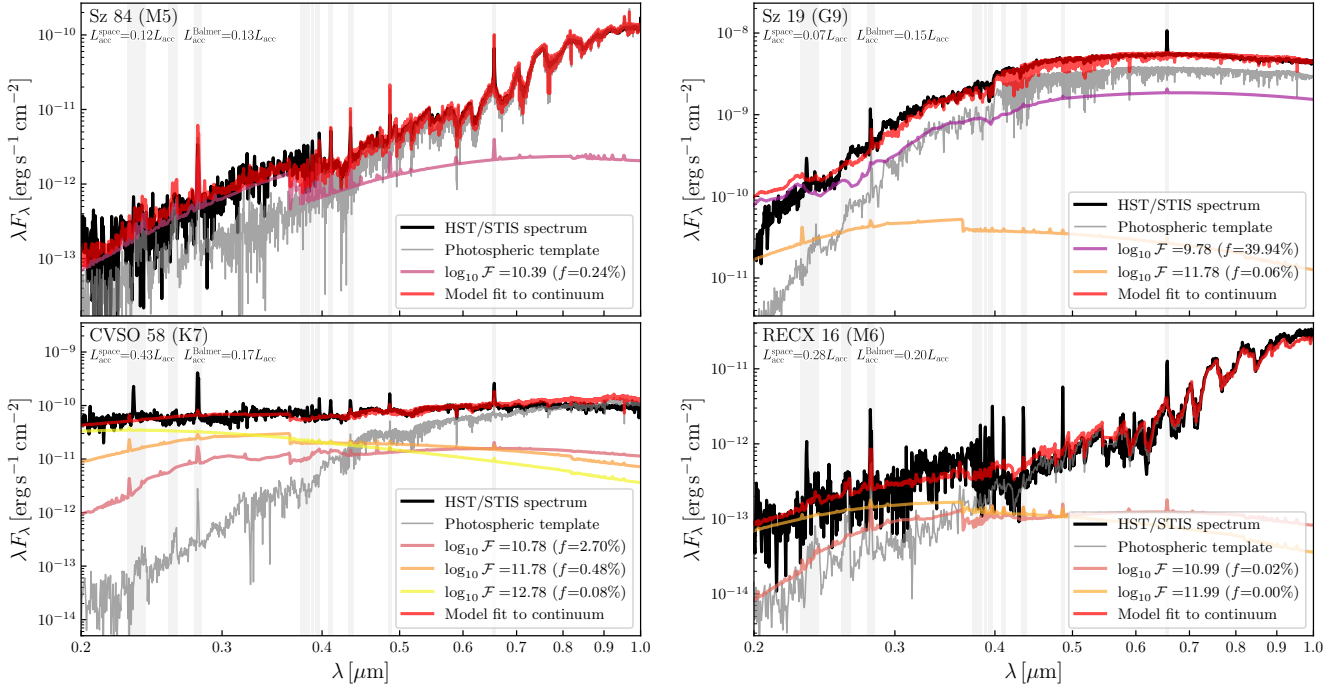

    \digitalasset
    \centering
    \includegraphics[trim=0 28pt 0 5pt, clip, width=0.49\textwidth]{sz84_fit.pdf}
    \includegraphics[trim=0 28pt 0 5pt, clip, width=0.49\textwidth]{sz19_fit.pdf}\\[-2.1mm]
    \includegraphics[trim=0 0 0 4pt, clip, width=0.49\textwidth]{cvso58_fit.pdf}
    \includegraphics[trim=0 0 0 4pt, clip, width=0.49\textwidth]{recx16_fit.pdf}
    \caption{Example shock model fits to NUV-NIR continua in Sz~84, Sz~19, CVSO~58, and RECX~16. \hst/STIS data are in black, and non-accreting TTS template spectra are in gray. Accretion shock model spectra are scaled by their associated surface coverages and shown as solid lines in the plasma colorscale, where more purple spectra are lower $\curf$ and more yellow spectra are higher $\curf$. Energy flux densities $\curf$ and filling factors $f$ are indicated in the legends. Spectral regions excluded from the fit are shaded in light gray. The percentage of \lacc\ that is accessible only from space is in the upper left corner, along with the percentage in the Balmer jump between 0.34 and 0.45~\micron. The complete figure set (74 images) is available in the online journal.
    }
    \label{fig:shock_short}
\end{figure*}

We perform accretion shock modeling of 67 unique CTTS, with DM~Tau and Sz~45 having multiple epochs of observation originally presented in \cite{re19}, producing a total of 74 CTTS observations.
Figure~\ref{fig:shock_short} shows 4 example fits, and the remaining fit figures are in the online figure set material. \figsetstart
\figsetnum{2}
\figsettitle{Accretion shock model fits to HST/STIS spectra}
\figsetgrpstart
\figsetgrpnum{2.1}
\figsetgrptitle{2MASS J11432669-7804454}
\figsetplot{figures/ShockFits/2massj11432669-7804454_fit.pdf}
\figsetgrpnote{Shock model fit to 2MASS J11432669-7804454.}
\figsetgrpend

\figsetgrpstart
\figsetgrpnum{2.2}
\figsetgrptitle{AA Tau}
\figsetplot{figures/ShockFits/aatau_fit.pdf}
\figsetgrpnote{Shock model fit to AA Tau.}
\figsetgrpend

\figsetgrpstart
\figsetgrpnum{2.3}
\figsetgrptitle{CHX18N}
\figsetplot{figures/ShockFits/chx18n_fit.pdf}
\figsetgrpnote{Shock model fit to CHX18N.}
\figsetgrpend

\figsetgrpstart
\figsetgrpnum{2.4}
\figsetgrptitle{CS Cha}
\figsetplot{figures/ShockFits/cscha_fit.pdf}
\figsetgrpnote{Shock model fit to CS Cha.}
\figsetgrpend

\figsetgrpstart
\figsetgrpnum{2.5}
\figsetgrptitle{CV Cha}
\figsetplot{figures/ShockFits/cvcha_fit.pdf}
\figsetgrpnote{Shock model fit to CV Cha.}
\figsetgrpend

\figsetgrpstart
\figsetgrpnum{2.6}
\figsetgrptitle{CVSO 58}
\figsetplot{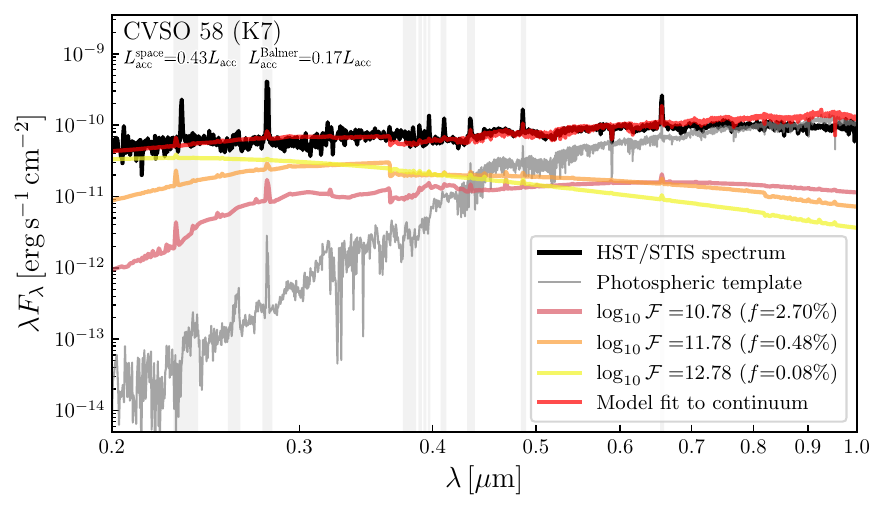}
\figsetgrpnote{Shock model fit to CVSO 58.}
\figsetgrpend

\figsetgrpstart
\figsetgrpnum{2.29}
\figsetgrptitle{RECX 16}
\figsetplot{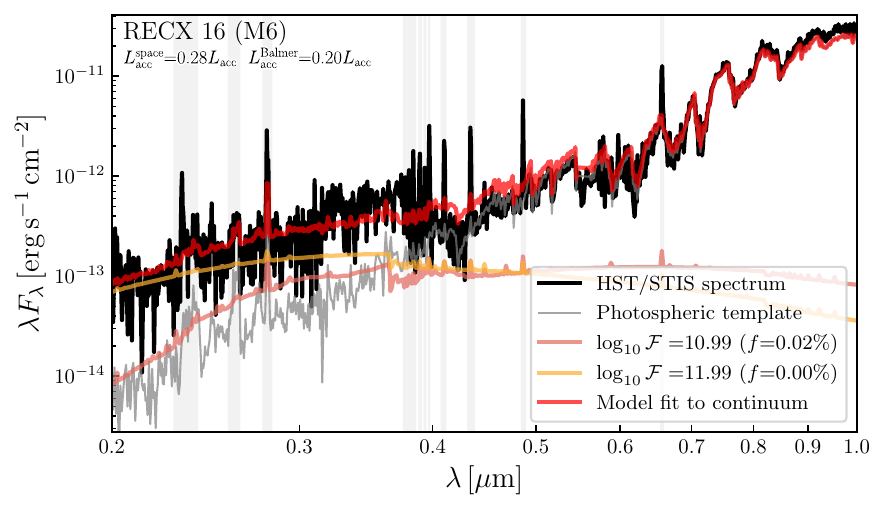}
\figsetgrpnote{Shock model fit to RECX 16.}
\figsetgrpend

\figsetgrpstart
\figsetgrpnum{2.40}
\figsetgrptitle{Sz 19}
\figsetplot{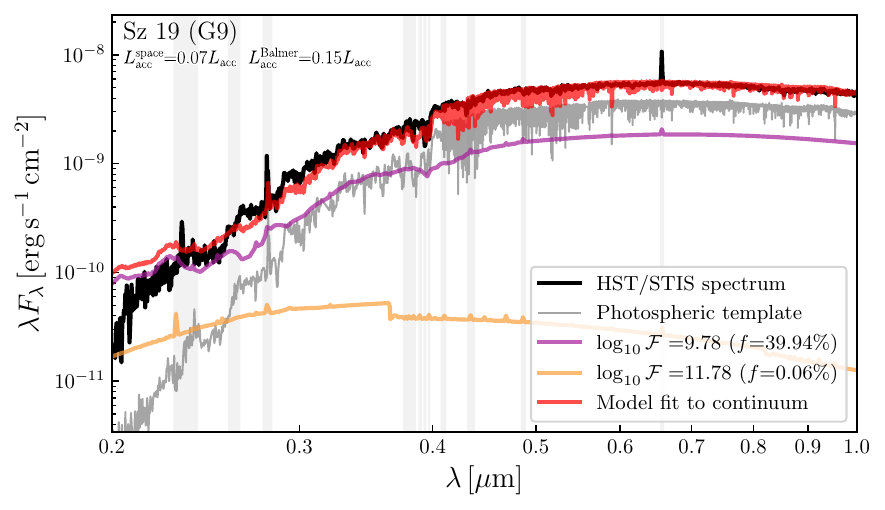}
\figsetgrpnote{Shock model fit to Sz 19.}
\figsetgrpend

\figsetgrpstart
\figsetgrpnum{2.55}
\figsetgrptitle{Sz 84}
\figsetplot{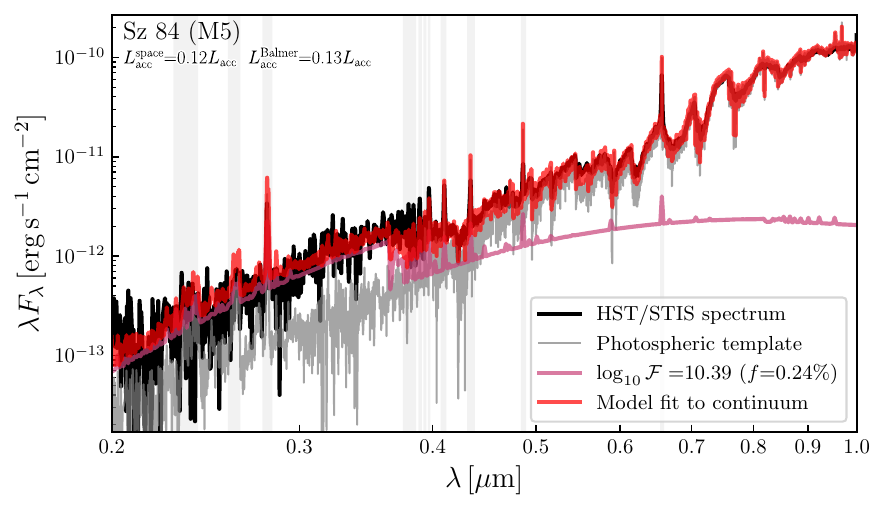}
\figsetgrpnote{Shock model fit to Sz 84.}
\figsetgrpend

\figsetend

Accretion rate and hotspot structure results are shown in Figure~\ref{fig:Mdot_barchart}. We find a large accretion rate range of $-9.5<\log_{10}(\dot{M}/{\rm M_\odot/yr})<-6.4$, consistent with the sample's diversity in stellar parameters, host regions, and ages.
We also find a large range in spectral energy distributions; some CTTS emit only 3\% of their total \lacc\ in the UV (and the rest is in the optical and infrared), whereas others emit up to 48\% of \lacc\ in the UV. All shock model results for the accretion rate \mdot,  extinction \av, filling factors $f_{\rm low,med,high}$, energy flux densities $\curf_{\rm low,med,high}$, and shock temperatures \tshock\ are given in Table~\ref{tab:shockresults} in Appendix~\ref{sec:app_shock}.

As discussed in Section~\ref{subsec:shockmodel}, we tailored the $\curf_{\rm low,med,high}$ and \ri\ for all systems using the results of the accretion flow model. 
While this ensures consistency and ideally gives a more accurate representation of specific system properties, we are interested in determining whether this added effort is significant for the sample-wide conclusions. 
We compared the shock model results that come from \emph{Step 1}, which used uniform $\curf$ and \ri\ ($\curf_{\rm low,med,high}=$10$^{10}$, 10$^{11}$, and 10$^{12}$ \escm\ and $R_{\rm i}=3.5$~\rstar), with those from \emph{Step 3}, which used measured values from the flow modeling. The $\log_{10}$(\mdot/\msunyr) distribution is statistically unchanged, with \emph{Steps 1} and \emph{3} having a median and standard deviation of $-8.2\pm0.6$. A slightly larger change occurs during the final iteration in \emph{Step 4}, in which we use an updated stellar radius derived using the extinction measured from the shock modeling. In that case, the distribution has a median and standard deviation of $-8.3\pm0.7$, a change of about 0.1~dex. However, accretion rates for individual CTTS can change appreciably, with a standard deviation of 0.26~dex between \emph{Step 1} and \emph{Step 3} and 0.15~dex between \emph{Step 3} and \emph{Step 4}. 

The mean energy flux density (\Fmean, calculated as the weighted average of $\curf_{\rm low,med,high}$ using the associated filling factors as weights) changes more significantly between iterations. In \emph{Step 1}, $\log_{10}$(\Fmean/\escm) has a median and standard deviation of $10.45\pm0.60$. In \emph{Step 3}, this distribution shifts down by 0.4~dex and becomes broader, with a median and standard deviation of $10.05\pm1.01$. This reflects the expanded parameter space explored by choosing $\curf_{\rm low,med,high}$ for each target. For individual CTTS, the typical difference between \emph{Step 3} and \emph{Step 1} is $-0.29\pm0.74$~dex. From \emph{Step 3} to \emph{Step 4}, $\log_{10}$(\Fmean/\escm) changes minimally. The final \emph{Step 4} distribution has a median and standard deviation of $10.08\pm0.92$, and object-level changes are around $-0.01\pm0.48$~dex. The most important sample-wide impact of the iterative modeling procedure is the change in the \Fmean\ distribution that results from using measurements from the flow model.

The statistical uncertainties from the MCMC fit, taken as the 16th and 84th percentiles of the posterior distributions, tend to be low. In dex, the median lower/upper uncertainties are 0.02/0.02 (\mdot), 0.06/0.07 ($f_{\rm low}$), 0.38/0.66 ($f_{\rm med}$), 0.06/0.07 ($f_{\rm high}$), and 0.03/0.03 (\av). However, propagating conservative uncertainties in \rstar\ (0.1~dex in our work), \mstar\ (0.1~dex in \citealt{manara21}), \lacc\ (0.25~dex in \citealt{manara21}), and \ri\ (0.04~dex in our work), our accretion rates should have a typical uncertainty of 0.38~dex.

\subsubsection{Hotspot structure} \label{subsec:hotspot_structure}

\begin{figure}
    \centering
    \includegraphics[width=0.8\linewidth]{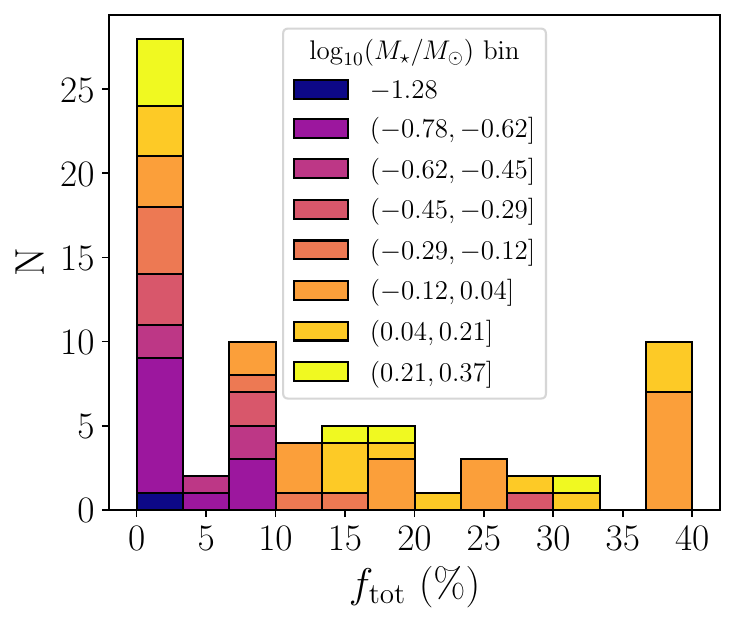}
    \caption{Histogram of the total hotspot surface coverage given by the accretion shock model. Bars are broken down into logarithmically-spaced stellar mass bins as indicated in the legend.}
    \label{fig:ftot_hist}
\end{figure}

We find that the logarithmic temperature difference between the accretion shock and stellar photosphere (i.e., $\log_{10}T_{\rm shock}-\log_{10}T_{\rm eff}$) is strongly anti-correlated with the total filling factor, with a Pearson correlation coefficient $r=-0.7$. This means that higher-energy hotspots cover smaller areas, as expected from theoretical predictions \citep[e.g.,][]{Zhu2025}.
As shown in Figure~\ref{fig:ftot_hist}, the total filling factors of the accretion hotspots tend to be low, with a median value of 9\%, a median absolute deviation (MAD) of 8\%, and a standard deviation of 14\%. This agrees with the typical 5-10\% surface coverage found by \cite{venuti15}. However, there are eleven stars with high veilings for which the surface coverage reaches the upper limit of the allowed range of 40\%: CHX18N, CVSO~107, CVSO~146, DK~Tau~A, SY~Cha, Sz~19, Sz~75, Sz~82, Sz~129, TX~Ori, and VZ~Cha. These stars are in the higher mass range of the ULLYSES sample, all having masses above 0.5~\msun. We compared their hotspot temperatures \tshock\ to their stellar temperatures \teff\, and the median value of \tshock$-$\teff\ is 600~K. This is a small difference that typically results from the accretion shock emission being dominated by the $\curf_{\rm low}$ column.

\cite{venuti15} also find a non-negligible distribution of stars with small $|$\tshock$-$\teff$|$ and large hotspot coverage and suggest that hotspot models may be ill-suited for explaining these observations. Perhaps a combination of accretion hotspots and starspots (whose emission peaks at longer wavelengths than the photosphere) could explain the emission of these sources. \cite{PerezPaolino2025}, for example, reproduce NIR excess spectra of CTTS using starspot models that cover $>$40\% of the stellar surface. Their starspot temperatures are in the same range as those obtained from fitting a blackbody to some of the $\curf_{\rm low}$ columns (though, it is important to note, the shock models are not blackbodies). Therefore, our $\curf_{\rm low}$ hotspot models and the \cite{PerezPaolino2025} starspot models may be accounting for similar observational signatures.

\begin{figure*}
    \centering
    \includegraphics[width=0.95\textwidth]{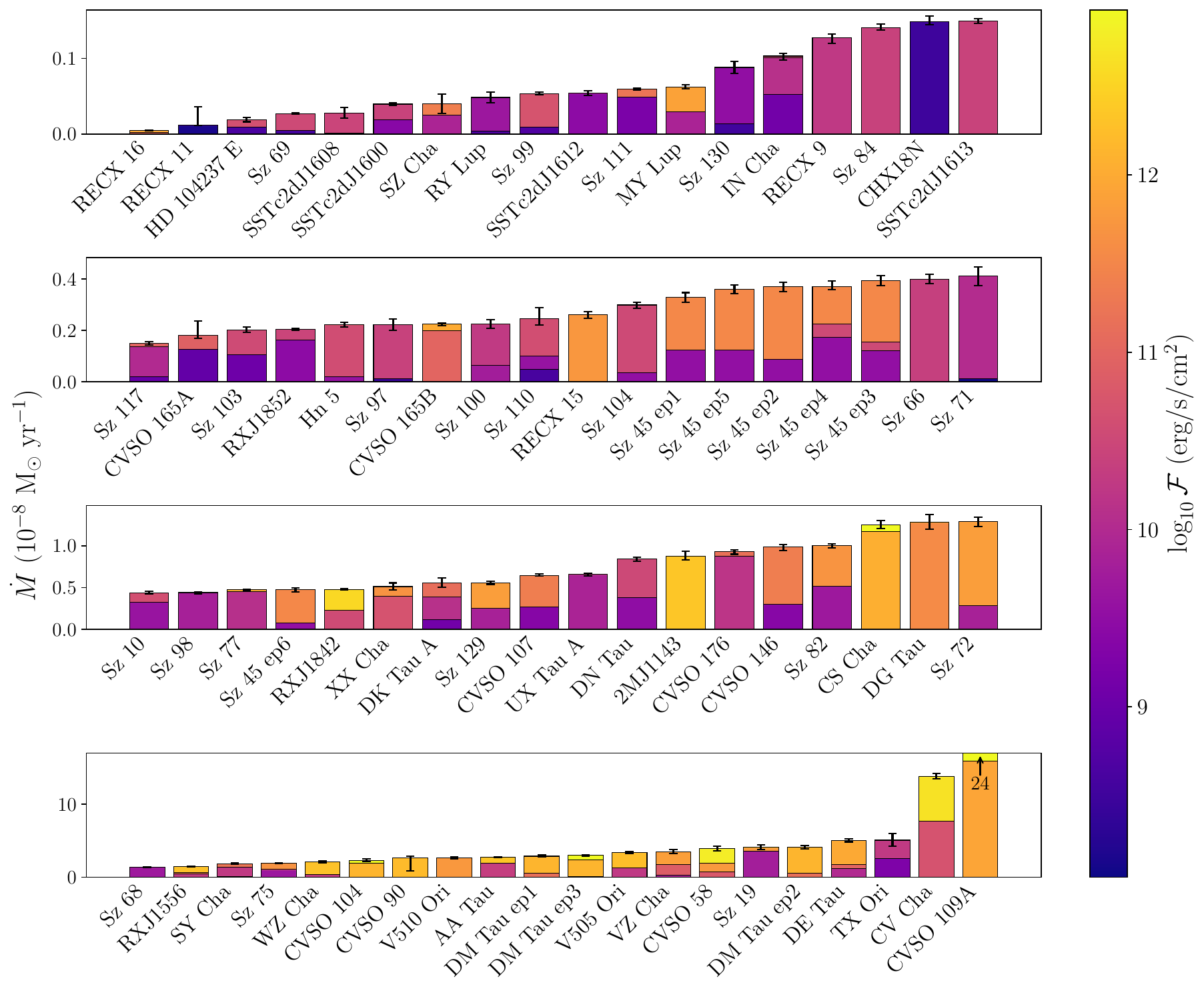} 
    \caption{Accretion rate results for each CTTS, ordered from low to high \mdot\ (note that the $y$-axis limits change between the subfigures). Each bar shows the contribution of accretion columns with energy flux densities $\curf$ given by the colorbar. 
    }
    \label{fig:Mdot_barchart}
\end{figure*}

The multi-column accretion shock model gives information on the non-uniform hotspot structure.
Figure~\ref{fig:Mdot_barchart} shows the hotspot configuration for each modeled \hst\ observation. We can divide results into one-, two-, and three-column models based on the number of $\curf_{\rm low, med, high}$ columns that significantly contribute to the total model flux. A given column is considered significant if it covers at least 0.1\% of the stellar surface in the case of $\log_{10}(\curf/{\rm erg/s/cm^2})\leq 10$ or at least 0.01\% in the case of higher $\curf$. 

Figure~\ref{fig:Mdot_barchart} indicates that many accretion excesses can be explained using a two-column model, with 44/74 observations having contributions from only two different $\curf$ accretion columns (e.g., Sz~19 and RECX~16 in Figure~\ref{fig:shock_short}). In many cases (27/44), the two-column model is a combination of $\curf_{\rm low}$ and $\curf_{\rm high}$. This indicates that emission from $\curf_{\rm med}$, which is typically offset from $\curf_{\rm low}$ and $\curf_{\rm high}$ by 1~dex, can often be accounted for by a combination of lower and higher energy emission in the model fitting. Our inspection of the MCMC corner plots shows a strong anti-correlation between $f_{\rm med}/f_{\rm tot}$ and $(f_{\rm low}+f_{\rm high})/f_{\rm tot}$, supporting this conclusion. 

One-column models are the second most common, accounting for 19/74 observations (e.g., Sz~84 in Figure~\ref{fig:shock_short}). Three-column models are the least common, accounting for only 11/74 cases  (e.g., CVSO~58 in Figure~\ref{fig:shock_short}). The majority (95\%) of all observations have significant contributions from the $\curf_{\rm low}$ column, and 82\% have significant contributions from $\curf_{\rm high}$. Only 64\% of observations have some of their total excess flux coming from $\curf_{\rm med}$. In cases with insignificant $\curf_{\rm med}$ contributions, the specific values of $f_{\rm med}$ in the MCMC chain can vary widely between the model's lower limit of $e^{-16}$ and the significance limit of 0.0001. This explains the large standard deviation of $f_{\rm med}$, which is typically over 0.3~dex higher than those of $f_{\rm low}$ and $f_{\rm high}$.

Two primary factors contribute to the observed hotspot configurations. First, physical density gradients seen in both 3D MHD models \citep[e.g.,][]{romanova04} and time-series observations \citep[e.g.,][]{espaillat21,Singh2024} require multiple emission components to explain the excesses. Second, the hotspot's azimuthal location at the phase at which the system is observed determines the visibility of the hotspot.
If the hotspot is at a quarter phase relative to the observer, for example, lower-density regions may be visible while higher-density regions are obscured. 
For sufficiently inclined stars, the hotspot can be completely occulted by the star, with the surrounding disk blocking any emission from the opposite hemisphere.

To examine whether the hotspot contributions we observe may be due to rotational modulation effects, we simulate hotspot rotation for this sample. For all stars, we assume an elliptical hotspot at a colatitude of 30\degrees\ with a surface coverage of 10\%.
It has an aspect ratio of 5 with its semi-major axis lying along the azimuthal direction. Comparing to the CTTS magnetic field models of \cite{Robinson2021}, the limiting latitudes and surface coverage of this assumed hotspot configuration would most closely correspond to a system with equal polar strengths of the dipolar and octupolar magnetic field components.
This remains consistent with the dipole approximation made by the accretion flow model, as the octupolar field weakens much faster than the dipole component as a function of distance. Observed magnetic field configurations vary significantly between different CTTS \citep[see][and references therein]{Gregory2010}, so a detailed consideration of the magnetic field configuration is beyond the scope of this work.

Points in the central 1\% of the hotspot have a dimensionless flux of 100; points in the next 10\% of the hotspot have a flux of 10; and the remaining outer ring of 89\% of points have a flux of 1. These approximate the $\curf_{\rm high,med,low}$ regions, respectively. Fluxes of all points outside the hotspot are 0. We incline each system using the measured magnetospheric inclinations presented in Table~\ref{tab:flowresults}. Finally, we rotate the sphere through phases of 0-360\degrees\ in steps of 1\degrees\ and project the fluxes onto the plane of the sky. Points facing the observer directly contribute more than those with a surface normal pointed further away (by Lambert's cosine law); this is to account for the reduced effective area of regions near the stellar horizon.

We create four categories of observed hotspot emission: \textit{0$\curf$} contains observations in which the hotspot is fully out of view on the hidden hemisphere;
\textit{1$\curf$} contains those in which up to 15\% of the maximum hotspot flux is observed, approximating the observation of the $\curf_{\rm low}$ region alone;
\textit{2$\curf$} contains those in which 15$-$50\% of the maximum hotspot flux is observed, approximating the observation of both the $\curf_{\rm low}$ and $\curf_{\rm med}$ regions; and
\textit{3$\curf$} contains observations of over 50\% of the total hotspot flux, approximating the observation of all three regions together. We place the simulated observations at each inclination/phase pair in these categories.

The simulated observations are categorized as follows: 3\% fall into category \textit{0$\curf$}, 19\% into \textit{1$\curf$}, 24\% into \textit{2$\curf$}, and 54\% into \textit{3$\curf$}.
Given that 3D MHD models often predict multiple hotspots at similar latitudes but different phases \citep[e.g.,][]{KulkarniRomanova2009,zhu2024,Zhu2025}, we test the effect of adding a second hotspot covering 5\% of the stellar surface and positioned 90\degrees\ from the initial spot in longitude. In this case, we find that 0\% of observations fall into category \textit{0$\curf$}, 19\% into \textit{1$\curf$}, 49\% into \textit{2$\curf$}, and 31\% into \textit{3$\curf$}.

These results indicate that we should expect no more than 3\% (2.2/74) of our CTTS to have been observed when none of the hotspot was in view. Cases of this could be SSTc2dJ161243.8-381503 and Sz~130. For these, the best-fit accretion shock emission is fainter than the photospheric emission at all wavelengths, and the hotspot temperatures found by fitting a blackbody to the shock model spectra are cooler than those of the stellar photospheres (by 14~K and 60~K, respectively). This is true for these 2 CTTS alone. Both targets show broad \halpha\ emission in excess of their chromospheric emission, so they are not WTTS. Thus, they were likely observed when the hotspot was obscured behind the star. 

The model prediction that 31$-$54\% of observations should fall into the \textit{3$\curf$} category does not align with our finding that only 11/74 observations have significant contributions from all three accretion columns.
Given our earlier proposition that $\curf_{\rm med}$ may be accounted for by the combination of $\curf_{\rm low}$ and $\curf_{\rm high}$ at the model-fitting stage, the two-column accretion shock results composed of $\curf_{\rm low}$ and $\curf_{\rm high}$ may correspond to the majority of the hotspot being in view (the \textit{3$\curf$} category). If so, we can add the 11 three-column models to the 27 $\curf_{\rm low+high}$ two-column models to find an adjusted \textit{3$\curf$} occurrence rate of 51\%. The remaining two-column models (consisting of 13 cases of $\curf_{\rm low+med}$ and 4 cases of $\curf_{\rm med+high}$) would imply an adjusted \textit{2$\curf$} occurrence rate of 23\%. This would bring our findings into agreement with the simulation predictions for the observed hotspot configurations.

\subsubsection{Hotspot structure variability} \label{subsec:hotspot_variability}

\begin{figure*}
    \centering
    \includegraphics[width=\linewidth]{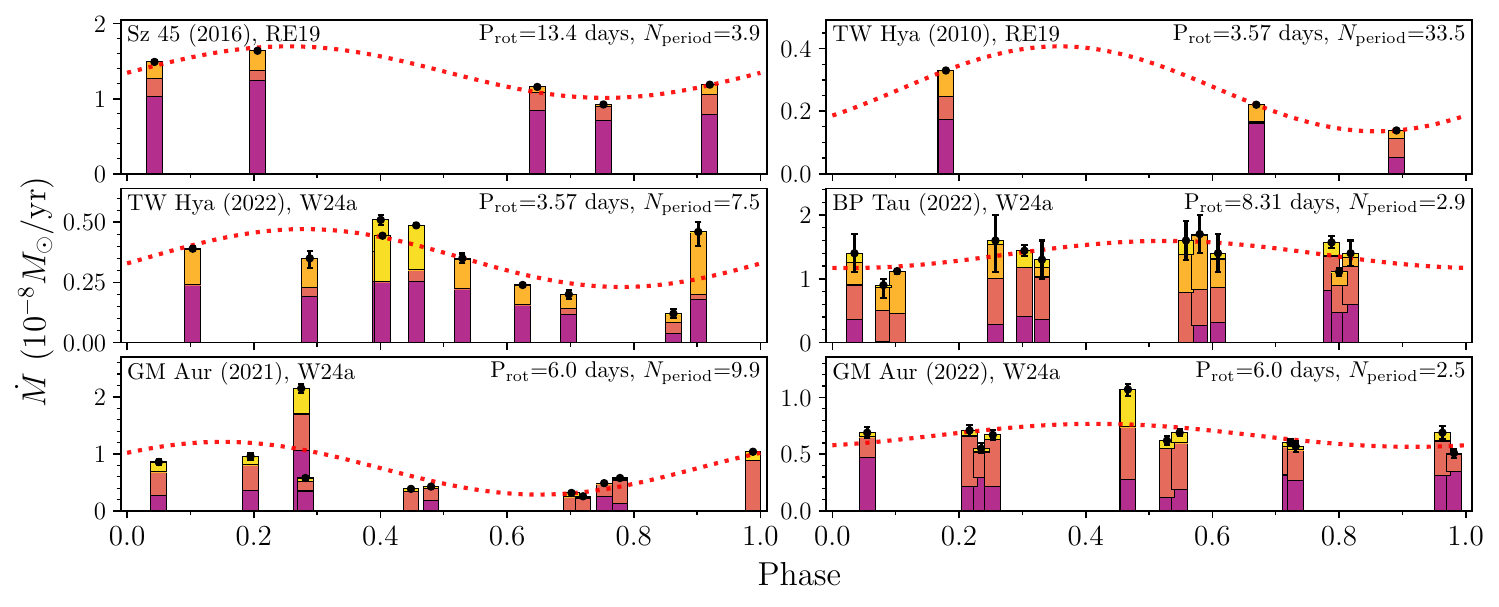} 
    \caption{Phase-folded accretion rates (black circles with errorbars) divided into their constituent $\curf$ components (colored bars) for multi-epoch accretion shock modeling presented by \citet[RE19]{re19} and \citet[W24a]{wendeborn24a}. These results are not included in Figure~\ref{fig:Mdot_barchart}, as the modeling was performed by the aforementioned authors and is simply phase-folded here. The year of observation is shown in parentheses. The assumed rotation period and total number of periods spanned are in the upper right corners. The colorbar is the same as Figure~\ref{fig:Mdot_barchart}, with the four $\curf$ values being 10$^{10}$, 10$^{11}$, 10$^{12}$, and 10$^{12.5}$~\escm. Sine function fits to \mdot\ are shown as dotted red lines to guide the eye.}
    \label{fig:phasefolded_Mdot}
\end{figure*}

Rotational modulation of accretion emission has been detected through photometric, spectroscopic, and spectropolarimetric observing campaigns and is often the dominate source of accretion variability in individual CTTS \citep[e.g.,][]{Donati2007,Donati2008,venuti14,Costigan2014,Rigon2017,espaillat21,CampbellWhite2021,bouvier2023}.
Most observations presented here are single snapshots, so they cannot easily be linked to the stellar rotation phase. However, we can use the multi-epoch analyses by \cite{re19} and \cite{wendeborn24a} to shed light on the typical rotational modulation of the hotspot emission. 

\cite{re19} presented multi-epoch accretion shock modeling of \hst/STIS spectra, and their Sz~45 and TW~Hya observations are sufficiently close in time to phase-fold their modeling results.
\cite{wendeborn24a} performed accretion shock model fits to \hst/COS spectra of the ULLYSES monitoring targets (TW~Hya, RU~Lup, BP~Tau, and GM~Aur). Each star was observed about 24 times: four times per rotation period, for three consecutive rotations, once in 2021 and once in 2022.
Then, \cite{wendeborn24b} completed periodogram analysis of multi-band light curves for the monitoring targets, and the following showed strong periodicity at the stellar rotation period for $u'$-, $B$-, and $g$-bands: TW~Hya 2022, BP~Tau 2022, and GM~Aur 2021 and 2022. These three targets have magnetic obliqities estimated around 10\degrees--20\degrees\ \citep{Donati2008,Donati2011b,McGinnis2020,donati2024}, which should produce non-axisymmetric hotspots \citep{KulkarniRomanova2013}. We choose these periodic epochs look for evidence of rotational hotspot modulation. 
For TW~Hya, BP~Tau, and GM~Aur, we use the \prot\ adopted by \citet[3.57d, 8.31d, and 6.0d, respectively]{wendeborn24a,wendeborn24b}, and for Sz~45, we use the period of 13.4~days found in the TESS analysis to be presented in C. Pittman et al. (2025b, in preparation). BP~Tau's periodicity ranges from 8.19 to 8.31 days; we chose the latter because it was the stronger signal in 2022.

Figure~\ref{fig:phasefolded_Mdot} shows the phase-folded accretion rates broken into their constituent $\curf$ components, similar to Figure~\ref{fig:Mdot_barchart}. \cite{re19} used $\curf=$10$^{10}$, 10$^{11}$, and 10$^{12}$~\escm\ accretion columns, and \cite{wendeborn24a} added a fourth with $\curf=$10$^{12.5}$~\escm. Each of the six panels shows evidence of periodicity in \mdot\ and hotspot $\curf$ distribution modulation. Sine curves are plotted to guide the eye, but we note that deviations from perfect sinusoids are expected given the effect of intrinsic accretion variability. 

Using the Akaike Information Criterion (AIC), which accounts for model complexity, we found that a sine curve model better fits the data than the null hypothesis (a constant accretion rate, $\dot{M}=\bar{\dot{M}}$) in all cases except GM~Aur (2022). The AIC indicates that the sine model is between 2.6 (TW~Hya, 2022) and 200 (Sz~45, 2016) times more probable than the null hypothesis.
Interestingly, there is no clear correlation between the magnetospheric inclinations measured from the flow model and the strength of sinusoidal variability.

Particularly notable is the rotational modulation of energy flux densities $\curf$. The 2022 observations of TW Hya\footnote{Though TW Hya's disk is viewed nearly face-on, the 18\degrees\ stellar inclination \citep{AlencarBatalha2002} and 20\degrees\ magnetic obliquity \citep[][]{donati2024} can explain the rotational modulation of its accretion signatures, as hotspots tend to be near the magnetic pole \citep[][]{McGinnis2020}. It has also been found to show episodic accretion at low latitudes, which would increase rotational modulation \citep{Donati2011b,argiroffi2017}.} and GM Aur, in particular, exhibit a concentration of strong $\curf=10^{12.5}$~\escm\ contributions around phases of 0.4 and 0.5, respectively. BP~Tau also shows a clustering of higher $\curf=10^{12}$~\escm\ surface coverage near phase 0.6. These observations suggest that the accretion shock models are in fact tracing the $\curf$ distributions of shock-induced hotspots as they rotate with the star. While it might seem unexpected that the observed accretion rates in Figure~\ref{fig:phasefolded_Mdot} remain non-zero near sine curve minima, this is consistent with the hotspot simulations presented in Section~\ref{subsec:hotspot_structure}. For the magnetospheric inclinations of these targets \citep[ranging from 16.7\degrees\ in TW~Hya to 53.1\degrees\ in GM~Aur in][]{wendeborn24c}, the simulations indicate that the hotspot should always be partially in view. Further, the hotspot shape predicted by \cite{KulkarniRomanova2013} for 10\degrees\ to 20\degrees\ magnetic obliquities is an azimuthally-stratified ring (i.e., a higher-energy arc linked by a lower-energy band), resulting in non-zero emission at all phases.

The TESS light curves of TW~Hya, GM~Aur, and BP~Tau presented in \cite{wendeborn24b} are also consistent with a hotspot rotating in and out of view. \cite{wendeborn24b} classified all contemporaneous TESS light curves of GM~Aur and TW~Hya as bursty, meaning that brightness changes from the mean level tend to be positive rather than negative or symmetric. The 2022 TESS light curve of BP~Tau also tends towards bursts, though not strongly enough to be considered a ``burster'' by definition of \cite{Cody14}. Burstiness is to be expected if the primary change to the optical brightness over time comes from a hotspot emitting above the undisturbed photosphere more strongly at certain phases.

The phase-folded accretion rates span at least 2.5 rotation periods for each CTTS. Sz~45 was observed over 52 days -- four observations over 6 days, then a gap of 46 days until the final observation -- which is 3.9 rotation periods. In 2010, TW~Hya was observed three times over a period of 119 days -- two observations 6 days apart, then a gap of 113 days until the final observation -- which is 33.5 rotation periods. In 2022, TW~Hya was observed ten times over 27 days, or 7.5 rotation periods. In 2022, BP~Tau was observed twelve times over 24 days, or 2.9 rotation periods. Finally, GM Aur was observed 11 times in 2021 over 60 days and 11 times in 2022 over 15 days, corresponding to 9.9 and 2.5 rotation periods, respectively. Therefore, any periodicity of the hotspot signatures indicates somewhat long-lived accretion structures in these systems, as we discuss more in Section~\ref{sec:discussion}. This is consistent with the 2021$-$2022 observations of GM~Aur by \cite{bouvier2023}, which showed accretion-driven continuum and emission line modulation at the stellar rotation period that persisted for at least 15, and up to 30, rotational periods.

\subsection{Correlations} \label{subsec:correlations}
Here we discuss correlations and empirical relationships with our accretion results.


\subsubsection{\mdot$_{\rm flow}$ vs \mdot$_{\rm shock}$} \label{subsec:MdotflowMdotshock}
\begin{figure}
    \centering
    \includegraphics[width=\linewidth]{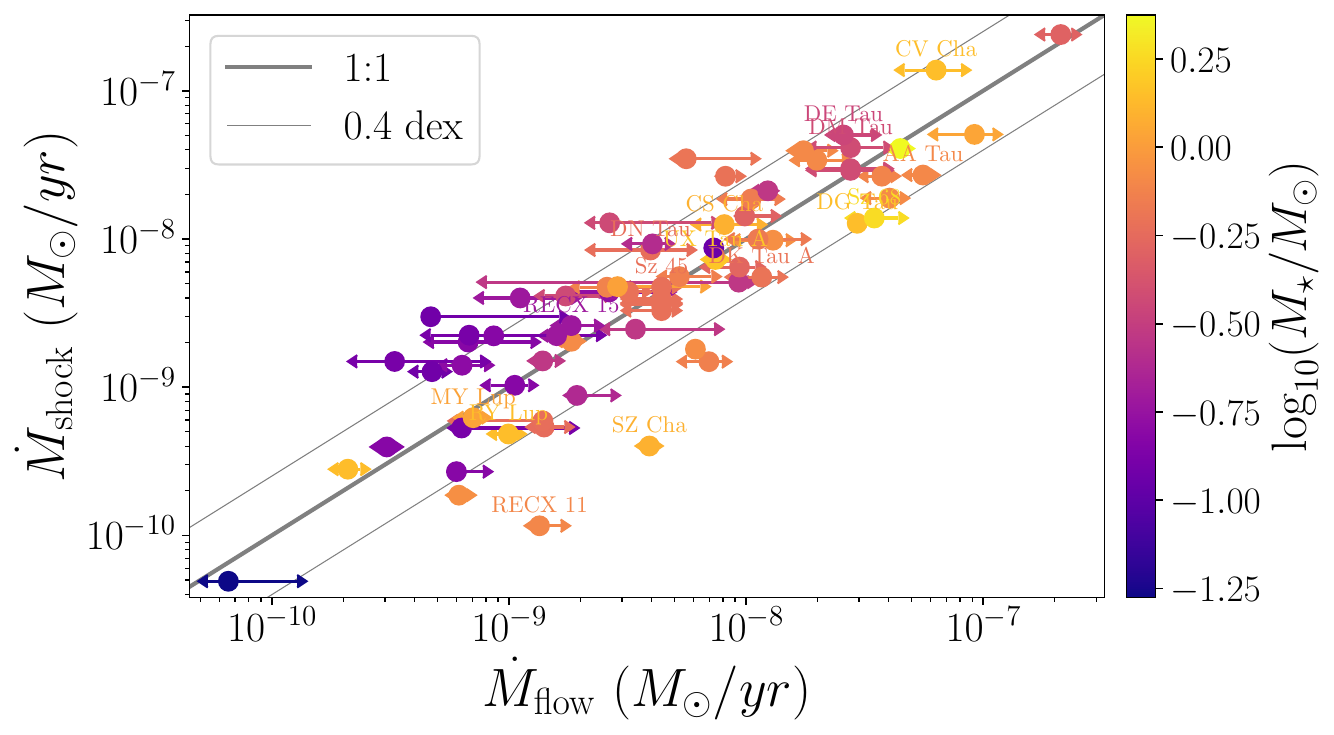}
    \caption{Comparison between accretion rates measured using the accretion shock and accretion flow models. Points are marked at the location of the weighted mean \mdot$_{\rm flow}$, and the arrows show the range between the minimum and maximum accretion rates found across all epochs of observation. The colorbar shows the stellar mass. Object names indicate the archival ULLYSES targets whose HST spectra were taken years before the \halpha\ spectra. The gray lines show a one-to-one relationship and a 0.4~dex deviation, which is the typical variability-induced spread of \mdot\ for a given CTTS \citep[][]{manara23PPVII}. The accretion rates are strongly correlated, with a Pearson correlation coefficient of 0.87. 
    }
    \label{fig:Mdot_comp}
\end{figure}

Figure~\ref{fig:Mdot_comp} shows that the accretion rates found from the flow model fits to \halpha\ profiles and shock model fits to UV-optical spectra are highly correlated, with a Pearson correlation coefficient of 0.87.
The 0.35~dex standard deviation between $\log_{10}$\mdot$_{\rm shock}$ and $\log_{10}$\mdot$_{\rm flow}$ is consistent with the expected variability-driven spread of 0.4 to 0.6~dex for a given CTTS \citep[][and references therein]{fischer23_ppvii,manara23PPVII,Betti2023}.
Previous work has found that this amplitude of variability occurs on timescales of a few days to weeks \citep[][]{Costigan2014,venuti15,Rigon2017} or months \citep[][]{Zsidi2022}. 
The timescales in this work are typically on the short end of the range, with a median time difference ($\Delta t$) between \halpha\ and UV observations of 6.1~days for the entire sample and 1.9 days for the new ULLYSES observations. This is also the typical timescale of CTTS stellar rotation \citep[$\sim$6.1 days,][]{venuti17}.

\begin{figure*}
    \centering
    \includegraphics[width=0.49\textwidth]{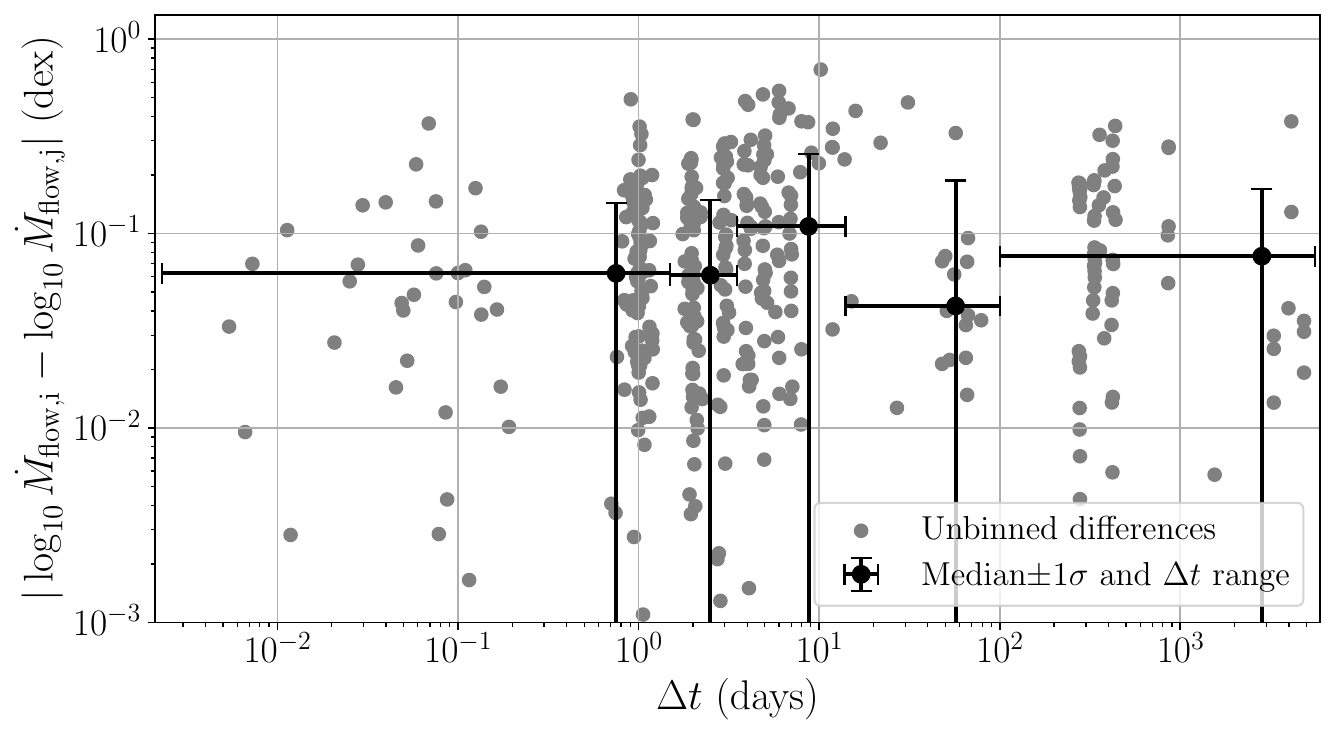}
    \includegraphics[width=0.49\textwidth]{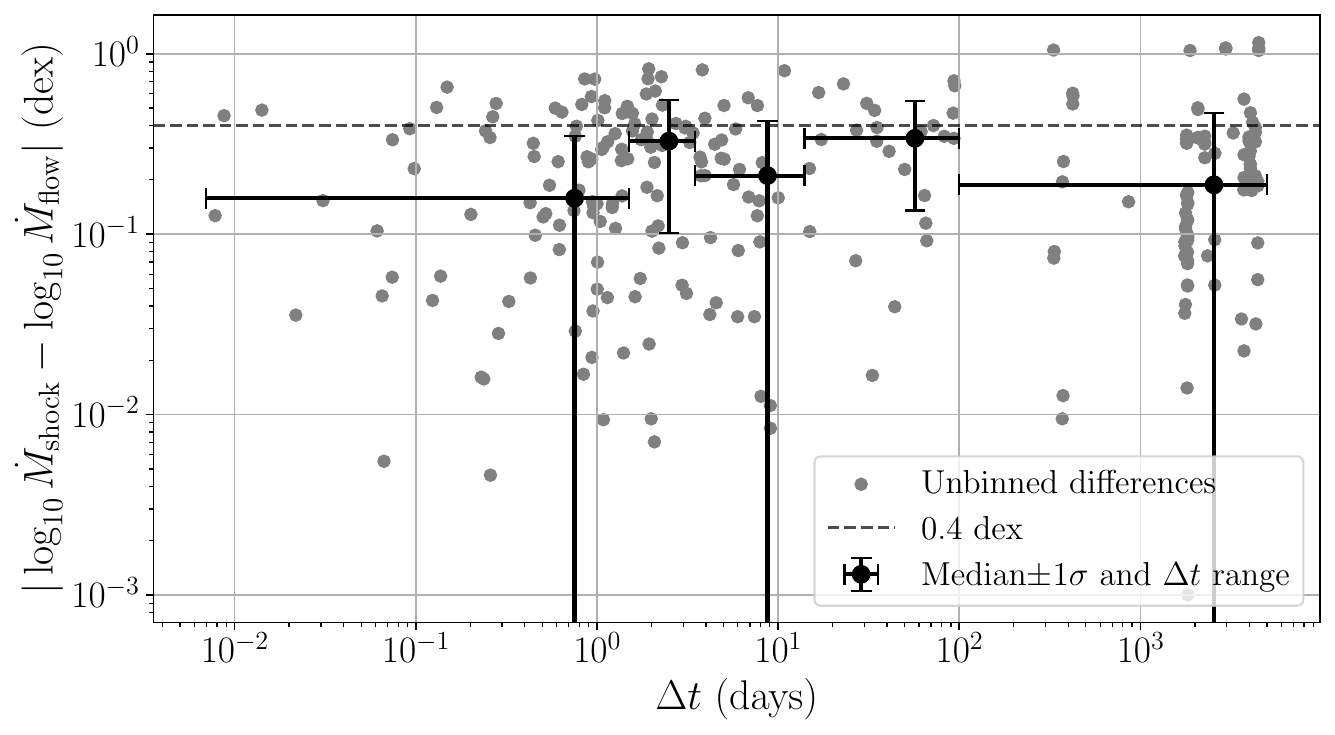} 
    \caption{Accretion variability over different timescales. \textit{Left}: Gray points show pairwise variability of accretion rates from the flow model fits to multiple epochs of \halpha\ observations. Black points show the median and 1$\sigma$ standard deviation within the time bins indicated by the $x$-axis errorbars. The median variability within $\sim$1 day is 0.06~dex with a standard deviation of 0.08~dex. \textit{Right}: Gray points show variability of accretion rates between the shock model fits to UV-optical spectra and flow model fits to multiple epochs of \halpha\ observations. Black points are the same as on the left figure. The baseline variability within $\sim$1 day is 0.16~dex with a standard deviation of 0.19~dex.
    }
    \label{fig:Mdot_deltat}
\end{figure*}

To determine whether the \mdot$_{\rm shock}$-\mdot$_{\rm flow}$ scatter originates solely from intrinsic variability, we look at the accretion rate variability as a function of $\Delta t$ between observations. First, we perform a pairwise comparison between the multiple epochs of flow model fits for each target, shown in Figure~\ref{fig:Mdot_deltat} (left). Most of the observations are within $\sim$3~days of each other according to the design of the PENELLOPE program (median$(\Delta t)=2.8$~days). The first bin contains the \mdot$_{\rm flow}$ pairs within $\sim$1~day of each other. For these, the median \mdot$_{\rm flow}$ difference is 0.06~dex.
Then, we compare the accretion rates from the accretion shock model with the multi-epoch flow model results in Figure~\ref{fig:Mdot_deltat} (right). In this case, the median difference between measurements with $\Delta t\lesssim1$~day is 0.16~dex.
This implies that there is a baseline difference between the accretion flow and shock models of 0.1~dex. The median difference of $\log_{10}$(\mdot$_{\rm shock}$/\mdot$_{\rm flow}$) is only +0.03~dex, so there is no strong trend for one method to produce systematically higher accretion rates than the other.

The remaining portion of the total 0.35~dex scatter (0.25~dex) likely comes from intrinsic variability. 
The \mdot$_{\rm flow}$ variability amplitude is consistent between the $\Delta t\lesssim1$~day and $1\lesssim\Delta t\lesssim2$~days bins, but it shows an increase on timescales out to 14~days as expected. The longer timescales are not as well sampled, but they do not show continued increases in the variability amplitude. The \mdot$_{\rm shock}$--\mdot$_{\rm flow}$ variability increases from 0.16~dex in the $\Delta t\lesssim1$~day bin to 0.33~dex in the $1\lesssim\Delta t\lesssim2$~days bin. It reaches its highest value of 0.34~dex in the months-scale bin of $14\lesssim\Delta t\lesssim100$~days. The longer timescales return to a lower baseline of 0.19~dex.

Some of the scatter between \mdot$_{\rm shock}$ and \mdot$_{\rm flow}$ may result from the different physical visibilities of the accretion flow and hotspot as they rotate with the star. The flow and shock models also make different assumptions about the geometry of accretion: the accretion flow model assumes that the footprint of the accretion flow is an axisymmetric ring on the stellar surface, whereas the accretion shock model is agnostic to the location or shape of the emitting region.

The overall agreement between the two accretion rates lends credence to the extinction correction performed in the shock modeling. If the \av\ measurements were introducing bias or producing erroneous accretion rates, $\dot{M}_{\rm shock}$ should be notably different
from the extinction-independent $\dot{M}_{\rm flow}$ measurements. Though the iterative nature of the modeling procedure may contribute to the agreement, the flow model grids were expanded as needed to ensure that the best-fit accretion rate was not at the edge of its allowed range.

\subsubsection{\mdot\ vs \mstar} \label{subsec:MdotvMstar}
\begin{figure}
    \centering
    \includegraphics[width=\linewidth]{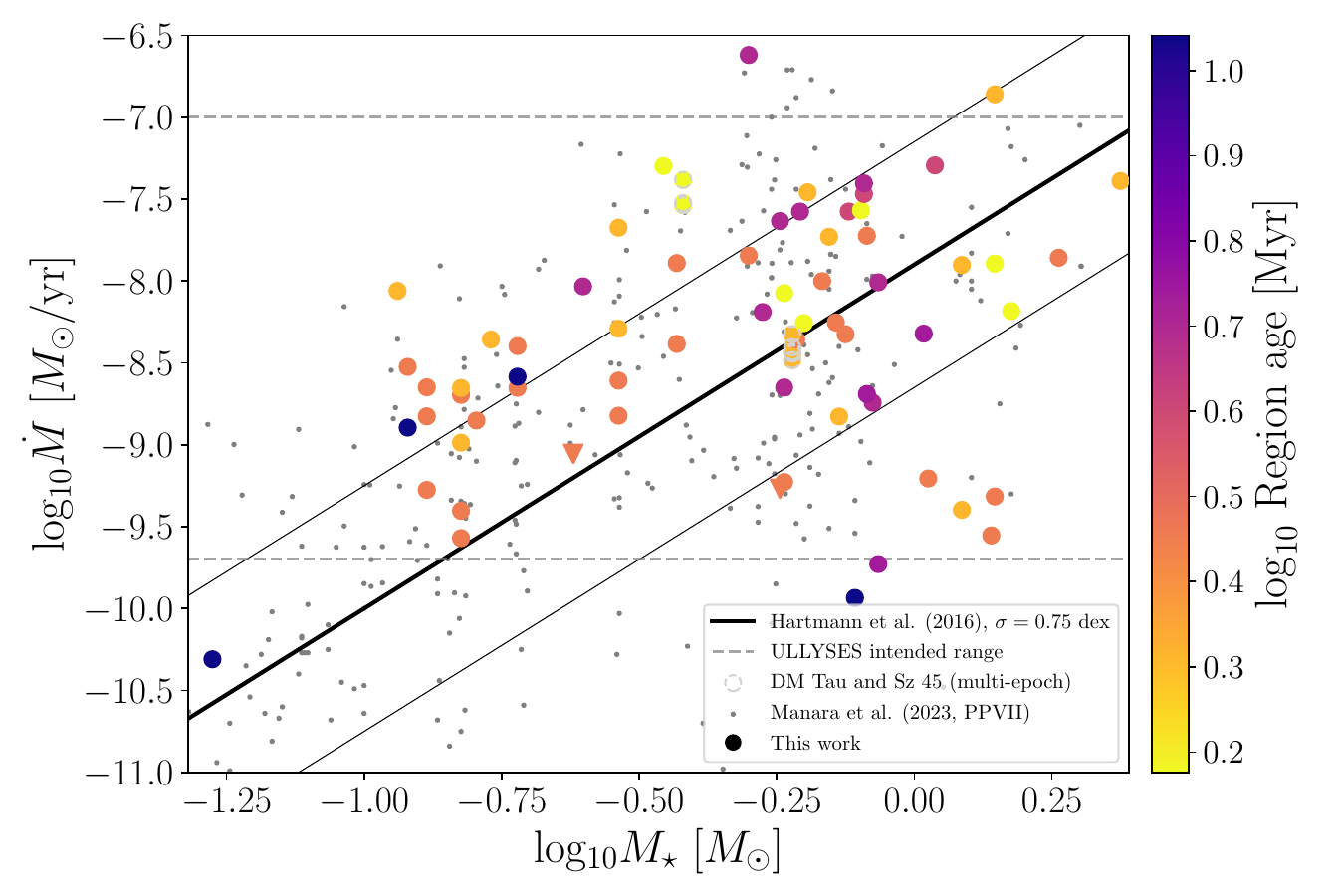}
    \caption{Accretion rate \mdot\ from the shock model vs stellar mass \mstar, with points colored by the median age of the host region. The two likely non-detections (SSTc2dJ161243.8-381503 and Sz~130) are marked as triangles. The \cite{hartmann16} \mdot-\mstar\ relation found from $U$-band excesses is plotted (thick black line), along with its standard deviation of 0.75~dex (thin black line). ULLYSES targeted CTTS with $-9.7<\dot{M}<-7$~\msunyr\ (dashed gray lines), and our accretion rates are consistent with that sample bias. Small gray points show the \mdot-\mstar\ pairs from \cite{manara23PPVII}, which are consistent with our results. 
    }
    \label{fig:MstarMdot}
\end{figure}

The relationship between \mdot\ and \mstar\ for the accretion shock model is shown in Figure~\ref{fig:MstarMdot}. The most conspicuous feature is the sparse region around $-0.71<\log_{10}M_\star/M_\odot<-0.31$. This is most likely an observational bias. The ULLYSES \mstar\ distribution is slightly bimodal, so there are only 10 CTTS in this central mass range range. 
Additionally, the ULLYSES Working Group specified a target range of $-9.7<\log_{10}\dot{M}/(M_\odot yr^{-1})<-7$, indicated by the dotted gray lines in Figure~\ref{fig:MstarMdot}.
Given this bias, this \mdot--\mstar\ distribution is broadly consistent with that presented in \cite{hartmann16}, whose fitted \mdot\ were derived from $U$-band spectra of 0.1-1.0~\msun\ CTTS. It is also consistent with the \mdot--\mstar\ distribution of the 278 CTTS compiled by \cite{manara23PPVII}.
A larger sample with better \mdot--\mstar\ coverage would be necessary to derive a new \mdot--\mstar\ relation because of the large intrinsic \mdot\ spread for a given \mstar\ \citep[e.g.,][]{venuti14}. As is, the correlation coefficient is weak, with $r=0.3$, and the inferred relationship would be shallow, with a slope of $\sim$0.7 rather than the values of 1.5-3.1 found in previous work on larger samples \citep[][and references therein]{hartmann16}.

\subsubsection{Correlations with \lacc} \label{subsec:LHalpha_LFUV}
While a significant fraction of the accretion luminosity (\lacc) in TTS can be at ultraviolet wavelengths (up to $\sim50$\% in this work), space-based UV observations are often unavailable. It is therefore helpful to determine empirical relationships between high-fidelity \lacc\ measurements and ground-based observables \citep[e.g.,][]{hh08,ingleby13,alcala14,alcala17}. The \halpha\ emission line, which is emitted by the accretion flow in excess of the stellar chromospheric emission, is readily-accessible from the ground and has been found to strongly correlate with \lacc\ \citep[e.g.,][]{ingleby13,alcala17,re19}.
\lacc\ is measured from the accretion shock model as $4\pi R_\star^2f\cdot{\cal F}$.
Here, we measure \halpha\ luminosity (\Lhalpha) from the same extinction-corrected \hst/STIS spectra used for the shock model, so \lacc\ and \Lhalpha\ are nearly simultaneous.\footnote{STIS/G230L and STIS/G430L spectra are taken consecutively, typically during the same orbit, producing an offset of less than 90 minutes.} To do so, we subtract a linear fit to the STIS continuum around \halpha, and then obtain the total flux $f_{H_\alpha}$ using trapezoidal integration. Finally, we convert this to luminosity using $L_{H_\alpha}=4\pi d^2f_{H_\alpha}$. The luminosity measurements are given in Table~\ref{tab:Lacc} in Appendix~\ref{Appsec:app_flowresults}.

\begin{figure}
    \centering
    \includegraphics[width=\linewidth]{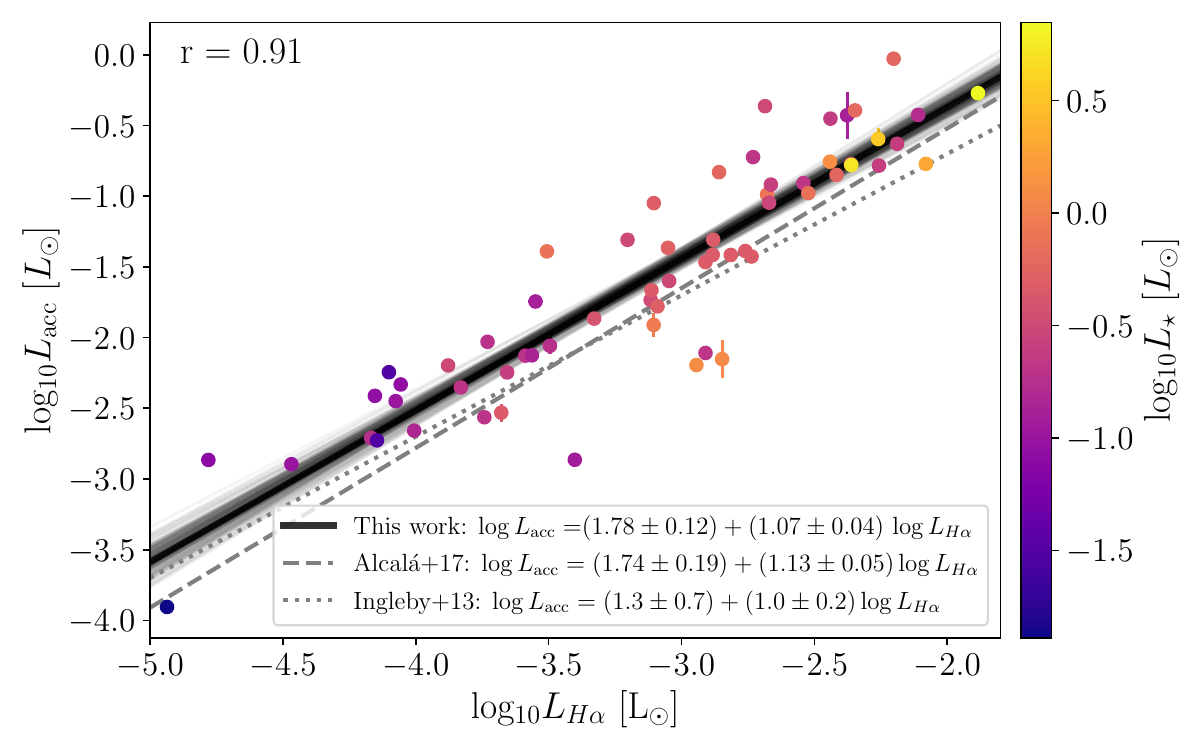} 
    \caption{
    \lacc\ and \Lhalpha\ are strongly correlated, with a Pearson correlation coefficient of 0.91. The colorbar indicates $\log_{10}$\lstar\ and shows that both \lacc\ and \Lhalpha\ are correlated with \lstar.
    The low-opacity solid gray lines show 200 bootstrapped linear fits, and the thick black line shows the median relation from this work. The dashed gray line shows the empirical \lacc-\Lhalpha\ relation from \cite{alcala17}, and the dotted gray line shows the relation from \cite{ingleby13}.
    The four outlier points below the fit lines are Sz~69, CHX18N, Sz~111, and SZ~Cha from lower left to upper right.
    } 
    \label{fig:Fmean_v_Lhalpha}
\end{figure}

As in previous work, we find that \lacc\ is strongly correlated with the \halpha\ luminosity, \Lhalpha, with a correlation coefficient of 0.92 as shown in Figure~\ref{fig:Fmean_v_Lhalpha}. This correlation is in large part driven by the dependence of \lacc\ on \lstar, as indicated by the colorbar. The best-fit relation to our 62 \lacc-\Lhalpha\ pairs is $\log_{10}{L_{\rm acc}}= (1.78 \pm 0.12) + (1.07 \pm 0.04)\log_{10}{L_{H\alpha}}$. The points have a standard deviation of 0.31~dex around the relation. Figure~\ref{fig:Fmean_v_Lhalpha} also includes the relations from \cite{alcala17}, who fit 87 TTS having a large \lacc\ range of $-5.2<L_{\rm acc}/L_\odot<-0.3$, and \cite{ingleby13}, who fit 10 CTTS having $-3.3<L_{\rm acc}/L_\odot<-0.7$. Our relation is shifted up from the other two, but all three are consistent within uncertainties.

\begin{figure*}
    \centering
    \includegraphics[width=\linewidth]{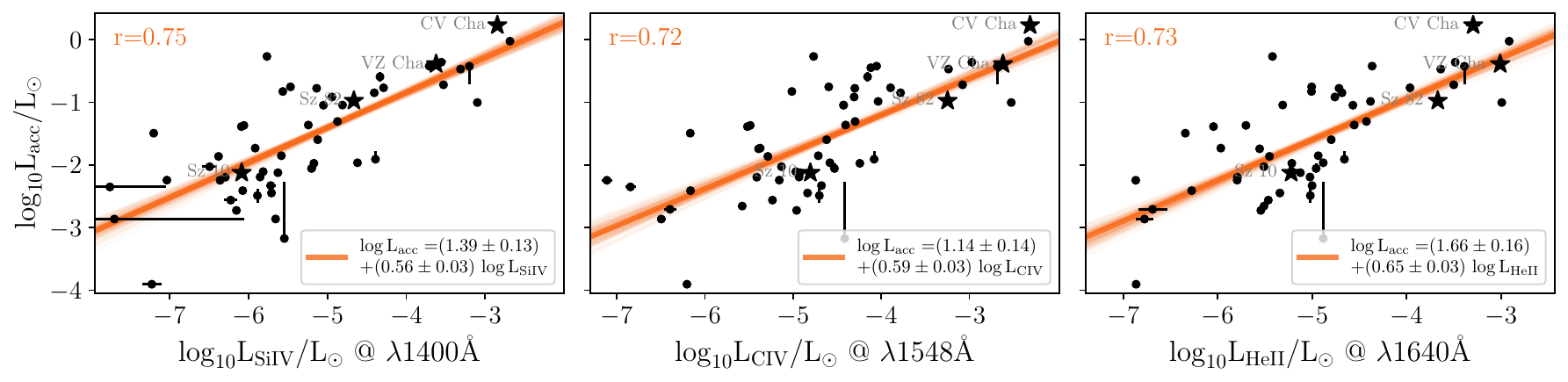}
    \includegraphics[width=\linewidth]{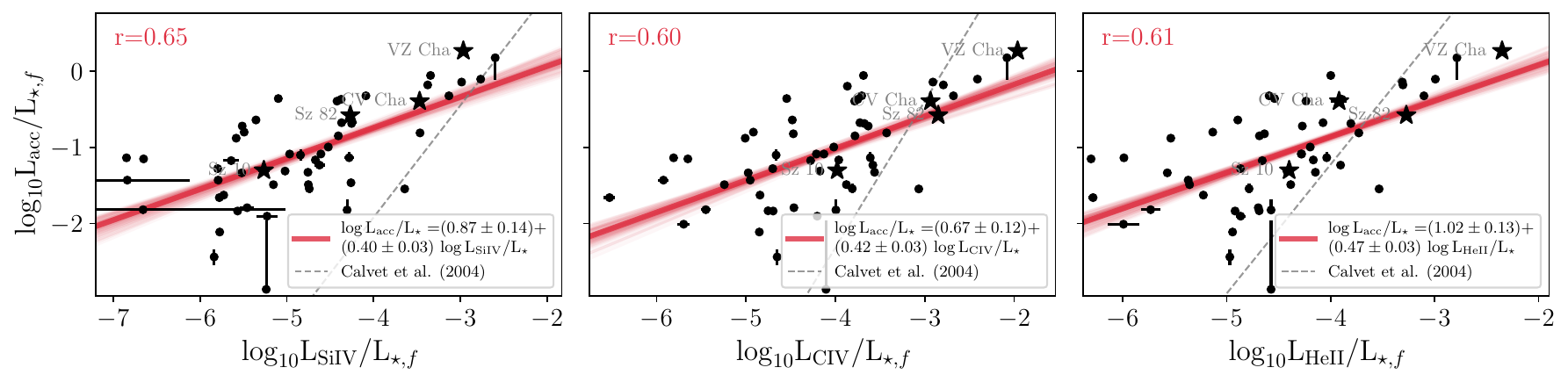}
    \includegraphics[width=0.37\linewidth]{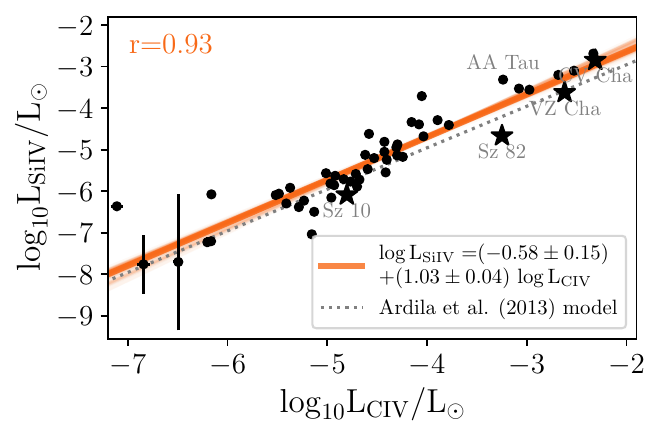}\includegraphics[width=0.37\linewidth]{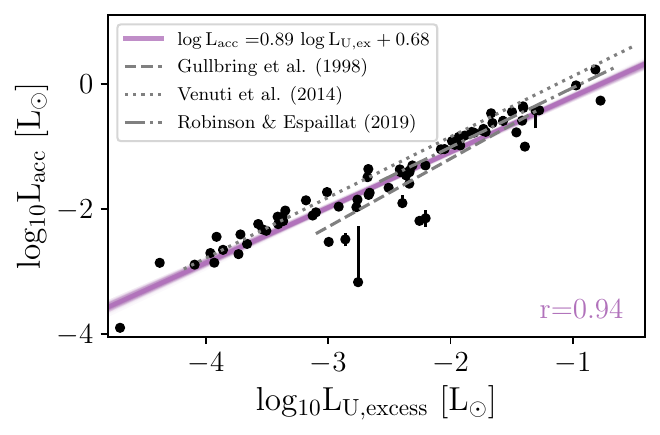}
    \caption{Empirical relationships between accretion, FUV line, and $U$-band continuum luminosities. In the orange and red subfigures, black circles indicate values taken from \cite{france23}, and black stars indicate fluxes measured in this work. In the purple subfigure (bottom right), all points are our measurements. Low-opacity colored lines show 100 bootstrapped linear fits, fitting 85\% of points at a time. Thick colored lines show the median best-fit line, with the equation indicated in the legend. The Pearson correlation coefficient is given in colored text.
    \textit{Top}: Accretion luminosity versus \siiv, \civ, and \heii\ emission line luminosities in units of \lsun. \textit{Middle}: Same as above, but in units of L$_{\star,f}$. The best-fit relations found from \cite{calvet04}, which were fit to a sample of 9 intermediate-mass TTS, are plotted as gray dashed lines for comparison.
    \textit{Bottom left}: ${\rm L_{SiIV}}$ versus ${\rm L_{CIV}}$ plotted with the line ratio model for the pre- and post-shock emission from \cite{Ardila2013}. AA~Tau, which was analyzed by \cite{Ardila2013}, is also marked.
    \textit{Bottom right}: ${\rm L_{acc}}$ versus excess $U$-band luminosity ${\rm L_{U, excess}}$ plotted along with relations from the literature. The extent of the lines indicates the ${\rm L_{U, excess}}$ range over which the relations were determined.
    }
    \label{fig:LaccLFUV}
\end{figure*}

We also find a strong correlation between excess $U$-band luminosity and \lacc, with r=0.94, as shown in Figure~\ref{fig:LaccLFUV} (bottom right). $\rm L_{U,ex}$ is found by convolving the dereddened CTTS and scaled template WTTS spectra with the Bessel $U$-band transmissivity, integrating under both, and then subtracting the WTTS contribution. Our best-fit relation, $\log_{10}{(L_{\rm acc}/L_\odot)}= (0.89\pm0.02)\log_{10}(\rm L_{U,ex}/L_\odot)+(0.68\pm0.04)$, is overall consistent with those from previous works \citep{Gullbring1998,venuti14,re19,wendeborn24b}. The points have an intrinsic scatter of 0.3~dex around the relation.
The relation of \cite{Gullbring1998} is the most different, but their relation was derived from a smaller sample of 17 CTTS with a smaller range of $\rm L_{U,ex}$ ($\rm -3.1 < \log_{10}L_{U,ex}/L_\odot <-1.3$), which likely explains the discrepancy. The consistency of our results with the ground-based work of \cite{Gullbring1998} and \cite{venuti14} reinforces the value of using $U$-band excesses to estimate \lacc.

Ultraviolet emission line luminosities have been found to correlate with accretion diagnostics, indicating a shared physical origin \citep{calvet04,ingleby13,Ardila2013,re19}. \cite{france23} studied the FUV continua and emission lines in 1200-1800~\AA\ \hst/COS spectra from ULLYSES. 
Here, we have taken their line flux measurements\footnote{\url{https://lasp.colorado.edu/cusp/projects/outflows-and-disks/stars-spectral-line-and-fuv-data/stars-data/}} for \siiv\ $\lambda1400$\AA, \civ\ $\lambda1548$\AA, and \heii\ $\lambda1640$\AA\ and adjusted them to our measured extinctions. Following the procedure in \cite{wendeborn24a}, we measured line fluxes for four additional stars not included in \cite{france23}: CV~Cha, Sz~10, Sz~82, and VZ~Cha. 

All three FUV line fluxes are strongly correlated with the accretion luminosities from our shock model, as shown in Figure~\ref{fig:LaccLFUV} (top row). They all have $r\ge0.72$ and standard deviations of $\sim$0.6~dex around the best-fit relations. The correlations remain even when any explicit dependence on stellar luminosity and distance is removed by normalizing \lacc\ and L$_{\rm line}$ by \lstar\ (Figure~\ref{fig:LaccLFUV} middle row, $r\ge0.60$). However, the linear fits differ from those found by \cite{calvet04}, which predict line luminosities to have a steeper correlation with \lacc. Because \cite{calvet04} studied intermediate-mass TTS between 1.5-3.7~\msun, this could indicate a stellar mass dependence of the relations. Alternatively, this could simply result from the much larger sample size available here (57 TTS) versus in \citet[9 TTS]{calvet04}.

\cite{Ardila2013} found a strong correlation between ${\rm L_{SiIV}}$ and ${\rm L_{CIV}}$. However, the slope of their best-fit relation\footnote{$\log_{10}{\rm (L_{SiIV}/L_\odot)}=(1.4\pm0.2)\log_{10}{\rm (L_{CIV}/L_\odot)}+(0.9\pm0.6)$} differed from their model of optically-thin \siiv\ and \civ\ emission from the pre- and post-shock regions, which predicts $\log_{10}{\rm L_{SiIV}}=1.0\log_{10}{\rm L_{CIV}}-0.95$ (equivalently, ${\rm L_{SiIV}}=0.111{\rm L_{CIV}}$). The authors noted that reasonable modifications to model assumptions could move the model line up by about 0.1~dex but would not change the slope. They proposed that the mismatch could be caused by some CTTS being silicon rich/carbon deficient (e.g., CV~Cha, which was above the model prediction line) and others being silicon poor/carbon rich (e.g., AA~Tau, which was below the model prediction line). \cite{Ardila2013} measured line fluxes for CV~Cha and AA~Tau using the same HST data as used here.

The bottom panel of Figure~\ref{fig:LaccLFUV} shows our best-fit $\rm L_{SiIV}$-$\rm L_{CIV}$ relation: $\log_{10}{\rm (L_{SiIV}/L_\odot)}=(1.03\pm0.04)\log_{10}{\rm (L_{CIV}/L_\odot)}-(0.58\pm0.15)$. This is remarkably consistent with the \cite{Ardila2013} pre- and post-shock model, especially if the model were shifted upwards by 0.1~dex. Here, CV~Cha no longer stands out from the model expectation, and silicon-richness/carbon-deficiency does not need to be invoked to explain its line ratio. AA~Tau is also overall consistent, in this case slightly above the model prediction in contrast to being silicon-poorer in \cite{Ardila2013}. There are two likely causes of the discrepancy.
First, the sample of \cite{Ardila2013} contained very few detections of accreting sources with $\log_{10}{\rm L_{CIV}/L_\odot}<-4$. Conversely, our sample spans $-7.1<\log_{10}{\rm L_{CIV}/L_\odot}<-2.2$. Second, as \cite{Ardila2013} discuss, their extinctions were taken from multiple literature sources, whereas ours are derived consistently as part of the accretion shock modeling. Thus, the larger sample and consistent extinction correction procedure seems to have reduced the scatter and revealed the expected line ratios from an accretion shock origin.

\subsubsection{Model degeneracies} \label{subsec:degeneracies}

To assess potential degeneracies between accretion flow model parameters, we analyze the top 100 models for each observation using corner plots and Pearson correlation coefficients ($r$). 
Consistent with previous work \citep{muzerolle01}, we find a strong anti-correlation between $\log_{10}\dot{M}$ and $\log_{10}T_{\rm max}$ (median $r=-0.7$, with 62\% of observations having a strong correlation of $|r|>0.6$). Despite this, the relatively low standard deviations of \mdot\ and \tmax\ indicate that this correlation does not hinder model convergence.

We also observe a moderate correlation between \rw\ and \tmax\ (median $r=0.5$, with 38\% of observations having a strong correlation of $|r|>0.6$). To determine if the \rw$-$\tmax\ correlation is related to the \mdot$-$\tmax\ correlation, we check for co-occurrence within individual observations and find that they are statistically independent. Finally, we test whether these correlations remain when all targets and epochs are combined, weighting each model by its \chisq\ value and accounting for the number of observations per star. In conglomeration, only the \mdot$-$\tmax\ correlation remains (still with $r=-0.7$). Thus, \rw$-$\tmax\ may be degenerate in individual observations due to system-specific degeneracies, but only \mdot\ and \tmax\ are globally correlated.

We evaluate the degeneracy of accretion shock model parameters using the corner plots of the MCMC posterior distributions. 
As discussed in Section~\ref{subsec:hotspot_structure}, we see that the $f_{\rm med}$ filling factor is anti-correlated with $f_{\rm low+high}$. However, the overall distribution of accretion shock excesses is too varied to justify reducing to a two-column model.
The extinction \av\ is often degenerate with $f_{\rm med}$ and/or $f_{\rm high}$, as smaller \av\ corrections require less high-energy emission to account for the continuum excess. To check whether this produces an overall \mdot-\av\ degeneracy in the sample, we look for a correlation between $\log_{10}$\mdot\ and \av\ in the final shock model results. We find that they are correlated with $r=0.6$. However, this appears to be a stellar mass effect. Normalizing the accretion rate as $\log_{10}$(\mdot/$\rm M_\star^2$), the correlation disappears ($r=0.1$). We find that both $\log_{10}$\mstar\ and $\log_{10}$\teff\ are correlated with \av\ with $r=0.5$. Thus, either higher-mass stars are more commonly extincted, or the shock model fitting procedure preferentially chooses high \av\ and $f$ for more massive, hotter stars.

\section{Discussion} \label{sec:discussion}

Here, we discuss the importance of NUV data for characterizing the spectral energy distributions that irradiate the disks (Section~\ref{sec:opttest}) and accretion hotspot stability (Section~\ref{sec:hotspotlifetimes}).

\subsection{The importance of the NUV for discerning accretion shock structure} \label{sec:opttest}

ULLYSES is the most comprehensive set of UV CTTS spectra. It provides an unprecedented opportunity to quantify what is inaccessible in ground-based observations. In our shock modeling, we find that up to 48\% of the total accretion luminosity is at wavelengths shorter than 0.31~\micron. We tested the importance of having space-based UV observations by repeating the shock modeling while ignoring all data below 0.31~\micron\ in the fitting procedure. As before, we use the parameters of the accretion flow model as input and perform a first iteration to derive an updated stellar radius, $R_{\star,f}$. Then, we complete a second iteration with models calculated using $R_{\star,f}$.
This replicates the spectrum we would have observed from the ground as a result of atmospheric UV absorption.

\begin{figure}
    \centering
    \includegraphics[width=\linewidth]{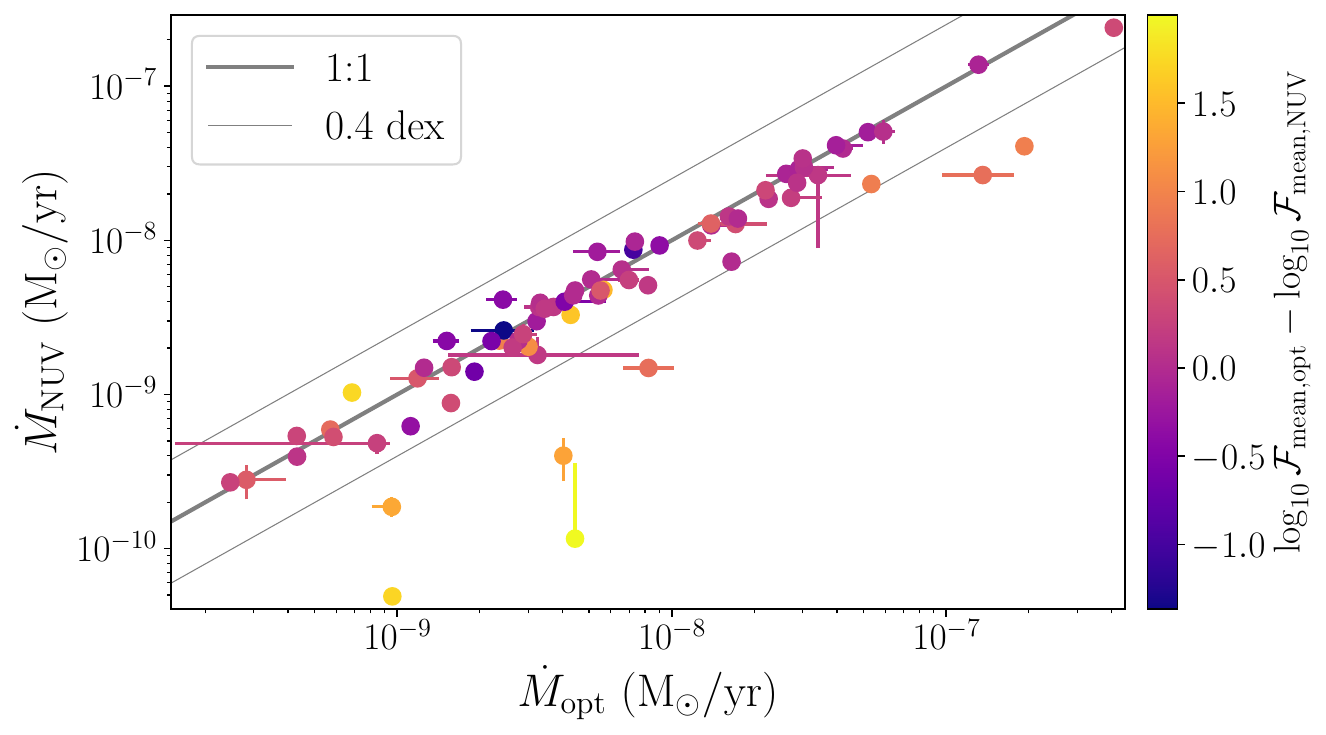}
    \caption{Comparison between accretion rates measured by fitting the shock models to the full 0.2-1.0~\micron\ region (\mdot$\rm _{NUV}$) and to the 0.31-1.0~\micron\ region accessible from the ground (\mdot$\rm _{opt}$). 
    The colorbar shows the logarithmic change of the mean energy flux density of the shock model. The gray lines show a one-to-one relationship and a 0.4~dex deviation, which is the typical variability-induced spread of \mdot\ for a given CTTS \citep[][]{manara23PPVII}. 
    }
    \label{fig:NUV_v_opt}
\end{figure}

Figure~\ref{fig:NUV_v_opt} plots the accretion rates from the full wavelength range ($\dot{M}_{\rm NUV}$) against those obtained ignoring wavelengths below 0.31~\micron\ ($\dot{M}_{\rm opt}$). The median difference $\dot{M}_{\rm NUV}$$-$$\dot{M}_{\rm opt}$ is $-$0.04~dex, indicating no significant bias toward one being larger than the other. The points show a spread of 0.31~dex around the one-to-one relationship. This is smaller than the typical $\sim$0.4~dex magnitudes of the uncertainty and intrinsic variability of accretion rates, which means that population-level accretion rate studies should not be significantly affected by the presence or lack of NUV data. However, at the individual object level, the measured accretion rate can change significantly, as shown by the outliers in Figure~\ref{fig:NUV_v_opt}. The \mdot\ uncertainties also increase when the NUV data are excluded, as indicated by the typically larger errorbars in $\dot{M}_{\rm opt}$.

\begin{figure}
    \centering
    \begin{tikzpicture}
        \node at (0,0) {\includegraphics[trim=0 40pt 0 5pt, clip, width=0.47\textwidth]{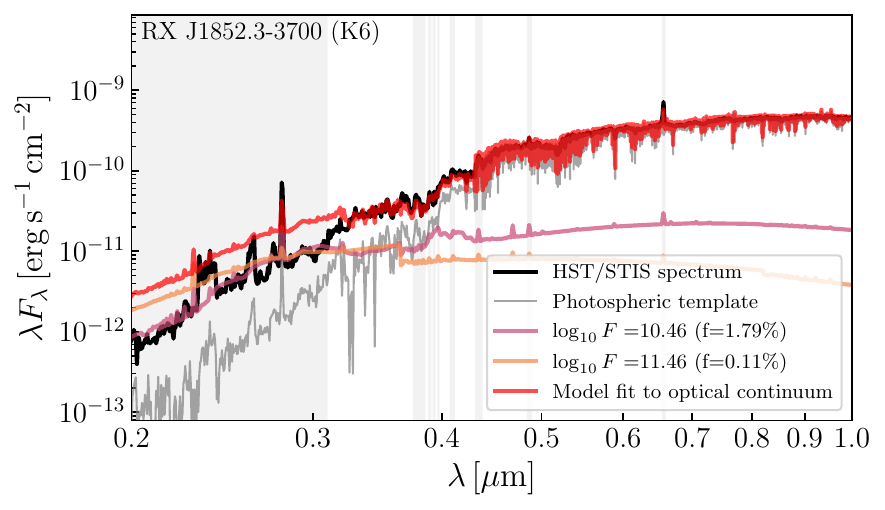}};
        \node[anchor=north east, font=\Large] at (3.8,1.9) {\textbf{a}};
        \node at (0,-3.95)
        {\includegraphics[trim=0 40pt 0 5pt, clip, width=0.47\textwidth]{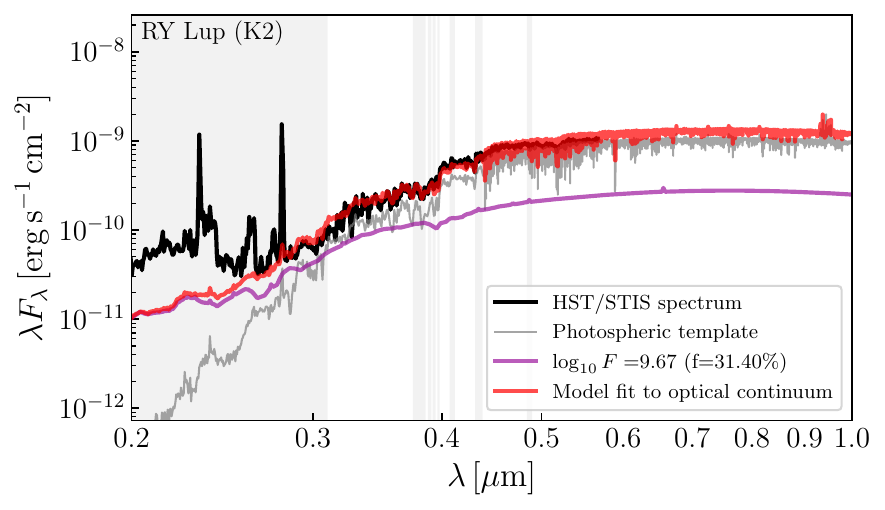}};
        \node[anchor=north east, font=\Large] at (3.8, -2.05) {\textbf{b}};
        \node at (0,-8.25)
        {\includegraphics[trim=0 5pt 0 5pt, clip, width=0.47\textwidth]{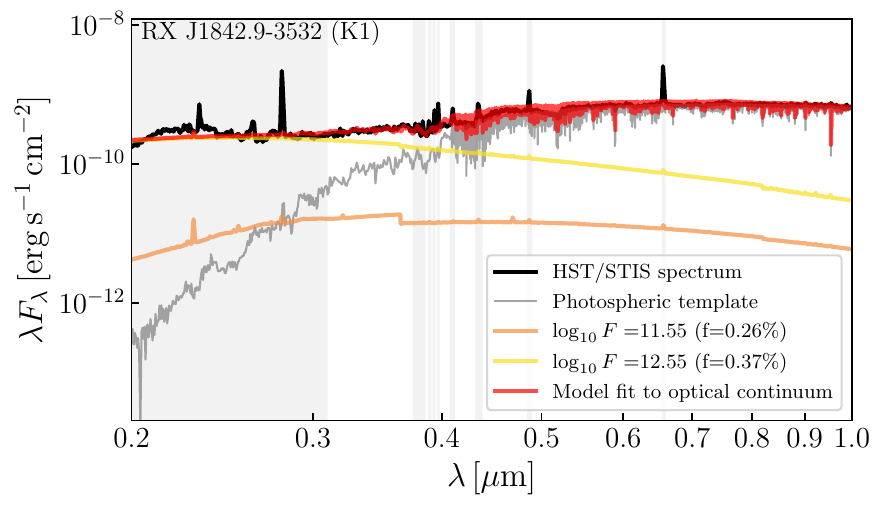}};
        \node[anchor=north east, font=\Large] at (3.8, -6) {\textbf{c}};
        \node at (0,-13.05)
        {\includegraphics[trim=0 5pt 0 6pt, clip, width=0.47\textwidth]{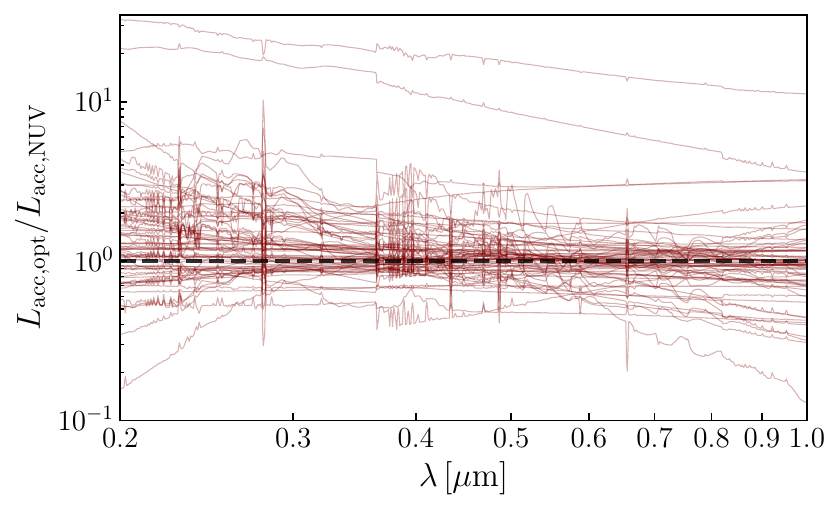}};
        \node[anchor=north east, font=\Large] at (3.8, -10.71) {\textbf{d}};
    \end{tikzpicture}
    \caption{Same as Figure~\ref{fig:shock_short}, but showing the results when only optical--NIR data are considered in the fit. NUV data are important for constraining the model fits across the full NUV-NIR range. When NUV data are ignored in the fit (indicated by the gray shaded region), the inferred model NUV fluxes can be (a) too high, (b) too low, and/or (c) the inferred model extinction corrections can be too strong.
    Panel (d) shows the ratio of $L_{\rm acc,opt}$ from the optical-only fit to $L_{\rm acc,NUV}$ from the full wavelength range, with a dashed grey line marking the line of equality.
    }
    \label{fig:optonly}
\end{figure}
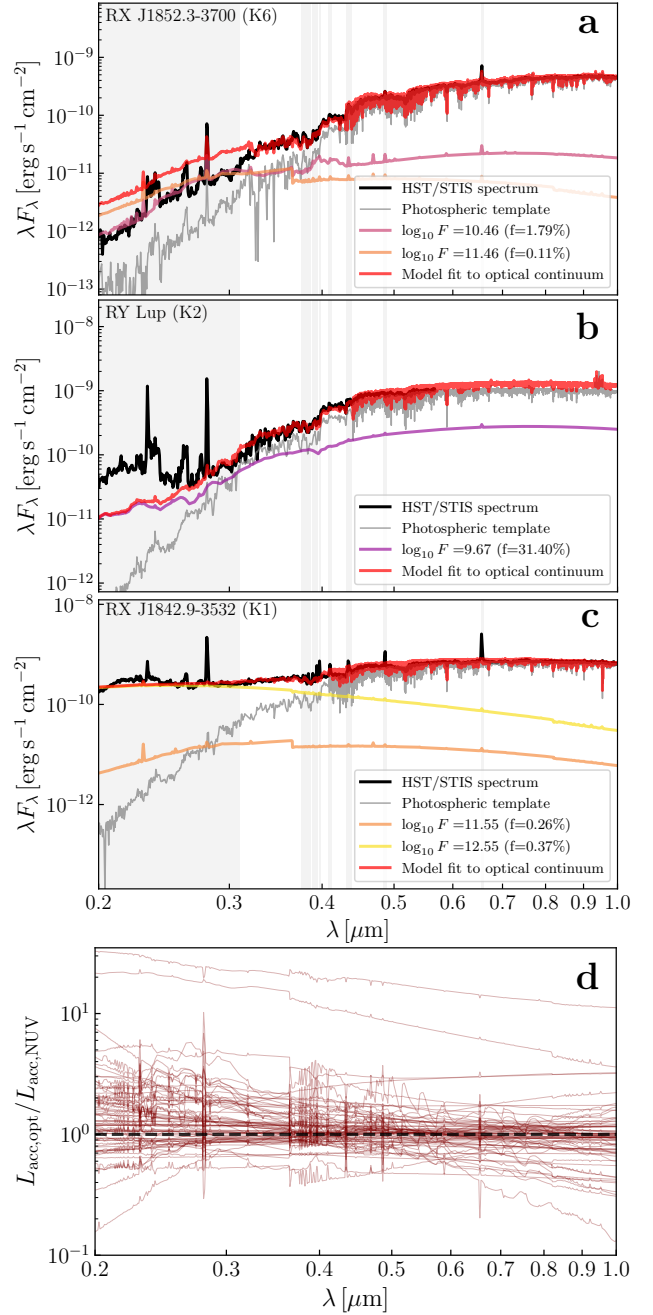

In 53/74 observations, the fit quality of the model that ignores the NUV is visibly worse than when including the full wavelength range. This includes cases where the model flux is notably higher than the observed flux (36/53; e.g., Figure~\ref{fig:optonly}a), lower than the observed flux (15/53; e.g., Figure~\ref{fig:optonly}b), or a combination of the two across the NUV region (2/53).

While the overall accretion rate distribution does not change when the NUV is excluded, the filling factor for the highest energy emission ($f_{\rm high}$) often changes significantly. The standard deviation of $f_{\rm high,opt}-f_{\rm high,NUV}$ is 1.5~dex, and over half (39/74) of the observations show increases greater than 15\%. Visual inspection of these cases shows that the models typically overestimate the real data.
The common increase in mean energy flux density, ${\cal F}_{\rm mean}$, is evident in Figure~\ref{fig:NUV_v_opt} (colorbar).
The median difference $\log_{10}{\cal F}_{\rm mean,opt}-\log_{10}{\cal F}_{\rm mean,NUV}$ is 0.13~dex.

The best-fit extinction can also change when the NUV is excluded. This makes sense, given that the 2200~\AA\ bump provides a constraint on the maximum \av\ that can be consistent with the shape of the observed spectrum. The median \av\ increases by 0.12~dex, from 0.42~mag to 0.55~mag, when the NUV is ignored.
When the extinction correction goes too high, the model fluxes can be too low in the NUV to account for the highly extinction-corrected data (e.g., Figure~\ref{fig:optonly}b), or $f_{\rm high}$ can increase too much (e.g., Figure~\ref{fig:optonly}c).. 

This test shows that the information contained in the optical--infrared portion of stellar spectra is often insufficient for predicting the NUV spectra. Therefore, when concerned with the actual spectral energy distribution of a specific object, it is not advisable to use the higher-$\curf$ spectra whose emission peaks in the NUV.

\subsection{Multi-epoch NUV spectra shed light on hotspot structure and lifetimes} \label{sec:hotspotlifetimes}
\begin{figure*}
    \centering
    \includegraphics[width=0.35\linewidth]{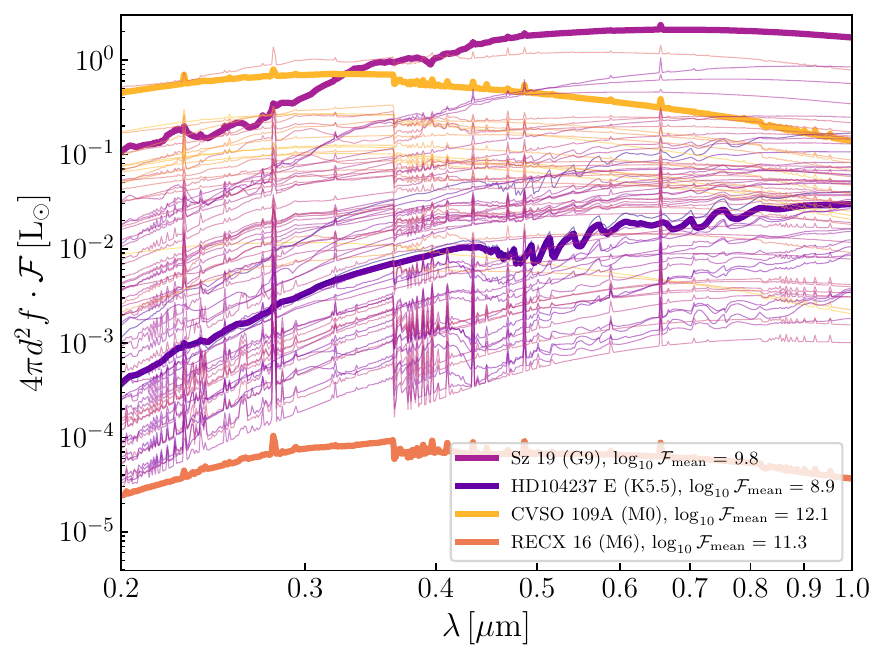}
    \includegraphics[width=0.31\linewidth]{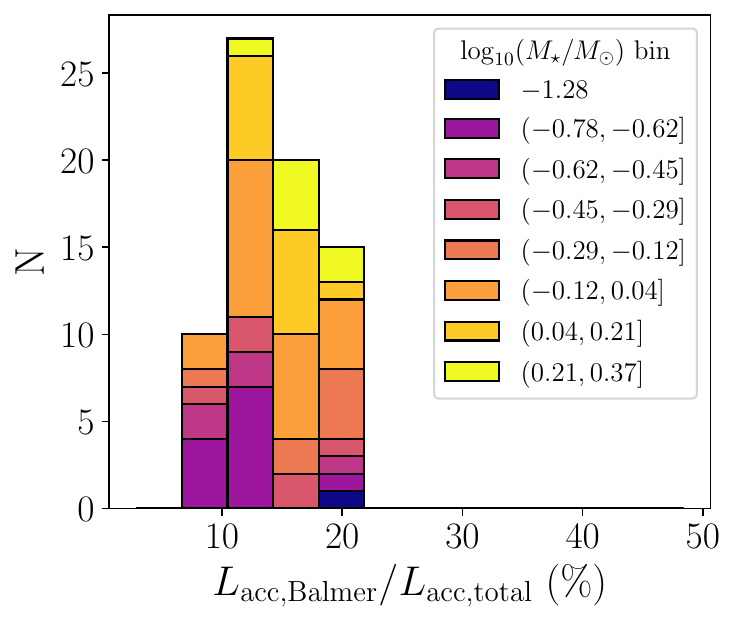}
    \includegraphics[width=0.31\linewidth]{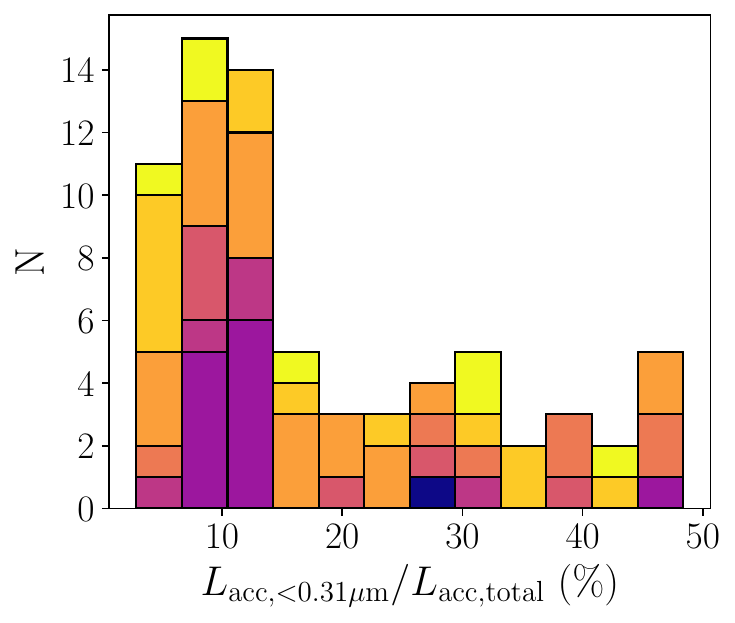}
    \caption{The leftmost panel shows the spectral energy distributions of the excess accretion shock luminosity for the entire sample.
    We have highlighted four CTTS, ranging from SpT of G9 to M6, that demonstrate the range in both luminosity and shape of accretion excess spectra. Line colors indicate the mean energy flux density using the same colorbar as Figure~\ref{fig:Mdot_barchart}.
    The middle panel shows the percentage of the total accretion luminosity, $L_{\rm acc,total}$ that is in the Balmer jump region between $0.34-0.45$~micron, and the rightmost panel shows percentage that is below 0.31~\micron\ (and therefore accessible only from space). The bars are divided by stellar mass as indicated in the legend.
    }
    \label{fig:fdotF} 
\end{figure*}

The measurement of hotspot configurations and rotational modulation presented in Section~\ref{subsec:shockresults} is possible because of the availability of NUV data, without which the high energy regions of the hotspots cannot be constrained. Figure~\ref{fig:Mdot_barchart} indicates that hotspot configurations are diverse across a large sample of CTTS, which means different disks will be irradiated by very different energy distributions. Figure~\ref{fig:fdotF} (left) shows all accretion excess spectra in units of \lsun. We have highlighted four CTTS that emphasize the range in both the magnitudes and shapes of the disk-irradiating spectra. 
The absolute accretion luminosity is strongly correlated with the stellar luminosity, as exemplified by the highest- and lowest-\lstar\ members of the sample (Sz~19 and RECX~16) occupying the uppermost- and lowermost regions of the figure. This supports the well-known correlations between accretion and disk parameters with the stellar luminosity (see, e.g., Figure~\ref{fig:Fmean_v_Lhalpha}).
However, the accretion strength also plays a non-negligible role. CVSO~109~A is a full spectral class later than Sz~19, but, as the highest accretor in the sample, its 0.2~\micron\ excess exceeds that of Sz~19 by almost a factor of 4.
These spectra demonstrate the broad range of irradiation experienced by protoplanetary disks as they are in the process of planet formation.


The phase-folded accretion rates of Sz~45, TW~Hya, BP~Tau, and GM~Aur in Figure~\ref{fig:phasefolded_Mdot} show that hotspots rotate with the star and/or inner disk and can last from $\sim$3 up to ten or more rotational periods \citep[consistent with recent work by][]{bouvier2023}. This produces constant variability of the irradiation received at the disk, even if the hotspot itself remains relatively unchanged. At the same time, neither TW~Hya and BP~Tau showed strong periodicity in their $u'$-, $B$-, or $g$-band light curves \citep{wendeborn24b} or phase-folded accretion rates in 2021, indicating that the presence of long-lived accretion funnels is variable \citep[in accord with recent work by][]{romanova2025}. Multi-epoch, contemporaneous HST and spectro-polarimetric observations of CTTS would be valuable to relate hotspot variability to magnetic field variability.

The variable presence of periodicity is consistent with the 3D MHD models of \cite{blinova16}, which found that both ordered and chaotic accretion can occur in a given system over time. They found that the stability of accretion flows is most strongly determined by the relationship between the corotation radius ($R_{\rm co}$) and the magnetospheric truncation radius. Specifically, the ``fastness parameter'' \omegas$=(R_{\rm i}/R_{\rm co})^{3/2}$ \citep[][]{Ghosh2007} demarcates the boundaries between the propeller, stable, unstable chaotic, and unstable ordered accretion regimes. 
Here, we can measure \omegas\ for Sz~45, TW~Hya, BP~Tau, and GM~Aur using their \ri\ values from the accretion flow modeling and their corotation radii using \rco$=(GM_\star P_{\rm rot}^2/4\pi^2)^{1/3}$. Doing this, we find \omegas$=$0.11 for Sz~45, and $0.21<\omega_s<0.27$ for all epochs of observation of TW~Hya, BP~Tau, and GM~Aur. These all fall into the unstable ordered regime (where \omegas$\lesssim$0.45), in which one to two ordered tongues can persist for periods of time even if the global accretion system is unstable over longer timescales.

3D MHD models also support the hotspot lifetimes found in this work. For systems accreting from stable funnel flows, \cite{KulkarniRomanova2009} and \cite{Kurosawa2013} found hotspot locations and shapes to be relatively stable throughout the duration of their simulations (ten and three stellar rotation periods, respectively).
For unstable systems with shorter-lived accretion tongues, the hotspot lifetimes of \cite{KulkarniRomanova2009} typically range from 8.2 down to $\sim$2 rotation periods. The longest timescale probed in our work (33.5 rotation periods for TW~Hya in 2010) contains too few points to draw a conclusion, as its three points can be perfectly explained by a single-phase sinusoid (which has three unknowns). More data are needed to explore this longer timescale.

The ULLYSES program provided an unprecedented, consistently-reduced UV$-$NIR dataset, enabling the characterization of accretion shock diversity across a large sample of TTS as well as the temporal variability of individual systems.
The present work lays the foundation for future investigations of the impact of accretion on the planet-forming environment, both through angular momentum transport and variable disk irradiation. Forthcoming work will study stellar rotation and accretion stability in the ULLYSES survey targets (C. Pittman et al. 2025b, in preparation) and spatial correlations of magnetospheric inclinations (C. Pittman et al. 2025c, in preparation).

\section{Summary} \label{sec:summary}

We have presented accretion modeling results for 67 CTTS from the HST ULLYSES DDT program, performed as part of the ODYSSEUS collaboration. This is the largest and most consistent UV through optical magnetospheric accretion study to date.

\begin{enumerate}
    \item Using an accretion flow model, we measured accretion rates, flow temperatures, magnetospheric truncation radii, truncation region widths, and magnetospheric inclinations using multiple epochs of \halpha\ observations. We found typical truncation radii to be about half the typically-assumed value of 5 stellar radii.
    \item Using an accretion shock model, we measured accretion rates, extinctions, and hotspot structures from \hst\ spectra. We found that continuum excesses can be explained by a single accretion column in 26\% of our CTTS, by two columns in 59\%, and by three columns in the remaining 15\%.
    \item We phase-folded multi-epoch accretion shock model results and found periodicity in the accretion rate and hotspot energy distributions. This is evidence of rotation-modulated hotspot emission that remains stable for at least 3 stellar rotation periods.
    \item The accretion rates from the flow and shock models are consistent with one another within about 0.16~dex for observations separated by $\sim1$~day or less, and within about 0.34~dex for larger time differences. This means that the flow model can be effective for measuring accretion rates when UV spectra are unavailable, though UV data remain necessary for constraining the SED of the accretion emission.
    \item We measured \halpha\ luminosities (\Lhalpha) from the same STIS spectra used to measure the accretion luminosity (\lacc) and derived an empirical relation: $\log_{10}{L_{\rm acc}}= (1.78 \pm 0.12) + (1.07\pm 0.04)\log_{10}{L_{H\alpha}}$. We also derive a relation between \lacc\ and the excess $U$-band luminosity $\rm L_{U,ex}$: $\log_{10}{(L_{\rm acc}/L_\odot)}= (0.89\pm0.02)\log_{10}(\rm L_{U,ex}/L_\odot)+(0.68\pm0.04)$. Both are generally consistent with previous work, providing validation for ground-based \lacc\ measurements that use \halpha\ or excess $U$-band luminosities.
    \item We measure empirical relationships between \lacc\ and FUV line luminosities (\siiv\ $\lambda1400$\AA, \civ\ $\lambda1548$\AA, and \heii\ $\lambda1640$\AA). We find a strong correlation between $\log_{10}{\rm L_{SiIV}}$ and $\log_{10}{\rm L_{CIV}}$ that is consistent with an origin in the pre- and post-shock regions without needing to invoke silicon richness or carbon deficiency.
    \item We show that space-based NUV observations are vital for accurately reproducing the NUV continuum, measuring extinction, and constraining the high-energy accretion emission. When the NUV data are ignored, 
    the median \av\ increases by 31\%, and the $\curf_{\rm high}$ filling factor increases by at least 15\% in over half of the sample, corresponding to a common increase in the mean energy flux density. However, the \mdot\ distribution does not change significantly, so population-level accretion rate surveys should be minimally affected.
\end{enumerate}

\begin{acknowledgments}

We wish to recognize the work of Will Fischer, who passed away in 2024. His dedication at STScI ensured the successful implementation of ULLYSES, and his kindness made him a friend to many. 

We thank the anonymous reviewer, as well as ODYSSEUS collaboration members Juan Alcalá, Suzan Edwards, Jochen Eisl\"offel, Eleonora Fiorellino, and P. Christian Schneider, for their insightful feedback that improved the manuscript. We thank P\'eter \'Abrah\'am, Pauline McGinnis, and Thomas Sperling for work on the PENELLOPE data reduction. C.P. thanks Uma Gorti for many helpful discussions throughout the duration of this work.

This work was supported by HST AR-16129 and benefited from discussions with the ODYSSEUS team \citep[\url{https://sites.bu.edu/odysseus/},][]{espaillat22}. C.P. acknowledges funding from the NSF Graduate Research Fellowship Program under grant No. DGE-1840990. C.F.M., K.M., and J.C.-W. acknowledge funding by the European Union under the Horizon Europe Research \& Innovation Programme 101039452 (WANDA). Views and opinions expressed are, however, those of the author(s) only and do not necessarily reflect those of the European Union or the European Research Council. Neither the European Union nor the granting authority can be held responsible for them. J.F.G. was supported by Funda\c{c}\~ao para a Ci\^encia e Tecnologia (FCT) through the research grants UIDB/04434/2020 and UIDP/04434/2020.  J.H. acknowledges support from the research projects UNAM-DGAPA-PAPIIT IG-101723 and CONAHCyT 86372. This work was also supported by the NKFIH NKKP grant ADVANCED 149943 and the NKFIH excellence grant TKP2021-NKTA-64. Project no.149943 has been implemented with the support provided by the Ministry of Culture and Innovation of Hungary from the National Research, Development and Innovation Fund, financed under the NKKP ADVANCED funding scheme.

Based on observations obtained with the NASA/ESA Hubble Space Telescope, retrieved from the Mikulski Archive for Space Telescopes (MAST) at the Space Telescope Science Institute (STScI). STScI is operated by the Association of Universities for Research in Astronomy, Inc. under NASA contract NAS 5-26555. The ULLYSES data are hosted on MAST at \dataset[https://doi.org/10.17909/t9-jzeh-xy14]{https://doi.org/10.17909/t9-jzeh-xy14} \citep[][]{ULLYSESdoi}.
    
Based on observations collected at the European Southern Observatory under ESO programme 106.20Z8 and on data obtained from the ESO Science Archive Facility hosted at \dataset[https://doi.org/10.18727/archive/88]{https://doi.org/10.18727/archive/88} under Program IDs 090.C-0050(A), 093.C-0658(A), 094.C-0805(A), and 108.22M8.001. 


\end{acknowledgments}

\facilities{HST(STIS), HST(COS), VLT(X-Shooter), VLT(ESPRESSO), VLT(UVES), CTIO:1.5m(Chiron), LCOGT(NRES)}

\software{astropy \citep{2013A&A...558A..33A,2018AJ....156..123A,Astropy2022},  
          Cloudy \citep{2013RMxAA..49..137F}, emcee \citep{emcee2013}
          }
          

\appendix
\section{Accretion flow model results} \label{Appsec:app_flowresults}

\begin{figure*}
    \centering
    \digitalasset
    \includegraphics[width=0.95\textwidth]{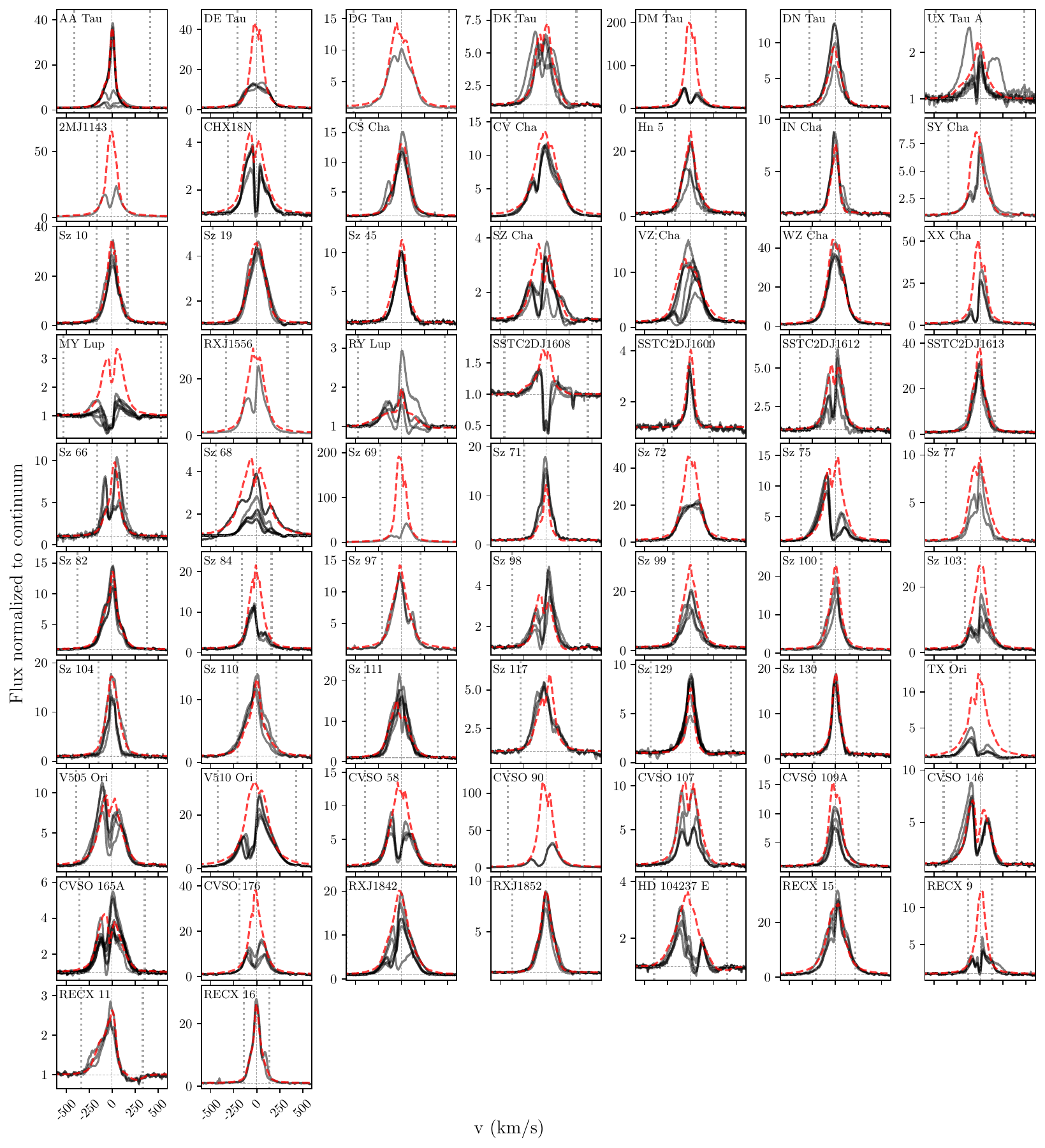}
    \caption{\halpha\ profiles for all epochs of observation used to derive the accretion flow properties (low-opacity gray lines). The dotted gray lines show the freefall velocity for that CTTS. The weighted-mean model described in Section~\ref{sec:models} is overlaid in dashed red to demonstrate that the ``typical'' accretion structures we derive successfully reproduce the typical \halpha\ profiles we observe. The corresponding weighted-mean parameters are given in Table~\ref{tab:flowresults}. The model fits to individual epochs are stronger, and these are included in the complete figure set (242 figures) in the online journal.}
    \label{fig:flowfits}
\end{figure*}

Here we present the flow modeling results for the ULLYSES sample. Figure~\ref{fig:flowfits} overlays all \halpha\ observations with the weighted-mean model representative of the ``typical'' accretion configuration for that system, calculated in \emph{Step 5}. The online material contains the model fits to individual epochs of observation, calculated in \emph{Step 2}. Table~\ref{tab:flowresults} gives the weighted-mean model parameters, and the online version of the table includes the results for individual epochs as well.

\startlongtable
\begin{deluxetable*}{>{\raggedright\arraybackslash}p{3.6cm}lllllp{1.5cm}p{1.5cm}}
\tabletypesize{\footnotesize}
\tablecaption{Weighted-mean accretion flow model results \label{tab:flowresults}}
\tablehead{
\colhead{Object} & \colhead{\ri} & \colhead{\rw} & \colhead{\mdot$_{\rm flow}$} & \colhead{\tmax} & \colhead{$i$} & \colhead{$A$} & \colhead{$\sigma$}\\
\colhead{} & \colhead{(R$_{\star,f}$)} & \colhead{(R$_{\star,f}$)} & \colhead{($10^{-8}$\msun/yr)} & \colhead{(K)} & \colhead{(\degrees)} & \colhead{(Norm. flux)} & \colhead{(km/s)}
}
\startdata
\multicolumn{8}{c}{\textbf{Taurus}} \\
AA Tau (2008-2009)\tablenotemark{\textdagger} & $5.70 \pm 0.85$ & $0.52 \pm 0.18$ & $7.46 \pm 1.68$ & $6686 \pm 127$ & $59 \pm 5$ & \dots & \dots \\
AA Tau (2021)\tablenotemark{\textdagger} & $3.47 \pm 0.27$ & $1.65 \pm 0.16$ & $5.58 \pm 0.90$ & $7540 \pm 79$ & $42 \pm 2$ & $25.61 \pm 0.18$ & $22.78 \pm 0.18$ \\
DE Tau\tablenotemark{a} & $3.23 \pm 0.15$ & $0.18 \pm 0.01$ & $2.58 \pm 0.30$ & $7193 \pm 29$ & $45 \pm 2$ & \dots & \dots \\
DG Tau A & $4.29 \pm 0.47$ & $0.91 \pm 0.56$ & $2.95 \pm 1.42$ & $7893 \pm 483$ & $52 \pm 6$ & \dots & \dots \\
DK Tau A & $1.81 \pm 0.09$ & $1.88 \pm 0.11$ & $1.17 \pm 0.14$ & $8589 \pm 114$ & $50 \pm 2$ & \dots & \dots \\
DM Tau & $7.57 \pm 0.01$ & $0.43 \pm 0.00$ & $2.76 \pm 0.02$ & $7350 \pm 2$ & $22 \pm 3$ & \dots & \dots \\
DN Tau & $1.47 \pm 0.06$ & $0.41 \pm 0.05$ & $0.40 \pm 0.08$ & $8603 \pm 148$ & $38 \pm 2$ & $3.76 \pm 0.01$ & $28.87 \pm 0.11$ \\
UX Tau A & $2.81 \pm 0.34$ & $0.48 \pm 0.15$ & $0.77 \pm 0.15$ & $8186 \pm 204$ & $79 \pm 2$ & $0.84 \pm 0.02$ & $29.10 \pm 0.98$ \\
\hline \multicolumn{8}{c}{\textbf{Cha I}} \\
2MASSJ11432669-7804454 & $5.13 \pm 0.88$ & $0.88 \pm 0.46$ & $0.60 \pm 0.20$ & $7667 \pm 217$ & $60 \pm 9$ & \dots & \dots \\
CHX18N & $1.54 \pm 0.07$ & $1.15 \pm 0.19$ & $0.70 \pm 0.17$ & $9000 \pm 194$ & $58 \pm 2$ & \dots & \dots \\
CS Cha\tablenotemark{b} & $4.04 \pm 0.08$ & $0.04 \pm 0.00$ & $0.81 \pm 0.18$ & $8108 \pm 157$ & $43 \pm 2$ & \dots & \dots \\
CV Cha & $4.14 \pm 0.10$ & $0.59 \pm 0.12$ & $6.34 \pm 1.10$ & $7944 \pm 102$ & $54 \pm 2$ & \dots & \dots \\
Hn 5 & $2.43 \pm 0.02$ & $1.91 \pm 0.09$ & $0.09 \pm 0.00$ & $10764 \pm 186$ & $78 \pm 1$ & $2.70 \pm 0.06$ & $19.87 \pm 0.49$ \\
IN Cha & $3.53 \pm 0.41$ & $0.98 \pm 0.17$ & $0.11 \pm 0.02$ & $8641 \pm 170$ & $69 \pm 3$ & $2.10 \pm 0.03$ & $22.63 \pm 0.50$ \\
SY Cha & $1.79 \pm 0.12$ & $1.34 \pm 0.18$ & $1.05 \pm 0.25$ & $8434 \pm 268$ & $26 \pm 4$ & \dots & \dots \\
Sz 10 & $1.93 \pm 0.02$ & $0.14 \pm 0.01$ & $0.27 \pm 0.03$ & $7779 \pm 52$ & $54 \pm 0$ & $5.05 \pm 0.09$ & $16.82 \pm 0.33$ \\
Sz 19\tablenotemark{b} & $2.59 \pm 0.12$ & $0.06 \pm 0.01$ & $4.26 \pm 0.72$ & $7349 \pm 106$ & $34 \pm 2$ & \dots & \dots \\
Sz 45 & $2.86 \pm 0.22$ & $0.58 \pm 0.12$ & $0.44 \pm 0.07$ & $8030 \pm 171$ & $58 \pm 3$ & $6.14 \pm 0.02$ & $32.29 \pm 0.09$ \\
SZ Cha & $2.58 \pm 0.33$ & $1.40 \pm 0.19$ & $0.39 \pm 0.12$ & $9776 \pm 320$ & $56 \pm 3$ & \dots & \dots \\
VZ Cha & $2.47 \pm 0.08$ & $0.89 \pm 0.08$ & $0.56 \pm 0.04$ & $8525 \pm 68$ & $53 \pm 2$ & \dots & \dots \\
WZ Cha & $2.89 \pm 0.01$ & $0.20 \pm 0.00$ & $1.24 \pm 0.14$ & $7353 \pm 54$ & $41 \pm 0$ & \dots & \dots \\
XX Cha & $4.65 \pm 0.17$ & $1.14 \pm 0.35$ & $0.93 \pm 0.39$ & $7824 \pm 431$ & $45 \pm 3$ & \dots & \dots \\
\hline \multicolumn{8}{c}{\textbf{Lupus}} \\
MY Lup & $7.17 \pm 0.48$ & $0.98 \pm 0.19$ & $0.07 \pm 0.01$ & $10120 \pm 319$ & $89 \pm 0$ & \dots & \dots \\
RX J1556.1-3655 & $3.78 \pm 0.33$ & $0.47 \pm 0.56$ & $0.99 \pm 0.47$ & $7779 \pm 443$ & $64 \pm 6$ & \dots & \dots \\
RY Lup & $1.78 \pm 0.14$ & $0.73 \pm 0.11$ & $0.10 \pm 0.01$ & $10389 \pm 262$ & $82 \pm 1$ & $0.86 \pm 0.01$ & $23.08 \pm 0.37$ \\
SSTc2dJ160000.6-422158 & $4.45 \pm 0.56$ & $1.13 \pm 0.30$ & $0.03 \pm 0.01$ & $9080 \pm 415$ & $72 \pm 2$ & $1.64 \pm 0.01$ & $23.38 \pm 0.24$ \\
SSTc2dJ161243.8-381503 & $5.61 \pm 0.42$ & $1.43 \pm 0.22$ & $0.14 \pm 0.03$ & $9228 \pm 217$ & $73 \pm 2$ & \dots & \dots \\
SSTc2dJ161344.1-373646 & $2.34 \pm 0.04$ & $1.29 \pm 0.09$ & $0.03 \pm 0.00$ & $11560 \pm 273$ & $58 \pm 1$ & $4.67 \pm 0.10$ & $18.79 \pm 0.42$ \\
SSTc2dJ160830.7-382827 & $1.58 \pm 0.09$ & $0.07 \pm 0.01$ & $0.02 \pm 0.00$ & $10631 \pm 513$ & $56 \pm 4$ & \dots & \dots \\
Sz 66 & $1.78 \pm 0.11$ & $0.23 \pm 0.03$ & $0.11 \pm 0.02$ & $8220 \pm 178$ & $79 \pm 2$ & \dots & \dots \\
Sz 68 & $2.82 \pm 0.25$ & $0.83 \pm 0.19$ & $3.48 \pm 0.45$ & $8263 \pm 156$ & $69 \pm 3$ & \dots & \dots \\
Sz 69 & $4.05 \pm 0.15$ & $0.98 \pm 0.67$ & $0.06 \pm 0.01$ & $9919 \pm 979$ & $22 \pm 15$ & \dots & \dots \\
Sz 71 & $1.64 \pm 0.10$ & $1.10 \pm 0.18$ & $0.17 \pm 0.03$ & $9189 \pm 239$ & $50 \pm 3$ & $8.84 \pm 0.04$ & $24.69 \pm 0.11$ \\
Sz 72\tablenotemark{a} & $3.50 \pm 0.00$ & $0.18 \pm 0.00$ & $0.27 \pm 0.01$ & $8509 \pm 25$ & $51 \pm 0$ & \dots & \dots \\
Sz 75 & $2.83 \pm 0.14$ & $3.16 \pm 0.00$ & $4.04 \pm 0.27$ & $8407 \pm 30$ & $52 \pm 1$ & \dots & \dots \\
Sz 77 & $2.44 \pm 0.74$ & $0.72 \pm 0.44$ & $0.26 \pm 0.16$ & $8939 \pm 733$ & $50 \pm 9$ & $2.70 \pm 0.02$ & $19.86 \pm 0.17$ \\
Sz 82 & $2.86 \pm 0.12$ & $0.94 \pm 0.09$ & $1.12 \pm 0.17$ & $7987 \pm 113$ & $62 \pm 1$ & $7.17 \pm 0.02$ & $25.85 \pm 0.07$ \\
Sz 84 & $1.90 \pm 0.12$ & $1.44 \pm 0.18$ & $0.06 \pm 0.01$ & $10535 \pm 329$ & $70 \pm 2$ & \dots & \dots \\
Sz 97 & $2.63 \pm 0.14$ & $2.58 \pm 0.20$ & $0.16 \pm 0.04$ & $11360 \pm 408$ & $83 \pm 3$ & $1.58 \pm 0.11$ & $25.58 \pm 2.89$ \\
Sz 98 & $2.94 \pm 1.28$ & $1.50 \pm 0.38$ & $0.25 \pm 0.21$ & $9143 \pm 613$ & $58 \pm 8$ & \dots & \dots \\
Sz 99 & $2.94 \pm 0.01$ & $2.14 \pm 0.12$ & $0.06 \pm 0.00$ & $10208 \pm 153$ & $82 \pm 1$ & $4.85 \pm 0.05$ & $29.01 \pm 0.36$ \\
Sz 100 & $2.11 \pm 0.12$ & $0.73 \pm 0.16$ & $0.07 \pm 0.05$ & $9914 \pm 571$ & $75 \pm 3$ & $2.95 \pm 0.10$ & $25.39 \pm 1.23$ \\
Sz 103 & $2.05 \pm 0.07$ & $0.57 \pm 0.05$ & $0.07 \pm 0.01$ & $10611 \pm 115$ & $82 \pm 1$ & \dots & \dots \\
Sz 104 & $1.92 \pm 0.01$ & $1.03 \pm 0.10$ & $0.05 \pm 0.00$ & $10151 \pm 176$ & $88 \pm 0$ & $7.32 \pm 0.03$ & $29.60 \pm 0.16$ \\
Sz 110 & $1.67 \pm 0.00$ & $1.82 \pm 0.26$ & $0.34 \pm 0.11$ & $9375 \pm 221$ & $77 \pm 2$ & $6.25 \pm 0.05$ & $29.03 \pm 0.25$ \\
Sz 111 & $3.33 \pm 0.00$ & $1.77 \pm 0.00$ & $0.14 \pm 0.00$ & $10250 \pm 3$ & $65 \pm 0$ & $0.01 \pm 0.00$ & $0.01 \pm 0.00$ \\
Sz 117 & $4.09 \pm 0.54$ & $1.45 \pm 0.31$ & $0.14 \pm 0.02$ & $8941 \pm 237$ & $82 \pm 1$ & \dots & \dots \\
Sz 129 & $3.06 \pm 0.25$ & $0.96 \pm 0.17$ & $0.52 \pm 0.08$ & $8129 \pm 169$ & $62 \pm 2$ & $5.89 \pm 0.01$ & $33.96 \pm 0.08$ \\
Sz 130 & $3.16 \pm 0.20$ & $0.69 \pm 0.21$ & $0.19 \pm 0.03$ & $8167 \pm 272$ & $75 \pm 2$ & $13.75 \pm 0.04$ & $28.80 \pm 0.10$ \\
\hline \multicolumn{8}{c}{\textbf{$\sigma$ Ori}} \\
TX Ori & $2.02 \pm 0.23$ & $2.46 \pm 0.20$ & $9.20 \pm 2.77$ & $8436 \pm 302$ & $52 \pm 5$ & \dots & \dots \\
V505 Ori & $2.78 \pm 0.11$ & $1.70 \pm 0.15$ & $1.99 \pm 0.26$ & $8392 \pm 123$ & $62 \pm 2$ & \dots & \dots \\
V510 Ori & $5.29 \pm 0.00$ & $0.43 \pm 0.00$ & $3.74 \pm 0.25$ & $7908 \pm 50$ & $33 \pm 0$ & \dots & \dots \\
\hline \multicolumn{8}{c}{\textbf{Orion OB1b}} \\
CVSO 58 & $3.78 \pm 0.17$ & $0.35 \pm 0.10$ & $1.52 \pm 0.28$ & $7690 \pm 197$ & $60 \pm 3$ & \dots & \dots \\
CVSO 90 & $7.49 \pm 0.13$ & $0.66 \pm 0.01$ & $0.82 \pm 0.08$ & $8089 \pm 94$ & $42 \pm 1$ & \dots & \dots \\
CVSO 107 & $2.19 \pm 0.07$ & $2.08 \pm 0.09$ & $0.94 \pm 0.15$ & $8742 \pm 203$ & $58 \pm 1$ & \dots & \dots \\
CVSO 109 A\tablenotemark{b} & $2.73 \pm 0.25$ & $0.28 \pm 0.10$ & $21.27 \pm 3.20$ & $6437 \pm 88$ & $33 \pm 5$ & \dots & \dots \\
CVSO 146 & $2.25 \pm 0.05$ & $2.73 \pm 0.07$ & $1.30 \pm 0.09$ & $9107 \pm 31$ & $60 \pm 1$ & \dots & \dots \\
CVSO 165 A\tablenotemark{b} & $1.76 \pm 0.21$ & $1.56 \pm 0.36$ & $0.61 \pm 0.25$ & $9400 \pm 417$ & $62 \pm 6$ & \dots & \dots \\
CVSO 176 & $1.99 \pm 0.02$ & $1.50 \pm 0.15$ & $0.40 \pm 0.05$ & $9464 \pm 220$ & $54 \pm 2$ & \dots & \dots \\
\hline \multicolumn{8}{c}{\textbf{CrA}} \\
RX J1842.9-3532 & $5.47 \pm 0.03$ & $1.17 \pm 0.06$ & $0.29 \pm 0.02$ & $8781 \pm 126$ & $50 \pm 1$ & \dots & \dots \\
RX J1852.3-3700 & $1.75 \pm 0.06$ & $0.07 \pm 0.01$ & $0.18 \pm 0.02$ & $8076 \pm 114$ & $29 \pm 1$ & $4.25 \pm 0.01$ & $29.98 \pm 0.08$ \\
\hline \multicolumn{8}{c}{\textbf{$\epsilon$ Cha}} \\
HD 104237 E & $1.99 \pm 0.10$ & $0.76 \pm 0.08$ & $0.06 \pm 0.01$ & $11374 \pm 362$ & $66 \pm 2$ & \dots & \dots \\
\hline \multicolumn{8}{c}{\textbf{$\eta$ Cha}} \\
RECX 9 & $1.50 \pm 0.05$ & $0.13 \pm 0.04$ & $0.05 \pm 0.00$ & $9035 \pm 265$ & $69 \pm 2$ & \dots & \dots \\
RECX 11 & $1.50 \pm 0.11$ & $2.29 \pm 0.12$ & $0.13 \pm 0.01$ & $11910 \pm 152$ & $70 \pm 2$ & $1.40 \pm 0.01$ & $34.37 \pm 0.19$ \\
RECX 15 & $2.57 \pm 0.00$ & $0.10 \pm 0.01$ & $0.18 \pm 0.04$ & $8449 \pm 186$ & $59 \pm 0$ & \dots & \dots \\
RECX 16 & $3.83 \pm 0.25$ & $1.88 \pm 0.16$ & $0.006 \pm 0.002$ & $11759 \pm 486$ & $90 \pm 0$ & $11.78 \pm 0.31$ & $19.34 \pm 0.60$ \\
 \enddata
 \tablenotetext{}{
Accretion flow model results calculated as the weighted mean of the individual epochs. A table with results for all individual epochs is available online in machine-readable format.
Columns are the inner magnetospheric truncation radius \ri, the width of the flow at the disk \rw, the accretion rate \mdot, the maximum temperature of the flow \tmax, the magnetospheric inclination $i$, and the amplitude (A) and width ($\sigma$) of the Gaussian used to represent the chromosphere, when applicable. 
Objects are sorted first by region, and then by name. $a$: Line center masked due to saturation. $b$: Unresolved binary modeled as described in Section~\ref{subsec:multiplicity}. \textit{\textdagger}: The \halpha\ profiles of AA~Tau show two distinct morphologies between the 2008-2009 UVES observations and the 2021 PENELLOPE observations. Here we separate the results and will discuss this in more detail in C. Pittman et al. (2025b, in preparation). We use the 2021 results in the iterative modeling procedure.
 }
\end{deluxetable*}

\figsetstart
\figsetnum{14}
\figsettitle{Accretion flow model fits to \halpha\ profiles}

\figsetend

\section{Accretion shock model results} \label{sec:app_shock}

Here we present the accretion shock model results for the ULLYSES sample in Table~\ref{tab:shockresults} and the accretion, emission line, and $U$-band luminosity measurements in Table~\ref{tab:Lacc}.

\begin{longrotatetable}
\begin{deluxetable*}{>{\raggedright\arraybackslash}p{2.3cm}llchlllcrc}
\tablecaption{Accretion shock model results \label{tab:shockresults}}
\tablehead{
\colhead{Object} & \colhead{$\dot{M}$} & \colhead{$A_V$} & \colhead{$\log_{10}\curf_{\rm low, med, high}$} & \nocolhead{$\log_{10}\curf_{\rm mean}$} & \colhead{$f_{\rm low}$} & \colhead{$f_{\rm med}$} & \colhead{$f_{\rm high}$} & \colhead{$f_{\rm tot}$} & \colhead{\tshock} & \colhead{$L_{\rm acc}^{\rm space}/L_{\rm acc}^{\rm tot}$} \\
\colhead{} & \colhead{($\rm 10^{-8}M_\odot/$yr)} & \colhead{(mag)} & \colhead{[\escm]} & \nocolhead{[\escm]} & \colhead{(\%)} & \colhead{(\%)} & \colhead{(\%)} & \colhead{(\%)} & \colhead{(K)} & \colhead{(\%)}
}
\startdata
\multicolumn{11}{c}{\textbf{Taurus}} \\
AA Tau & $2.70^{+0.04}_{-0.05}$ & $2.00^{+0.00}_{-0.00}$ & 10.34, 11.34, 12.34 & 10.00 & $32.8^{+0.6}_{-0.8}$ & 0 & $0.133^{+0.004}_{-0.005}$ & 32.94 & $5965\pm71$ & 24 \\
DE Tau & $5.04^{+0.21}_{-0.19}$ & $1.17^{+0.04}_{-0.03}$ & 9.99, 10.99, 11.99 & 9.07 & $2.7^{+0.4}_{-0.6}$ & $0.14^{+0.11}_{-0.06}$ & $0.080^{+0.008}_{-0.008}$ & 2.96 & $5423\pm358$ & 27 \\
DG Tau A & $1.28^{+0.09}_{-0.08}$ & $1.54^{+0.04}_{-0.04}$ & 10.56, 11.56, 12.56 & 9.74 & 0 & $1.54^{+0.10}_{-0.11}$ & 0 & 1.54 & $9776\pm259$ & 30 \\
DK Tau A & $0.55^{+0.06}_{-0.05}$ & $0.08^{+0.05}_{-0.04}$ & 9.12, 10.12, 11.12 & 9.29 & $32.2^{+1.5}_{-3}$ & $7.3^{+1.3}_{-1.3}$ & $0.46^{+0.08}_{-0.06}$ & 39.99 & $4182\pm118$ & 9 \\
DM Tau ep. 1 & $2.91^{+0.16}_{-0.15}$ & $1.16^{+0.04}_{-0.04}$ & 11.14, 12.14, 13.14 & 9.48 & $0.41^{+0.07}_{-0.07}$ & $0.177^{+0.018}_{-0.022}$ & 0 & 0.58 & $9536\pm383$ & 38 \\
DM Tau ep. 2 & $4.14^{+0.21}_{-0.23}$ & $1.21^{+0.03}_{-0.04}$ & 11.14, 12.14, 13.14 & 9.63 & $0.39^{+0.11}_{-0.09}$ & $0.272^{+0.024}_{-0.04}$ & 0 & 0.66 & $9913\pm519$ & 39 \\
DM Tau ep. 3 & $2.98^{+0.11}_{-0.13}$ & $1.21^{+0.02}_{-0.03}$ & 11.14, 12.14, 13.14 & 9.49 & $0.078^{+0.08}_{-0.007}$ & $0.174^{+0.004}_{-0.018}$ & 0 & 0.26 & $11185\pm957$ & 45 \\
DN Tau & $0.84^{+0.02}_{-0.03}$ & $0.01^{+0.00}_{-0.00}$ & 8.51, 9.51, 10.51 & 8.91 & 0 & $11.6^{+1.7}_{-2.2}$ & $1.40^{+0.15}_{-0.15}$ & 12.97 & $4519\pm235$ & 9 \\
UX Tau A & $0.66^{+0.02}_{-0.02}$ & $2.00^{+0.00}_{-0.01}$ & 9.89, 10.89, 11.89 & 9.17 & $19.2^{+0.5}_{-0.7}$ & 0 & 0 & 19.23 & $5754\pm70$ & 7 \\
\hline \multicolumn{11}{c}{\textbf{Cha I}} \\
2MASS  J11432669-7804454 & $0.87^{+0.06}_{-0.04}$ & $0.03^{+0.04}_{-0.01}$ & 10.34, 11.34, 12.34 & 8.75 & 0 & 0 & $0.0257^{+0.0018}_{-0.0016}$ & 0.03 & $12827\pm1535$ & 48 \\
CHX18N & $0.15^{+0.01}_{-0.00}$ & $0.40^{+0.02}_{-0.01}$ & 8.49, 9.49, 10.49 & 8.05 & $36.9^{+0.9}_{-1.4}$ & 0 & 0 & 36.89 & $4054\pm55$ & 3 \\
CS Cha & $1.25^{+0.05}_{-0.04}$ & $0.90^{+0.01}_{-0.01}$ & 11.06, 12.06, 13.06 & 9.37 & 0 & $0.193^{+0.007}_{-0.008}$ & 0 & 0.19 & $11608\pm212$ & 43 \\
CV Cha & $13.80^{+0.39}_{-0.31}$ & $1.37^{+0.02}_{-0.02}$ & 10.68, 11.68, 12.68 & 10.14 & $16.1^{+0.4}_{-0.7}$ & 0 & $0.130^{+0.010}_{-0.010}$ & 16.27 & $7020\pm101$ & 30 \\
Hn 5 & $0.22^{+0.01}_{-0.01}$ & $0.43^{+0.03}_{-0.04}$ & 9.59, 10.59, 11.09 & 8.15 & $0.340^{+0.3}_{-0.015}$ & $0.333^{+0.019}_{-0.04}$ & 0 & 0.67 & $5083\pm643$ & 11 \\
IN Cha & $0.10^{+0.00}_{-0.01}$ & $0.01^{+0.00}_{-0.00}$ & 9.12, 10.12, 10.62 & 7.57 & $1.5^{+0.3}_{-0.3}$ & $0.13^{+0.03}_{-0.03}$ & 0 & 1.60 & $3478\pm251$ & 8 \\
SY Cha & $1.86^{+0.06}_{-0.04}$ & $0.16^{+0.02}_{-0.02}$ & 9.26, 10.26, 11.26 & 9.78 & $15.4^{+1.1}_{-1.6}$ & $23.8^{+0.9}_{-1.1}$ & $0.86^{+0.07}_{-0.06}$ & 39.99 & $4972\pm79$ & 12 \\
SZ Cha & $0.04^{+0.01}_{-0.01}$ & $1.10^{+0.00}_{-0.00}$ & 9.89, 10.39, 11.39 & 8.38 & $1.9^{+1.1}_{-1.2}$ & 0 & $0.038^{+0.005}_{-0.006}$ & 1.95 & $5768\pm1125$ & 18 \\
Sz 10 & $0.44^{+0.02}_{-0.02}$ & $0.20^{+0.03}_{-0.03}$ & 9.62, 10.62, 11.12 & 8.30 & $3.55^{+0.26}_{-0.27}$ & $0.126^{+0.022}_{-0.021}$ & 0 & 3.68 & $3998\pm107$ & 11 \\
Sz 19 & $4.08^{+0.32}_{-0.32}$ & $1.75^{+0.05}_{-0.05}$ & 9.78, 11.78, 12.78 & 9.43 & $39.94^{+0.07}_{-0.16}$ & $0.06^{+0.04}_{-0.04}$ & 0 & 40.00 & $5496\pm21$ & 7 \\
Sz 45 ep. 1 & $0.33^{+0.02}_{-0.02}$ & $0.52^{+0.04}_{-0.05}$ & 9.53, 10.53, 11.53 & 9.18 & $16.8^{+2.0}_{-2.3}$ & 0 & $0.28^{+0.04}_{-0.03}$ & 17.05 & $4310\pm180$ & 16 \\
Sz 45 ep. 2 & $0.37^{+0.02}_{-0.02}$ & $0.38^{+0.04}_{-0.04}$ & 9.53, 10.53, 11.53 & 9.23 & $11.9^{+1.9}_{-2.6}$ & 0 & $0.39^{+0.04}_{-0.04}$ & 12.29 & $4628\pm282$ & 20 \\
Sz 45 ep. 3 & $0.39^{+0.02}_{-0.02}$ & $0.22^{+0.04}_{-0.04}$ & 9.53, 10.53, 11.53 & 9.26 & $16.4^{+1.8}_{-2.5}$ & $0.471^{+0.6}_{-0.029}$ & $0.33^{+0.04}_{-0.05}$ & 17.24 & $4485\pm250$ & 17 \\
Sz 45 ep. 4 & $0.37^{+0.02}_{-0.01}$ & $0.05^{+0.06}_{-0.03}$ & 9.53, 10.53, 11.53 & 9.23 & $23.7^{+1.8}_{-2.0}$ & $0.71^{+0.26}_{-0.3}$ & $0.197^{+0.05}_{-0.021}$ & 24.61 & $4242\pm140$ & 13 \\
Sz 45 ep. 5 & $0.36^{+0.02}_{-0.02}$ & $0.41^{+0.04}_{-0.04}$ & 9.53, 10.53, 11.53 & 9.22 & $16.8^{+1.7}_{-2.2}$ & 0 & $0.32^{+0.03}_{-0.03}$ & 17.11 & $4367\pm175$ & 17 \\
Sz 45 ep. 6 & $0.47^{+0.02}_{-0.03}$ & $0.63^{+0.03}_{-0.04}$ & 9.53, 10.53, 11.53 & 9.34 & $11.1^{+2.3}_{-2.6}$ & 0 & $0.54^{+0.05}_{-0.05}$ & 11.59 & $4890\pm300$ & 22 \\
VZ Cha & $3.48^{+0.32}_{-0.23}$ & $0.91^{+0.06}_{-0.05}$ & 9.86, 10.86, 11.86 & 10.37 & $24.5^{+1.4}_{-1.8}$ & $13.8^{+1.0}_{-1.7}$ & $1.63^{+0.4}_{-0.28}$ & 40.00 & $6481\pm193$ & 26 \\
WZ Cha & $2.12^{+0.12}_{-0.12}$ & $1.64^{+0.04}_{-0.04}$ & 10.29, 11.29, 12.29 & 9.86 & $7.2^{+0.8}_{-1.0}$ & 0 & $0.305^{+0.026}_{-0.025}$ & 7.49 & $6198\pm229$ & 39 \\
XX Cha & $0.51^{+0.04}_{-0.04}$ & $0.15^{+0.04}_{-0.04}$ & 9.68, 10.68, 11.68 & 8.86 & 0 & $1.19^{+0.03}_{-0.04}$ & $0.034^{+0.012}_{-0.010}$ & 1.23 & $7579\pm238$ & 19 \\
\hline \multicolumn{11}{c}{\textbf{Lupus}} \\
MY Lup & $0.06^{+0.00}_{-0.00}$ & $1.27^{+0.01}_{-0.01}$ & 9.89, 10.89, 11.89 & 8.72 & $3.2^{+0.6}_{-0.6}$ & 0 & $0.036^{+0.004}_{-0.004}$ & 3.25 & $5836\pm355$ & 15 \\
RX J1556.1-3655 & $1.43^{+0.05}_{-0.04}$ & $0.63^{+0.03}_{-0.02}$ & 10.29, 11.29, 12.29 & 9.60 & $6.85^{+0.24}_{-0.26}$ & $0.19^{+0.04}_{-0.04}$ & $0.117^{+0.010}_{-0.009}$ & 7.16 & $5752\pm90$ & 32 \\
RY Lup & $0.05^{+0.01}_{-0.01}$ & $0.53^{+0.06}_{-0.07}$ & 8.67, 9.67, 10.67 & 8.92 & $14.7^{+2.2}_{-2.5}$ & $16.5^{+2.8}_{-2.8}$ & 0 & 31.25 & $5132\pm221$ & 3 \\
SSTc2d J160000.6-422158 & $0.04^{+0.00}_{-0.00}$ & $0.01^{+0.00}_{-0.00}$ & 8.41, 9.41, 10.41 & 7.87 & 0 & $1.4^{+0.4}_{-0.4}$ & $0.15^{+0.03}_{-0.03}$ & 1.57 & $3797\pm292$ & 9 \\
SSTc2d J160830.7-382827 & $0.03^{+0.01}_{-0.01}$ & $0.19^{+0.02}_{-0.02}$ & 8.45, 9.45, 10.45 & 8.02 & $1.0^{+0.4}_{-0.7}$ & 0 & $0.36^{+0.09}_{-0.09}$ & 1.38 & $5499\pm1268$ & 9 \\
SSTc2d J161243.8-381503 & $0.05^{+0.00}_{-0.00}$ & $0.26^{+0.02}_{-0.02}$ & 9.47, 10.47, 11.47 & 8.40 & $8.5^{+0.5}_{-0.5}$ & 0 & 0 & 8.54 & $3686\pm72$ & 3 \\
SSTc2d J161344.1-373646 & $0.15^{+0.00}_{-0.00}$ & $0.01^{+0.00}_{-0.00}$ & 8.42, 9.42, 10.42 & 8.96 & 0 & 0 & $3.55^{+0.07}_{-0.08}$ & 3.55 & $5086\pm91$ & 11 \\
Sz 66 & $0.40^{+0.02}_{-0.02}$ & $0.52^{+0.04}_{-0.04}$ & 9.87, 10.37, 11.37 & 8.00 & 0 & $0.436^{+0.028}_{-0.15}$ & 0 & 0.44 & $5333\pm1642$ & 12 \\
Sz 68 & $1.39^{+0.02}_{-0.02}$ & $1.39^{+0.00}_{-0.00}$ & 9.72, 10.72, 11.72 & 9.03 & $20.38^{+0.24}_{-0.29}$ & 0 & 0 & 20.40 & $5204\pm30$ & 6 \\
Sz 69 & $0.03^{+0.00}_{-0.00}$ & $0.01^{+0.01}_{-0.00}$ & 8.54, 9.54, 10.54 & 8.24 & $8.5^{+1.4}_{-1.4}$ & 0 & $0.422^{+0.022}_{-0.025}$ & 8.95 & $3354\pm166$ & 9 \\
Sz 71 & $0.41^{+0.03}_{-0.04}$ & $0.35^{+0.02}_{-0.02}$ & 8.04, 9.04, 10.04 & 8.64 & $10.7^{+1.8}_{-3}$ & 0 & $3.9^{+0.3}_{-0.4}$ & 14.64 & $3552\pm230$ & 6 \\
Sz 72 & $1.29^{+0.06}_{-0.06}$ & $1.06^{+0.03}_{-0.03}$ & 9.84, 10.84, 11.84 & 9.59 & $12.4^{+1.0}_{-1.2}$ & 0 & $0.444^{+0.029}_{-0.028}$ & 12.88 & $5006\pm136$ & 28 \\
Sz 75 & $1.89^{+0.06}_{-0.06}$ & $0.21^{+0.03}_{-0.03}$ & 9.68, 10.68, 11.68 & 9.52 & $39.60^{+0.06}_{-0.15}$ & $0.0966^{+0.09}_{-0.0024}$ & $0.300^{+0.024}_{-0.024}$ & 40.00 & $4613\pm29$ & 13 \\
Sz 77 & $0.47^{+0.01}_{-0.01}$ & $0.35^{+0.01}_{-0.01}$ & 10.09, 11.09, 12.09 & 9.05 & $8.87^{+0.14}_{-0.15}$ & 0 & 0 & 8.87 & $4742\pm48$ & 10 \\
Sz 82 & $1.00^{+0.02}_{-0.02}$ & $0.47^{+0.02}_{-0.02}$ & 9.69, 10.69, 11.69 & 9.57 & $39.63^{+0.13}_{-0.3}$ & 0 & $0.369^{+0.017}_{-0.017}$ & 40.00 & $4582\pm22$ & 14 \\
Sz 84 & $0.14^{+0.00}_{-0.00}$ & $0.01^{+0.01}_{-0.01}$ & 9.39, 10.39, 10.89 & 7.76 & 0 & $0.238^{+0.007}_{-0.009}$ & 0 & 0.24 & $5343\pm396$ & 12 \\
Sz 97 & $0.22^{+0.02}_{-0.02}$ & $0.24^{+0.03}_{-0.04}$ & 8.39, 9.39, 10.39 & 8.58 & $8^{+3}_{-4}$ & 0 & $1.48^{+0.18}_{-0.22}$ & 9.50 & $3510\pm376$ & 8 \\
Sz 98 & $0.44^{+0.01}_{-0.01}$ & $0.47^{+0.01}_{-0.01}$ & 8.83, 9.83, 10.83 & 9.22 & 0 & $25.1^{+0.6}_{-0.9}$ & 0 & 25.09 & $4349\pm78$ & 7 \\
Sz 99 & $0.05^{+0.00}_{-0.00}$ & $0.01^{+0.01}_{-0.00}$ & 8.70, 9.70, 10.70 & 8.29 & $7.0^{+1.0}_{-1.3}$ & 0 & $0.328^{+0.012}_{-0.014}$ & 7.31 & $3624\pm187$ & 10 \\
Sz 100 & $0.22^{+0.02}_{-0.02}$ & $0.40^{+0.03}_{-0.03}$ & 9.76, 10.26, 10.76 & 8.07 & $0.60^{+0.7}_{-0.21}$ & $0.47^{+0.08}_{-0.18}$ & 0 & 1.06 & $4502\pm903$ & 12 \\
Sz 103 & $0.20^{+0.01}_{-0.01}$ & $0.53^{+0.04}_{-0.04}$ & 9.06, 10.06, 10.56 & 8.17 & $6.9^{+0.6}_{-0.8}$ & 0 & $0.20^{+0.04}_{-0.05}$ & 7.10 & $3384\pm150$ & 8 \\
Sz 104 & $0.30^{+0.01}_{-0.01}$ & $0.56^{+0.03}_{-0.04}$ & 9.54, 10.04, 10.54 & 8.33 & $0.74^{+0.7}_{-0.06}$ & 0 & $0.54^{+0.04}_{-0.08}$ & 1.28 & $4882\pm711$ & 12 \\
Sz 110 & $0.25^{+0.04}_{-0.03}$ & $0.19^{+0.04}_{-0.05}$ & 8.60, 9.60, 10.60 & 8.70 & $24.7^{+2.8}_{-6}$ & $2.7^{+3}_{-0.4}$ & $0.76^{+0.11}_{-0.12}$ & 28.17 & $3439\pm288$ & 7 \\
Sz 111 & $0.06^{+0.00}_{-0.00}$ & $0.14^{+0.02}_{-0.02}$ & 9.24, 10.24, 11.24 & 8.70 & $24.2^{+0.9}_{-0.9}$ & 0 & $0.052^{+0.008}_{-0.008}$ & 24.25 & $3563\pm48$ & 5 \\
Sz 117 & $0.15^{+0.01}_{-0.01}$ & $0.10^{+0.02}_{-0.02}$ & 8.99, 9.99, 10.99 & 8.67 & $6.0^{+1.4}_{-1.5}$ & $3.8^{+0.3}_{-0.4}$ & 0 & 9.85 & $3772\pm171$ & 9 \\
Sz 129 & $0.56^{+0.02}_{-0.02}$ & $0.76^{+0.02}_{-0.02}$ & 9.84, 10.84, 11.84 & 9.78 & $39.51^{+0.24}_{-0.6}$ & 0 & $0.488^{+0.029}_{-0.028}$ & 40.00 & $5118\pm32$ & 22 \\
Sz 130 & $0.09^{+0.01}_{-0.01}$ & $0.06^{+0.01}_{-0.01}$ & 8.52, 9.52, 10.52 & 8.00 & $4.5^{+1.0}_{-1.0}$ & $2.6^{+0.4}_{-0.4}$ & 0 & 7.08 & $3200\pm157$ & 5 \\
\hline \multicolumn{11}{c}{\textbf{$\sigma$ Ori}} \\
TX Ori & $5.08^{+0.92}_{-0.84}$ & $1.02^{+0.04}_{-0.04}$ & 9.27, 10.27, 11.27 & 9.11 & $35.4^{+2.8}_{-4}$ & $3.4^{+1.5}_{-1.6}$ & $0.0111^{+0.019}_{-0.0009}$ & 38.83 & $4579\pm193$ & 5 \\
V505 Ori & $3.40^{+0.07}_{-0.12}$ & $2.00^{+0.01}_{-0.02}$ & 10.25, 11.25, 12.25 & 9.95 & $18.5^{+0.7}_{-0.8}$ & 0 & $0.317^{+0.012}_{-0.020}$ & 18.83 & $6003\pm95$ & 32 \\
V510 Ori & $2.65^{+0.15}_{-0.16}$ & $0.49^{+0.04}_{-0.05}$ & 11.26, \dots, 11.76 & 10.09 & 0 & \dots & $2.16^{+0.13}_{-0.14}$ & 2.16 & $10260\pm247$ & 34 \\
\hline \multicolumn{11}{c}{\textbf{Orion OB1b}} \\
CVSO 58 & $3.95^{+0.34}_{-0.32}$ & $1.86^{+0.05}_{-0.06}$ & 10.78, 11.78, 12.78 & 9.96 & $2.7^{+0.4}_{-0.5}$ & $0.48^{+0.09}_{-0.08}$ & $0.079^{+0.010}_{-0.010}$ & 3.26 & $8559\pm420$ & 43 \\
CVSO 90 & $2.65^{+0.25}_{-1.76}$ & $1.04^{+0.06}_{-0.85}$ & 11.22, 12.22, 13.22 & 10.20 & 0 & $1.0^{+0.5}_{-0.4}$ & 0 & 0.97 & $12390\pm7294$ & 47 \\
CVSO 104 & $2.32^{+0.16}_{-0.15}$ & $1.24^{+0.03}_{-0.03}$ & 11.13, 12.13, 13.13 & 9.73 & 0 & $0.339^{+0.009}_{-0.011}$ & 0 & 0.35 & $11884\pm202$ & 46 \\
CVSO 107 & $0.65^{+0.02}_{-0.01}$ & $0.01^{+0.01}_{-0.00}$ & 9.37, 10.37, 11.37 & 9.33 & $38.5^{+0.8}_{-10}$ & 0 & $0.533^{+0.011}_{-0.009}$ & 39.03 & $4034\pm409$ & 13 \\
CVSO 109 A & $23.99^{+0.98}_{-1.04}$ & $1.30^{+0.02}_{-0.02}$ & 10.93, 11.93, 12.93 & 9.95 & 0 & $0.698^{+0.010}_{-0.013}$ & $0.035^{+0.005}_{-0.004}$ & 0.73 & $11732\pm136$ & 45 \\
CVSO 146 & $0.98^{+0.04}_{-0.04}$ & $0.58^{+0.02}_{-0.03}$ & 9.38, 10.38, 11.38 & 9.48 & $39.10^{+0.13}_{-0.4}$ & 0 & $0.89^{+0.05}_{-0.05}$ & 40.00 & $4644\pm27$ & 13 \\
CVSO 165 A & $0.18^{+0.06}_{-0.01}$ & $0.17^{+0.02}_{-0.02}$ & 8.92, 9.92, 10.92 & 8.43 & $22.7^{+0.7}_{-4}$ & 0 & $0.096^{+0.024}_{-0.04}$ & 22.79 & $4597\pm323$ & 4 \\
CVSO 165 B & $0.22^{+0.00}_{-0.01}$ & $0.01^{+0.00}_{-0.00}$ & 10.00, 11.00, 12.00 & 8.86 & 0 & $0.654^{+0.025}_{-0.027}$ & 0 & 0.66 & $7906\pm123$ & 22 \\
CVSO 176 & $0.93^{+0.02}_{-0.03}$ & $0.27^{+0.02}_{-0.02}$ & 10.18, 10.68, 11.18 & 8.47 & $1.85^{+0.10}_{-0.12}$ & 0 & $0.011^{+0.007}_{-0.007}$ & 1.86 & $4906\pm117$ & 10 \\
\hline \multicolumn{11}{c}{\textbf{CrA}} \\
RX J1842.9-3532 & $0.48^{+0.01}_{-0.01}$ & $0.60^{+0.00}_{-0.00}$ & 10.55, 11.55, 12.55 & 10.07 & $15.8^{+0.5}_{-0.6}$ & 0 & $0.174^{+0.003}_{-0.004}$ & 15.96 & $7170\pm87$ & 35 \\
RX J1852.3-3700 & $0.20^{+0.00}_{-0.00}$ & $0.01^{+0.00}_{-0.00}$ & 9.46, 10.46, 11.46 & 8.74 & $15.5^{+0.8}_{-0.9}$ & $0.39^{+0.07}_{-0.07}$ & 0 & 15.88 & $4403\pm91$ & 6 \\
\hline \multicolumn{11}{c}{\textbf{$\epsilon$ Cha}} \\
HD 104237 E & $0.02^{+0.00}_{-0.00}$ & $1.00^{+0.00}_{-0.00}$ & 8.64, 9.64, 10.64 & 8.39 & $28.3^{+1.6}_{-2.1}$ & 0 & $0.29^{+0.09}_{-0.11}$ & 28.63 & $4228\pm108$ & 4 \\
\hline \multicolumn{11}{c}{\textbf{$\eta$ Cha}} \\
RECX 9 & $0.13^{+0.00}_{-0.01}$ & $0.01^{+0.01}_{-0.00}$ & 9.74, 10.24, 11.24 & 7.69 & 0 & $0.284^{+0.010}_{-0.11}$ & 0 & 0.29 & $5068\pm675$ & 14 \\
RECX 11 & $0.01^{+0.02}_{-0.00}$ & $0.01^{+0.00}_{-0.00}$ & 8.12, 9.12, 10.12 & 7.30 & $15.39^{+0.27}_{-5}$ & 0 & 0 & 15.39 & $3960\pm434$ & 3 \\
RECX 15 & $0.26^{+0.01}_{-0.01}$ & $0.34^{+0.03}_{-0.03}$ & 9.74, 10.74, 11.74 & 8.93 & 0 & 0 & $0.156^{+0.008}_{-0.008}$ & 0.16 & $9774\pm2224$ & 31 \\
RECX 16 & $0.005^{+0.000}_{-0.000}$ & $0.10^{+0.00}_{-0.01}$ & 10.49, 10.99, 11.99 & 7.51 & 0 & $0.015^{+0.003}_{-0.003}$ & $0.0017^{+0.000}_{-0.000}$ & 0.02 & $8074\pm831$ & 28 \\
\enddata
 \tablenotetext{}{
Results from the final iteration of accretion shock model fitting to \hst/STIS continua for the ULLYSES sample. Objects are grouped by region in order of median region age. In most cases,  $\log_{10}\curf_{\rm med}$ is the mean $\curf$ calculated from the flow modeling ($\curf_{\rm flow}$) as described in Section~\ref{subsec:flowmodel}. For a few CTTS,  the models failed for those values, so the final $\log_{10}\curf_{\rm med}$ is shifted to a value that produces successful model calculations. When the best-fit surface coverage of an accretion hotspot $f$ is negligible,  we mark it as 0.
}
\end{deluxetable*}
\end{longrotatetable}

\startlongtable
\begin{deluxetable*}{lcccccc}
\tablecaption{Luminosity measurements \label{tab:Lacc}}
\tablehead{
\colhead{Object} & \colhead{$\log_{10}$\lacc} & \colhead{$\log_{10}$\Lhalpha} & \colhead{${\rm \log_{10}L_{SiIV}}$} & \colhead{${\rm \log_{10}L_{CIV}}$} & \colhead{${\rm \log_{10}L_{HeII}}$} & \colhead{${\rm \log_{10}L_{U,ex}}$} \\
\colhead{} & \colhead{[\lsun]} & \colhead{[\lsun]} & \colhead{[\lsun]} & \colhead{[\lsun]} & \colhead{[\lsun]} & \colhead{[\lsun]}
}
\startdata
\multicolumn{7}{c}{\textbf{Taurus}} \\
AA Tau & $-0.47^{+0.01}_{-0.01}$ & \dots & $-3.31\pm0.01$ & $-3.24\pm0.00$ & $-3.64\pm0.00$ & $-1.67\pm0.06$ \\
DE Tau & $-0.85^{+0.02}_{-0.02}$ & \dots & $-4.41\pm0.01$ & $-3.78\pm0.01$ & $-4.68\pm0.02$ & $-1.93\pm0.04$ \\
DG Tau A & $-0.59^{+0.03}_{-0.03}$ & \dots & $-4.48\pm0.01$ & $-3.42\pm0.00$ & $-3.82\pm0.01$ & $-1.41\pm0.01$ \\
DK Tau A & $-1.37^{+0.05}_{-0.04}$ & \dots & $-6.06\pm0.02$ & $-5.49\pm0.01$ & $-5.70\pm0.01$ & $-2.41\pm0.05$ \\
DM Tau ep. 1 & $-0.78^{+0.02}_{-0.02}$ & $-2.26\pm0.00$ & \dots & \dots & \dots & $-1.81\pm0.01$ \\
DM Tau ep. 3 & $-0.77^{+0.02}_{-0.02}$ & $-2.36\pm0.00$ & \dots & \dots & \dots & $-1.83\pm0.01$ \\
DM Tau ep. 2 & $-0.63^{+0.02}_{-0.02}$ & $-2.19\pm0.00$ & \dots & \dots & \dots & $-1.66\pm0.01$ \\
DN Tau & $-1.50^{+0.01}_{-0.01}$ & \dots & $-7.20\pm0.01$ & $-6.17\pm0.00$ & $-6.34\pm0.00$ & $-2.68\pm0.05$ \\
UX Tau A & $-1.01^{+0.01}_{-0.01}$ & \dots & $-3.10\pm0.01$ & $-2.53\pm0.00$ & $-2.99\pm0.01$ & $-1.52\pm0.13$ \\
\hline \multicolumn{7}{c}{\textbf{Cha I}} \\
2MASS J11432669-7804454 & $-1.74^{+0.03}_{-0.02}$ & $-3.55\pm0.03$ & \dots & $-5.39\pm0.01$ & $-5.56\pm0.02$ & $-2.66\pm0.13$ \\
CHX18N & $-2.19^{+0.02}_{-0.01}$ & $-2.94\pm0.01$ & $-5.85\pm0.02$ & $-4.93\pm0.01$ & $-5.02\pm0.01$ & $-2.25\pm0.04$ \\
CS Cha & $-0.77^{+0.02}_{-0.02}$ & $-2.08\pm0.00$ & $-4.29\pm0.01$ & $-3.89\pm0.01$ & $-3.96\pm0.01$ & $-1.71\pm0.08$ \\
CV Cha & $0.23^{+0.01}_{-0.01}$ & \dots & $-2.85\pm0.01$ & $-2.31\pm0.01$ & $-3.29\pm0.08$ & $-0.82\pm0.03$ \\
Hn 5 & $-2.35^{+0.02}_{-0.02}$ & $-3.83\pm0.00$ & $-7.76\pm0.71$ & $-6.84\pm0.06$ & \dots & $-3.51\pm0.12$ \\
IN Cha & $-2.71^{+0.02}_{-0.02}$ & $-4.17\pm0.01$ & \dots & $-6.40\pm0.07$ & $-6.69\pm0.15$ & $-3.96\pm0.16$ \\
SY Cha & $-0.83^{+0.02}_{-0.01}$ & $-2.86\pm0.00$ & $-5.57\pm0.01$ & $-5.01\pm0.01$ & $-5.01\pm0.01$ & $-1.88\pm0.01$ \\
Sz 10 & $-2.13^{+0.02}_{-0.02}$ & $-3.59\pm0.00$ & $-6.09\pm0.06$ & $-4.80\pm0.01$ & $-5.22\pm0.02$ & $-3.41\pm0.06$ \\
Sz 19 & $-0.27^{+0.03}_{-0.03}$ & $-1.88\pm0.00$ & $-5.77\pm0.02$ & $-4.77\pm0.01$ & $-5.42\pm0.02$ & $-0.77\pm0.04$ \\
Sz 45 ep. 1 & $-1.47^{+0.02}_{-0.03}$ & $-2.91\pm0.01$ & \dots & \dots & \dots & $-2.36\pm0.07$ \\
Sz 45 ep. 5 & $-1.43^{+0.02}_{-0.02}$ & $-2.74\pm0.00$ & \dots & \dots & \dots & $-2.37\pm0.06$ \\
Sz 45 ep. 2 & $-1.42^{+0.02}_{-0.02}$ & $-2.81\pm0.01$ & \dots & \dots & \dots & $-2.34\pm0.06$ \\
Sz 45 ep. 4 & $-1.41^{+0.03}_{-0.01}$ & $-2.88\pm0.01$ & \dots & \dots & \dots & $-2.41\pm0.04$ \\
Sz 45 ep. 3 & $-1.39^{+0.02}_{-0.02}$ & $-2.76\pm0.00$ & \dots & \dots & \dots & $-2.33\pm0.05$ \\
Sz 45 ep. 6 & $-1.31^{+0.02}_{-0.02}$ & $-2.88\pm0.00$ & \dots & \dots & \dots & $-2.20\pm0.05$ \\
SZ Cha & $-2.15^{+0.13}_{-0.14}$ & $-2.85\pm0.02$ & \dots & \dots & \dots & $-2.20\pm0.28$ \\
VZ Cha & $-0.39^{+0.04}_{-0.03}$ & $-2.37\pm0.00$ & $-3.62\pm0.01$ & $-2.62\pm0.00$ & $-3.01\pm0.01$ & $-1.41\pm0.04$ \\
WZ Cha & $-0.91^{+0.03}_{-0.02}$ & $-2.54\pm0.00$ & \dots & \dots & \dots & $-1.93\pm0.04$ \\
XX Cha & $-1.60^{+0.04}_{-0.03}$ & $-3.05\pm0.00$ & $-5.12\pm0.01$ & $-4.62\pm0.01$ & $-4.80\pm0.01$ & $-2.34\pm0.01$ \\
\hline \multicolumn{7}{c}{\textbf{Lupus}} \\
MY Lup & $-1.85^{+0.02}_{-0.02}$ & \dots & $-5.58\pm0.02$ & $-4.71\pm0.01$ & $-4.94\pm0.01$ & $-2.76\pm0.59$ \\
RX J1556.1-3655 & $-0.92^{+0.01}_{-0.01}$ & $-2.66\pm0.00$ & $-4.94\pm0.00$ & $-4.31\pm0.00$ & $-4.76\pm0.01$ & $-1.99\pm0.01$ \\
RY Lup & $-1.98^{+0.06}_{-0.06}$ & \dots & $-5.17\pm0.02$ & $-4.24\pm0.01$ & $-5.22\pm0.02$ & $-2.77\pm0.17$ \\
SSTc2dJ160830.7-382827 & $-2.49^{+0.10}_{-0.11}$ & \dots & $-5.89\pm0.04$ & $-4.70\pm0.01$ & $-5.02\pm0.01$ & $-2.86\pm0.47$ \\
SSTc2dJ160000.6-422158 & $-2.87^{+0.02}_{-0.02}$ & $-4.78\pm0.01$ & $-7.70\pm1.65$ & $-6.49\pm0.05$ & $-6.78\pm0.09$ & $-4.38\pm0.21$ \\
SSTc2dJ161243.8-381503 & $-2.20^{+0.02}_{-0.02}$ & $-3.88\pm0.01$ & $-6.30\pm0.02$ & $-5.41\pm0.01$ & $-5.79\pm0.01$ & $-3.37\pm0.09$ \\
SSTc2dJ161344.1-373646 & $-2.24^{+0.01}_{-0.01}$ & $-4.10\pm0.00$ & $-7.04\pm0.05$ & $-5.16\pm0.00$ & $-5.79\pm0.01$ & $-3.57\pm0.02$ \\
Sz 66 & $-2.25^{+0.02}_{-0.02}$ & $-3.66\pm0.00$ & $-6.36\pm0.02$ & $-7.11\pm0.06$ & $-6.87\pm0.04$ & $-3.41\pm0.07$ \\
Sz 68 & $-0.78^{+0.01}_{-0.01}$ & $-2.36\pm0.00$ & $-5.14\pm0.02$ & $-4.30\pm0.01$ & $-4.72\pm0.01$ & $-1.46\pm0.10$ \\
Sz 69 & $-2.86^{+0.02}_{-0.01}$ & $-3.40\pm0.00$ & $-5.66\pm0.03$ & \dots & \dots & $-3.93\pm0.09$ \\
Sz 71 & $-1.87^{+0.04}_{-0.04}$ & $-3.33\pm0.00$ & $-6.38\pm0.04$ & $-5.29\pm0.01$ & $-5.45\pm0.01$ & $-3.18\pm0.06$ \\
Sz 72 & $-1.05^{+0.02}_{-0.02}$ & $-2.67\pm0.00$ & $-4.81\pm0.01$ & $-4.43\pm0.01$ & $-5.31\pm0.02$ & $-2.05\pm0.01$ \\
Sz 75 & $-0.76^{+0.01}_{-0.01}$ & $-2.44\pm0.00$ & $-5.47\pm0.01$ & $-4.60\pm0.00$ & $-5.01\pm0.01$ & $-1.73\pm0.01$ \\
Sz 77 & $-1.36^{+0.01}_{-0.01}$ & $-3.05\pm0.00$ & $-5.24\pm0.03$ & $-4.40\pm0.01$ & $-4.55\pm0.01$ & $-2.67\pm0.04$ \\
Sz 82 & $-0.98^{+0.01}_{-0.01}$ & $-2.52\pm0.00$ & $-4.67\pm0.01$ & $-3.25\pm0.00$ & $-3.67\pm0.01$ & $-1.96\pm0.01$ \\
Sz 84 & $-2.66^{+0.01}_{-0.01}$ & $-4.01\pm0.00$ & \dots & $-5.58\pm0.02$ & $-5.51\pm0.02$ & $-3.86\pm0.07$ \\
Sz 97 & $-2.13^{+0.04}_{-0.04}$ & $-3.56\pm0.00$ & $-5.63\pm0.03$ & $-4.92\pm0.01$ & $-5.12\pm0.01$ & $-3.37\pm0.02$ \\
Sz 98 & $-1.39^{+0.01}_{-0.01}$ & $-3.51\pm0.01$ & $-6.09\pm0.01$ & $-5.52\pm0.01$ & $-6.05\pm0.01$ & \dots \\
Sz 99 & $-2.73^{+0.02}_{-0.01}$ & $-4.15\pm0.00$ & $-6.15\pm0.04$ & $-4.96\pm0.01$ & $-5.54\pm0.02$ & $-3.73\pm0.03$ \\
Sz 100 & $-2.45^{+0.03}_{-0.03}$ & $-4.08\pm0.00$ & $-5.71\pm0.04$ & $-4.84\pm0.01$ & $-5.34\pm0.02$ & $-3.91\pm0.06$ \\
Sz 103 & $-2.41^{+0.02}_{-0.02}$ & $-4.15\pm0.00$ & $-6.07\pm0.02$ & $-6.16\pm0.02$ & $-6.28\pm0.03$ & $-3.72\pm0.06$ \\
Sz 104 & $-2.33^{+0.02}_{-0.02}$ & $-4.06\pm0.00$ & $-5.72\pm0.06$ & $-4.68\pm0.01$ & $-5.00\pm0.02$ & $-3.53\pm0.04$ \\
Sz 110 & $-2.06^{+0.07}_{-0.05}$ & $-3.50\pm0.00$ & $-5.20\pm0.06$ & $-4.53\pm0.01$ & $-4.96\pm0.02$ & $-3.10\pm0.02$ \\
Sz 111 & $-2.11^{+0.01}_{-0.01}$ & $-2.91\pm0.00$ & $-5.82\pm0.03$ & $-4.97\pm0.01$ & $-5.22\pm0.01$ & $-3.13\pm0.06$ \\
Sz 117 & $-2.03^{+0.02}_{-0.02}$ & $-3.73\pm0.00$ & $-6.49\pm0.09$ & $-5.13\pm0.01$ & $-5.51\pm0.01$ & $-3.35\pm0.04$ \\
Sz 129 & $-1.05^{+0.01}_{-0.01}$ & $-3.10\pm0.01$ & $-5.05\pm0.00$ & $-4.43\pm0.00$ & $-4.57\pm0.00$ & $-2.08\pm0.02$ \\
Sz 130 & $-2.56^{+0.04}_{-0.04}$ & $-3.74\pm0.00$ & $-6.23\pm0.08$ & $-5.23\pm0.01$ & $-5.46\pm0.02$ & $-3.66\pm0.13$ \\
\hline \multicolumn{7}{c}{\textbf{$\sigma$ Ori}} \\
TX Ori & $-0.60^{+0.08}_{-0.07}$ & $-2.26\pm0.00$ & $-4.33\pm0.04$ & $-4.16\pm0.03$ & \dots & $-1.57\pm0.07$ \\
V505 Ori & $-0.45^{+0.01}_{-0.02}$ & $-2.44\pm0.00$ & \dots & $-4.12\pm0.04$ & \dots & $-1.50\pm0.12$ \\
V510 Ori & $-0.42^{+0.02}_{-0.03}$ & $-2.11\pm0.00$ & $-3.71\pm0.01$ & $-4.05\pm0.02$ & $-4.36\pm0.03$ & $-1.27\pm0.01$ \\
\hline \multicolumn{7}{c}{\textbf{Orion OB1b}} \\
CVSO 58 & $-0.36^{+0.04}_{-0.04}$ & $-2.69\pm0.00$ & $-3.56\pm0.02$ & $-2.97\pm0.01$ & $-3.48\pm0.01$ & $-1.41\pm0.03$ \\
CVSO 90 & $-0.43^{+0.04}_{-0.29}$ & $-2.38\pm0.00$ & $-3.20\pm0.02$ & $-2.68\pm0.01$ & $-3.39\pm0.01$ & $-1.30\pm0.01$ \\
CVSO 104 & $-0.72^{+0.03}_{-0.03}$ & $-2.64\pm0.00$ & $-3.53\pm0.02$ & $-3.08\pm0.01$ & $-3.50\pm0.02$ & $-1.73\pm0.03$ \\
CVSO 107 & $-1.31^{+0.01}_{-0.00}$ & $-3.20\pm0.00$ & $-4.87\pm0.01$ & $-4.30\pm0.01$ & $-4.43\pm0.01$ & $-2.31\pm0.02$ \\
CVSO 109 A & $-0.03^{+0.02}_{-0.02}$ & $-2.20\pm0.00$ & $-2.69\pm0.01$ & $-2.33\pm0.01$ & $-2.91\pm0.02$ & $-0.97\pm0.02$ \\
CVSO 146 & $-0.99^{+0.02}_{-0.02}$ & $-2.68\pm0.00$ & $-4.68\pm0.02$ & $-4.04\pm0.01$ & $-4.38\pm0.01$ & $-1.92\pm0.02$ \\
CVSO 165 A & $-1.91^{+0.14}_{-0.03}$ & $-3.10\pm0.01$ & $-4.39\pm0.01$ & $-4.08\pm0.01$ & $-4.66\pm0.02$ & $-2.39\pm0.06$ \\
CVSO 165 B & $-1.66^{+0.01}_{-0.01}$ & $-3.11\pm0.00$ & \dots & \dots & \dots & $-2.51\pm0.05$ \\
CVSO 176 & $-1.73^{+0.01}_{-0.01}$ & $-3.12\pm0.00$ & $-5.92\pm0.02$ & $-5.37\pm0.01$ & $-5.97\pm0.03$ & $-3.01\pm0.10$ \\
\hline \multicolumn{7}{c}{\textbf{CrA}} \\
RX J1842.9-3532 & $-0.85^{+0.01}_{-0.01}$ & $-2.42\pm0.00$ & \dots & \dots & \dots & $-1.92\pm0.03$ \\
RX J1852.3-3700 & $-1.78^{+0.01}_{-0.01}$ & $-3.09\pm0.00$ & \dots & \dots & \dots & $-2.67\pm0.04$ \\
\hline \multicolumn{7}{c}{\textbf{$\epsilon$ Cha}} \\
HD 104237 E & $-2.53^{+0.07}_{-0.06}$ & $-3.68\pm0.01$ & \dots & \dots & \dots & $-3.00\pm0.09$ \\
\hline \multicolumn{7}{c}{\textbf{$\eta$ Cha}} \\
RECX 9 & $-2.90^{+0.02}_{-0.03}$ & $-4.47\pm0.00$ & \dots & \dots & \dots & $-4.09\pm0.05$ \\
RECX 11 & $-3.18^{+0.90}_{-0.01}$ & \dots & $-5.55\pm0.01$ & $-4.42\pm0.00$ & $-4.88\pm0.00$ & $-2.76\pm0.05$ \\
RECX 15 & $-1.97^{+0.02}_{-0.02}$ & \dots & $-4.62\pm0.01$ & $-4.58\pm0.00$ & $-4.88\pm0.00$ & $-2.91\pm0.05$ \\
RECX 16 & $-3.91^{+0.02}_{-0.02}$ & $-4.94\pm0.00$ & $-7.23\pm0.12$ & $-6.20\pm0.02$ & $-6.87\pm0.04$ & $-4.70\pm0.09$ \\
\enddata
 \tablenotetext{}{
 Here we report accretion luminosity (\lacc) from the accretion shock model; \halpha\ luminosity (\Lhalpha) from dereddened HST/STIS spectra; $\rm {SiIV}$, $\rm {CIV}$, and $\rm {HeII}$ line luminosities measured from HST/COS spectra in \cite{france23}, adjusted to the extinction \av\ measured in the shock modeling; and excess $U$-band luminosity (${\rm L_{U,ex}}$) measured from dereddened HST/STIS spectra. FUV line luminosities for CV~Cha, VZ~Cha, Sz~10, and Sz~82 were measured in this work.
}
\end{deluxetable*}

\end{document}